\documentclass[fleqn]{annalen}

\newcommand{\ds}{{\vphantom{\dagger}}}
\newcommand{\etalia}{{\em et al.}}         % et al. 
\newcommand{\ia}{\emph{i.a.}}              % inter alia (among other things)
\newcommand{\ie}{i.e.}                     % that is
\newcommand{\eg}{e.g.}                     % for example  
\newcommand{\gc}{g.c.}                     % grand canonical
\newcommand{\Eq}[1]{Eq.~(\ref{#1})}        % Eq. (#) Equation         
\newcommand{\Eqs}[1]{Eqs.~(\ref{#1})}      % Eqs. (#) Equations
\newcommand{\Sec}[1]{Sec.~\ref{#1}}        % Sec. #  Section
\newcommand{\Secs}[1]{Secs.~\ref{#1}}      % Secs. # Sections
\newcommand{\Fig}[1]{Fig.~\ref{#1}}        % Fig. # Figure
\newcommand{\Figs}[1]{Figs.~\ref{#1}}      % Figs. # Figures
\newcommand{\Ref}[1]{Ref.~\cite{#1}}      % Ref. (#) Reference
\newcommand{\ii}{{\rm i}}    % imaginary i
\newcommand{\re}{{\mathop{\mathrm{Re}}}}   % Real part 
\newcommand{\ssg}{{\mathop{\mathrm{g}}}}   % math subscript g (gate)
\newcommand{\ssL}{{\mathop{\mathrm{L}}}}   % math subscript L (left)
\newcommand{\ssR}{{\mathop{\mathrm{R}}}}   % math subscript R (right)
\newcommand{\ssD}{{\mathop{\mathrm{D}}}}   % math subscript D (dot)
\newcommand{\typ}{{\mathop{\mathrm{typ}}}} % math sub. tun (typical)
\newcommand{\orb}{{\mathop{\mathrm{orb}}}} % math sub. orb (orbital)
\newcommand{\bulk}{{\mathop{\mathrm{bulk}}}} % bulk
\newcommand{\pert}{{\mathop{\mathrm{pert}}}} % pert (perturbative)
\newcommand{\can}{{\mathop{\mathrm{can}}}} % math subscript for canonical
\newcommand{\kF}{{\mathop{k_{\mathrm{F}}}}}% Fermi momentum k_F
\newcommand{\vF}{{\mathop{v_{\mathrm{F}}}}}% Fermi velocity v_F
\newcommand{\eF}{{\mathop{\varepsilon_{\mathrm{F}}}}} % Fermi ener. e_F
\newcommand{\kB}{{\mathop{k_{\mathrm{B}}}}}  % Boltzmann constant k_B
\newcommand{\Vol}{{\mathop{\mathrm{Vol}}}}   % abbreviation for Volume
\newcommand{\Ec}{{\mathop{E_{\mathrm{C}}}}}  % charging energy E_C
\newcommand{\ETh}{{\mathop{E_{\mathrm{Thouless}}}}}
\newcommand{\Epot}{{\mathop{E_{\mathrm{pot}}}}}% elec. pot. en. E_pot
\newcommand{\Vg}{{\mathop{V_{\mathrm{g}}}}}  % gage voltage V_g
\newcommand{\Tc}{{\mathop{T_{\mathrm{c}}}}}  % Tc, critical temperature
\newcommand{\muB}{{\mathop{\mu_{\mathrm{B}}}}} % Bohr magneton \mu_B 
\newcommand{\VD}{{\mathop{V_{\mathrm{D}}}}}  % potential on dot V_D
\newcommand{\Nex}{{\mathop{N_{\mathrm{ex}}}}}% excess elec. number N_ex
\newcommand{\Nexsq}{{\mathop{N^2_{\mathrm{ex}}}}}%ex.el.numb.sq. N^2_ex
\newcommand{\E}{{\mathcal{E}}}               % energy
\newcommand{\op}[1]{{\hat{#1}}}              % operator
\newcommand{\half}{{\textstyle {1 \over 2}}}
\newcommand{\ddr}{{\mathrm{d}}}              % infinitessimal d
\newcommand{\eer}{{\mathrm{e}}}              % exponential e
\newcommand{\iir}{{\mathrm{i}}}              % imaginary i
\newcommand{\bbalpha}{{\B}}         % bold alpha

\newcommand{\cond}{{\mathrm{cond}}}             % cond (condensation)
\newcommand{\PBCS}{{\mathrm{PBCS}}}          % Projected BCS
\newcommand{\PP}{{\mathrm{P}}}               % P (parity)
\newcommand{\pb}{{\mathrm{pb}}}              % pb (pair-breaking)
\newcommand{\GC}{{\mathrm{BCS}}}             % BCS
\newcommand{\BCS}{{\mathrm{BCS}}}            % BCS
\newcommand{\Pauli}{{\mathrm{Pauli}}}        % Pauli
\newcommand{\MF}{{\mathrm{gc}}}            % grand-canonical 
\newcommand{\G}{{\mathrm{G}}}              % G, for exact ground state
\newcommand{\F}{{\mathrm{F}}}              % F, for Fermi sea
\newcommand{\red}{{\mathrm{red}}}          % reduced 
\newcommand{\ren}{{\mathrm{ren}}}          % renormalized 
\newcommand{\ML}{{\mathrm{ML}}}            % Matveev-Larkin
\newcommand{\Vac}{{\mathrm{Vac}}}          % Vacuum
\newcommand{\ex}{{\mathrm{exact}}}         % exact
\newcommand{\omegaD}{{\mathop{\omega_{\mathrm{D}}}}} 
\newcommand{\baromegaD}{{\mathop{\bar \omega_{\mathrm{D}}}}} 
\newcommand{\CC}{{\mathrm{CC}}}      % Clogston-Chandrasekhar
\newcommand{\gol}{\stackrel{\scriptscriptstyle >}{\scriptscriptstyle
<}}                                        % ><
\newcommand{\N}{{\mathcal N}}          % calig. N: density of states
\newcommand{\I}{I}          % I: set of interacting levels
\newcommand{\B}{B}          % B: set of blocked levels
\newcommand{\U}{U}          % U: set of unblocked levels
\newcommand{\dbcsm}{discrete BCS model} % name of BCS model
\usepackage{amssymb}
\usepackage{amsfonts}
\usepackage{epsfig}

\begin{document}
%%%%%%%%%%%%%%%%%%%%%%%%%%%%%%%%%%%%%%%%%%%%%%%%%%%%%%%%%%%%%%%%%%%%%%%%%%%%%%
%%%%%%%% the following new commands will be completed by the publisher %%%%%%%%
%%%%%%%%%%%%%%%%%%%%%%%%%%%%%%%%%%%%%%%%%%%%%%%%%%%%%%%%%%%%%%%%%%%%%%%%%%%%%%
\newcommand{\volume}{10}             %sets current volume,
\newcommand{\xyear}{2001}            %sets year in header
\newcommand{\issue}{3}               %sets current issue,
\newcommand{\recdate}{01.08.2000}    %sets received date,
\newcommand{\revdate}{dd.mm.yyyy}    %sets revised date,     
\newcommand{\revnum}{0}              %number of revisions,
\newcommand{\accdate}{22.08.2000}    %sets accepted date,
\newcommand{\coeditor}{ue}           %sets (co)editor,
\newcommand{\firstpage}{1}           %first page number,  
\newcommand{\lastpage}{60}            %last page number,
\setcounter{page}{\firstpage}        %sets page counter to first page number 
%%%%%%%%%%%%%%%%%%%%%%%%%%%%%%%%%%%%%%%%%%%%%%%%%%%%%%%%%%%%%%%%%%%%%%%%%%%%%%
%%%%%%%%%%%%%%%%%%%%%%%%%%%%%%%%%%%%%%%%%%%%%%%%%%%%%%%%%%%%%%%%%%%%%%%%%%%%%%
%%%%%%%%%%%%%%%%%% please give up to three keywords here %%%%%%%%%%%%%%%%%%%%%
%%%%%%%%%%%%%%%%%%%%%%%%%%%%%%%%%%%%%%%%%%%%%%%%%%%%%%%%%%%%%%%%%%%%%%%%%%%%%%
\newcommand{\keywords}{superconductivity, metallic grains } 
%%%%%%%%%%%%%%%%%%%%%%%%%%%%%%%%%%%%%%%%%%%%%%%%%%%%%%%%%%%%%%%%%%%%%%%%%%%%%%
%%%%%%%%%%%%%%%% please give up to three PACS numbers here %%%%%%%%%%%%%%%%%%%
%%%%%%%%%%%%%%%%%%%%%%%%%%%%%%%%%%%%%%%%%%%%%%%%%%%%%%%%%%%%%%%%%%%%%%%%%%%%%%
\newcommand{\PACS}{74.20.Fg, 74.20.-z, 74.25.Ha, 74.80.Fp}
%%%%%%%%%%%%%%%%%%%%%%%%%%%%%%%%%%%%%%%%%%%%%%%%%%%%%%%%%%%%%%%%%%%%%%%%%%%%%%
%% please enter (First) Author (et al.) and short version of the title here %%
%%%%%%%%%%%% must not exceed 80 characters in length together %%%%%%%%%%%%%%%%
%%%%%%%%%%%%%%%%%%%%%%%%%%%%%%%%%%%%%%%%%%%%%%%%%%%%%%%%%%%%%%%%%%%%%%%%%%%%%%
\newcommand{\shorttitle}{Jan von Delft, Superconductivity
in ultrasmall metallic grains} %% sets the header on oddpage
%%%%%%%%%%%%%%%%%%%%%%%%%%%%%%%%%%%%%%%%%%%%%%%%%%%%%%%%%%%%%%%%%%%%%%%%%%%%%%
%%%%%%%%%%%%%%%%%%%%%%%% here comes the title group %%%%%%%%%%%%%%%%%%%%%%%%%%
%%%%%%%%%%%%%%%%%%%%%%%%%%%%%%%%%%%%%%%%%%%%%%%%%%%%%%%%%%%%%%%%%%%%%%%%%%%%%%
\title{Superconductivity in ultrasmall metallic grains
\footnote{Part of the
author's habilitation thesis, approved by the Faculty of Physics of the
University of Karlsruhe, July 2000.}}
%%%%%%%%%%%%%%%%%%%%%%%%%%%%%%%%%%%%%%%%%%%%%%%%%%%%%%%%%%%%%%%%%%%%%%%%%%%%%%
\author{Jan von Delft}
%%%%%%%%%%%%%%%%%%%%%%%%%%%%%%%%%%%%%%%%%%%%%%%%%%%%%%%%%%%%%%%%%%%%%%%%%%%%%%
\newcommand{\address}
  {Institut f\"ur Theoretische Festk\"orperphysik \\
    Universit\"at   Karlsruhe \\  76128 Karlsruhe \\ Germany 
% \\ \hspace*{0.5mm} 
}
%%%%%%%%%%%%%%%%%%%%%%%%%%%%%%%%%%%%%%%%%%%%%%%%%%%%%%%%%%%%%%%%%%%%%%%%%%%%%%
\newcommand{\email}{\tt vondelft@th.physik.uni-bonn.de} 
\maketitle
%%%%%%%%%%%%%%%%%%%%%%%%%%%%%%%%%%%%%%%%%%%%%%%%%%%%%%%%%%%%%%%%%%%%%%%%%%%%%
\begin{abstract}
  We review recent experimental and theoretical work on superconductivity in
  ultrasmall metallic grains, i.e.\ grains sufficiently small that the
  conduction electron energy spectrum becomes discrete.  The discrete
  excitation spectrum of an individual grain can be measured by the technique
  of single-electron tunneling spectroscopy, and reveals parity effects
  indicative of pairing correlations in the grain. After introducing the
  discrete BCS model that has been used to model such grains, we review a
  phenomenological, grand-canonical, variational BCS theory describing the
  paramagnetic breakdown of these pairing correlations with increasing
  magnetic field.  We also review recent canonical theories that have
  been developed to describe how
  pairing correlations change during the crossover, with decreasing grain
  size, from the bulk limit to the limit of few electrons, and compare their
  results to those obtained using Richardson's exact solution
of the discrete BCS model.
\end{abstract}
%%%%%%%%%%%%%%%%%%%%%%%%%%%%%%%%%%%%%%%%%%%%%%%%%%%%%%%%%%%%%%%%%%%%%%%%%%%%%

%\newpage

\tableofcontents

\newpage

\section{Introduction}
\label{sec:introduction}

Since its discovery  by Kammerlingh Onnes in 1911,
superconductivity has become one of the most-studied
phenomena in condensed matter physics, and its microscopic explanation via the
highly successful pairing theory proposed in 1957 by Bardeen, Cooper and
Schrieffer (BCS) \cite{BCS-57} is one of the landmark achievements of 20th
century physics. Yet, despite the long history of the subject, to this day
experimental advances in sample fabrication and measurement techniques
continue to reveal novel aspects of superconductivity, which often require
extensions or modifications of the existing theoretical framework.

The subject of this review, superconductivity in ultrasmall metallic grains,
is a case in point: In the mid 1990's, Ralph, Black and Tinkham (RBT)
succeeded for the first time to directly measure the \emph{discrete excitation
  spectrum of individual ultrasmall metallic grains} (of radii $r
\lesssim 5$~nm and mean level spacings $d \gtrsim 0.1$~meV) using a technique
called single-electron-tun\-ne\-ling spectroscopy: by attaching such a grain
via oxide tunnel barriers to two leads they constructed a single-electron
transistor having the grain as central island, and showed that a
well-resolved, discrete excitation spectrum could indeed be extracted from the
conductance \vphantom{
\cite{rbt95,rbt96a,rbt96b,black-thesis,ralph-curacao,rbt97}}
\cite{rbt95}-\cite{rbt97}.

This opened up a new frontier in the study of electron correlations in metals,
since the ability to resolve discrete energy levels allows the nature of
electron correlations to be studied in unprecedented detail.  Since 1995,
RBT's technique has been used to probe superconducting pairing correlations in
Al grains \cite{rbt96a,rbt97}, nonequilibrium excitations
\vphantom{\cite{rbt97,agam97a,agam97b,agam98}}
\cite{rbt97}-\cite{agam98}
and spin-orbit interactions
\cite{rbt95,rbt96b,salinas99,davidovich99,davidovich00} in normal grains, and
ferromagnetic correlations in Co grains \cite{desmicht98,gueron99}.  A
comprehensive survey of all experimental and theoretical developments (up to
March 2000) relating to spectroscopic studies of ultrasmall metallic grains
may be found in the review of von Delft and Ralph \cite{PR-VDR}. The present
review is an excerpt  of \Ref{PR-VDR}, and
is devoted exclusively to \emph{superconductivity} (more precisely, to
pairing correlations) in ultrasmall metallic grains.

For several reasons, RBT's experiments on superconducting pairing correlations
in ultrasmall Al grains attracted quite some attention
\vphantom{\cite{vondelft96,braun97,braun99,smith96,balian-short,balian-long,%
bonsager98,matveev97,Rossignoli-98,Rossignoli-99a,Rossignoli-99b,%
Rossignoli-00,mastellone98,berger98,braun98,dukelsky99a,dukelsky99b,%
braun-vieweg,sierra99,vondelft-ankara99,dukelsky99c,tian99,%
tanaka99,dilorenzo99}} 
\cite{vondelft96}-\cite{dilorenzo99}:  

First, for largish ($r \gtrsim 5$ nm) Al grains, RBT's measurements revealed a
rather striking \emph{parity effect} \cite{rbt96a,rbt97,vondelft96}:
a grain with an even number of electrons
had a distinct spectroscopic gap $(\gg d)$ but an odd grain did not. This is
clear evidence for the presence of \emph{superconducting pairing correlations}
in these grains, and indicates that a BCS-like theory would be appropriate for
their description.

Second, the spectroscopic gap for even grains was driven to zero by an
applied magnetic field, hence the \emph{paramagnetic breakdown of pairing
correlations} could be studied in detail \cite{braun97,braun99}.

Third, even the ``largish'' grains were so small that standard
grand-canonical BCS mean field theory is no longer applicable: (a) the
single-particle mean level spacing $d = 1 / \N (\varepsilon_F) \sim 1 /
\mbox{Vol}$ [where $\N (\varepsilon)$ is the density of states per spin
species] was comparable to the bulk  superconducting 
gap, which we shall denote by $\tilde \Delta$, so
that a \emph{mean field approach is no longer
reliable} (it requires $d \ll  \tilde \Delta$); 
and (b), the number of electrons on such a grain is well-defined,
hence a \emph{canonical} theory is required. RBT's experiments 
stimulated the development of  corresponding extensions of BCS theory.

Fourth, in RBT's smallest grains ($r \lesssim 3$ nm), the distinct
spectroscopic gap observed for largish even grains could no longer be
unambiguously discerned.  This observation revived an old but
fundamental question: \emph{What is the lower size limit for the
  existence of superconductivity in small grains?}\/: Anderson had
addressed this question already in 1959 \cite{anderson59}, arguing
that if the sample is so small that its electronic eigenspectrum
becomes discrete, ``superconductivity would no longer be possible''
when its mean level spacing $d$ becomes larger than the bulk gap
$\tilde \Delta$. Heuristically, this is obvious (see \Fig{fig:v2u2-prb97} below):
$\tilde \Delta / d$ is the number of free-electron states that
pair-correlate (those with energies within $\tilde \Delta $ of $\eF$),
i.e. the ``number of Cooper pairs'' in the system; when this becomes
$\lesssim 1$, it clearly no longer makes sense to call the system
``superconducting''.

Although Anderson's answer is correct in general, it generates further
questions: What, precisely, does ``superconductivity'' mean in ultrasmall
grains, for which many of the standard criteria such as zero resistivity,
Meissner effect and Josephson effect, are not relevant\footnote{For an
  isolated nm-scale grain, (i) its resistivity is not defined, since electron
  motion is ballistic and the mean free path is boundary-limited; (ii) the
  grain radius is smaller than the penetration depth, so that no Meissner
  effect occurs; and (iii) the electron number is fixed, so that the order
  parameter cannot have a well-defined phase.}?  What happens in the regime $d
\gtrsim \tilde \Delta$ in which superconductivity has broken down? Is the
breakdown parity dependent?  How is it influenced by a magnetic field?  This
review attempts to provide detailed answers to these and related questions.

The review is divided into two distinct parts.  Part I
(Secs.~\ref{subsec:ultrasmallset}
 to \ref{sec:BCS-parity}) is devoted to
RBT's experiments and their detailed theoretical interpretation. After briefly
discussing the experimental setup and summarizing the main experimental
results, we analyze and qualitatively explain the latter in the framework of a
phenomenological theory by Braun \etalia\ \cite{braun97,braun99,braun-thesis}.
This theory offers a simple intuitive picture for visualizing the pairing
correlations and how these change when the grain size is decreased.

Part II (Secs.~\ref{sec:sc-richardson} to \ref{sec:finite-T}) is
devoted to further theoretical developments, inspired by RBT's experiments but
not directly concerned with their interpretation
\vphantom{\cite{vondelft96,braun97,braun99,smith96,balian-short,balian-long,%
bonsager98,matveev97,Rossignoli-98,Rossignoli-99a,Rossignoli-99b,%
Rossignoli-00,mastellone98,berger98,braun98,dukelsky99a,dukelsky99b,%
braun-vieweg,sierra99,vondelft-ankara99,dukelsky99c,tian99,%
tanaka99,dilorenzo99}} 
\cite{vondelft96}-\cite{dilorenzo99}. 
In particular,  RBT's new experiments stimulated a number of
theoretical attempts to quantitatively describe the \emph{crossover}
from the bulk limit $d \ll \tilde\Delta$, where superconductivity is
well-developed, to the fluctuation-dominated regime of $d \gg \tilde
\Delta$, where pairing correlations survive only in the form of weak
fluctuations.  Describing this crossover constituted a conceptual
challenge, since the standard grand-canonical mean-field BCS treatment
of pairing correlations
\vphantom{\cite{vondelft96,braun97,braun99,smith96,balian-short,balian-long,%
  bonsager98,matveev97}} 
\cite{vondelft96}-\cite{matveev97} 
breaks down for $d \gtrsim \tilde \Delta$.
This challenge elicited a series of increasingly sophisticated
canonical treatments of pairing correlations
\vphantom{\cite{mastellone98,berger98,braun98,%
dukelsky99a,dukelsky99b,braun-vieweg,sierra99,%
vondelft-ankara99,dukelsky99c}}
\cite{mastellone98}-\cite{dukelsky99c}, 
based on a simple reduced
BCS Hamiltonian for discrete energy levels, which showed that the
crossover is completely smooth, but, interestingly, depends on the
parity of the number of electrons on the grain 
\cite{vondelft96}.  Very recently, the main conclusions
of these works were confirmed \cite{sierra99} using an exact solution
of the discrete-level BCS model, discovered by Richardson in the context of
nuclear physics in the 1960s 
\vphantom{\cite{richardson63a,richardson63b,richardson64,%
  richardson65a,richardson65b,richardson66,richardson66-b,richardson67,%
  richardson77}}
\cite{richardson63a}-\cite{richardson77}. 
 (The existence of this solution came as a surprise
-- in the form of a polite letter from its inventor -- to those
involved with ultrasmall grains, since hitherto it had apparently
completely escaped the attention of the condensed-matter community.)

A detailed outline of the contents of the two
parts may be found in the opening paragraph  of each,
or in the table of contents. 

\newpage

{\large {\bf Part I: Experiment and phenomenological
theory}} \\
 
\label{sec:superconductivity}
%%AA%% change label?

Part I of this review 
is devoted to  RBT's experiments and their
detailed theoretical interpretation. It is organized
as follows: section  \vspace*{-3mm}
\begin{itemize}
\item[(\ref{subsec:ultrasmallset})]
is devoted to experimental details;  \vspace*{-3mm}
\item[(\ref{sec:gap-in-spectrum})]
summarizes RBT's main experimental results; \vspace*{-3mm}
\item[(\ref{sec:model})]  proposes a phenomenological model
for an isolated ultrasmall grain; \vspace*{-3mm}
\item[(\ref{sec:canonical-pair-mixing})] discusses how pairing
  correlations can be visualized in a fixed-$N$ system and explains
  when and in what sense it can be called ``superconducting''; \vspace*{-3mm}
\item[(\ref{sec:generalBCS})] presents  a generalized variational BCS
  approach for calculating the eigenenergies of various variational
  eigenstates of general spin $s$; \vspace*{-3mm}
\item[(\ref{sec:CC-transition})] discusses how an increasing magnetic
  field induces a transition from a pair-correlated state to a normal
  paramagnetic state; \vspace*{-3mm}
\item[(\ref{sec:tunneling-spectra-prb97})]
  presents theoretical tunneling spectra of the RBT type,
  which are in qualitative agreement with RBT's measurements; \vspace*{-3mm}
%\item[(\ref{sec:time-reversed})] explains how RBT's experiments give
%  direct evidence for the dominance of purely time-reversed states in
%  the pairing interaction; \vspace*{-3mm}
\item[(\ref{sec:BCS-parity})] discusses various parity effects that
  are expected to occur in ultrasmall grains.
\end{itemize}

\section{Experimental details}
\label{subsec:ultrasmallset}

\begin{figure}[b]
\centerline{\epsfig{figure=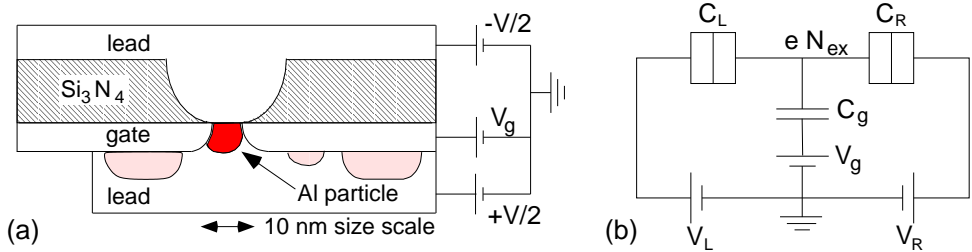,%
width=0.98\linewidth}}
  \caption[Schematic depiction of an ultrasmall SET]{
   (a) Schematic cross section of the ultrasmall SETs 
       studied by RBT in \protect\cite{rbt97}, and (b)  
   the corresponding circuit diagram.
 }
   \label{fig:sample} \label{fig:circuit-diagram}
\end{figure}

In the first generation of RBT's experiments of 1995
\cite{rbt95,rbt96a,rbt96b}, a grain made from Al (a superconducting
material) was connected to two metal leads via high-resistance tunnel
junctions, with capacitances $C_\ssL$ and $C_\ssR$, say.  In the next
generation of 1997 \cite{rbt97}, the grain was also coupled
capacatively to a gate, with capacitance $C_\ssg$.  The resulting
device, schematically depicted in \Fig{fig:sample}(a), has the
structure of a SET, with the grain as central island.  The circuit
diagram for an SET is shown in \Fig{fig:circuit-diagram}(b).  Applying
a bias voltage $V$ between the two leads causes a tunnel current $I$
to flow between the leads through the grain, via incoherent sequential
tunneling through the tunnel junctions.  The current can be influenced
by changing the gate voltage $\Vg$ (hence the name ``transistor''),
which tunes the electrostatic potential on the grain and thereby also
its average number of electrons $N$.  (For devices without a gate
these two quantities cannot be tuned and instead have some
sample-dependent, fixed value. For such devices, set $C_\ssg = 0$ in
all formulas below.)

The physics of SETs had been clarified in the early 1990s
\cite{curacao} through extensive studies of lithographically defined
SETs of {\em mesoscopic\/} size, \ie\ with {\em micron}-scale central
islands. The fundamentally new aspect of RBT's work was that their
SETs, made by a novel fabrication technique (described in
\Ref{rbt95}), were {\em nanoscopic\/} in size:
they had ultrasmall grains with radii between 15nm and 2nm as central
islands, which were thus several orders of magnitude smaller in volume
than in previous experiments.  This had two important consequences: 
\vspace*{2mm}

%\begin{enumerate} 
%\item 
1. {\em The grain's charging energy\/} $\Ec \equiv e^2/2C$ {\em was
    much larger than for mesoscopic SETs\/}, ranging roughly between 5
  and 50~meV (where $C \equiv C_\ssL + C_\ssR + C_\ssg$). $\Ec$ is the
  scale that determines the energy cost for changing the electron 
number $N$ by one.
  Since for ultrasmall grains it far exceeds all other typical energy
  scales of the SET, such as those set by the bias voltage ($V
  \lesssim 1$~mV), the temperature ($T \lesssim 4.2$~K) and the bulk
  superconducting gap for Al ($\Delta_\bulk = 0.18$~meV), fluctuations
  in electron number are strongly suppressed. \vspace*{2mm}

%\item 
2. {\em Discrete eigenstates of the conduction electron energy
    spectrum became resolvable\/} -- their mean level spacing $d$
  ranged from 0.02 to 0.3~meV. This agrees in order of magnitude
with the  estimate $d = 1/ \N(\eF)$
obtained using the  free-electron expression for 
the density of states $ \N(\eF)$ at the Fermi surface of a 3D grain,
\begin{eqnarray}
  \label{eq:d-estimate}
d = {2 \pi^2 \hbar^2
\over m \kF \Vol} = {1.50 \, 
 \mbox{eV} . \mbox{nm}^2 \over  \kF \Vol} \; ,
\end{eqnarray}
where, for example, $\kF = 17.5$ $ {\rm nm}^{-1}$ for Al.  The measured
$d$-values are much larger than $\kB T$ for the lowest temperatures attained
(around $T \simeq 30 \mbox{mK}$), but on the order of $\Delta_\bulk$.
However, the number of conduction electrons for grains of this size is still
rather large (between $10^4$ and $10^5$). \vspace*{2mm}
%\end{enumerate}

Since the two scales $\Ec$ and $d$ differ by at least an order of
magnitude, they manifest themselves in two distinct and easily
separable ways in the low-temperature $I$-$V$ curves of RBT's devices:
\vspace*{2mm}

\begin{figure}[t]
\centerline{\epsfig{figure=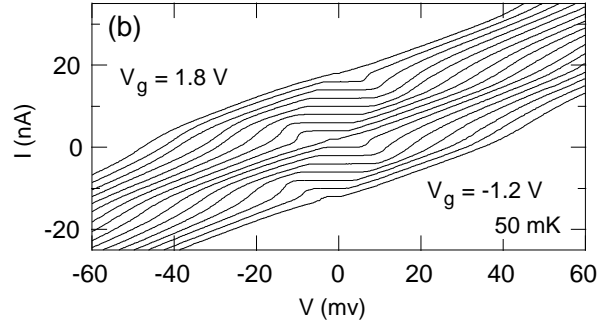,%
width=0.6\linewidth}}
  \caption[Current-voltage characteristics of ultrasmall Al 
  SETs]{Current-voltage curves for
    an ultrasmall SET \protect\cite{rbt97} at 50~mK, artificially
    offset on the vertical axis for a set of equally-spaced values of
    $\Vg$ between $-1.2$ and 1.8~V. The $I$-$V$ curves display
    Coulomb-staircase structure on a bias voltage scale of tens of mV.
    By fitting these to standard SET theory \cite{hanna91}, the SET's
    basic parameters can be determined: $R_\ssL = 3.5~{\rm M}\Omega$,
    $R_\ssR = 0.2~{\rm M}\Omega$, $C_\ssL = 3.5$~aF, $C_\ssR =
    9.4$~aF, $C_\ssg = 0.09$~aF, $\Ec = 46$~meV. Assuming
    the grain shape to be hemispherical and using
    the measured capacitances to estimate its surface areas
    (see \cite{rbt97,PR-VDR} for details), the grain's
    radius and mean level spacing were estimated
    as  $r \simeq 4.5$~nm and $d \simeq 0.45$~meV.}
\label{fig:generic-IV}
\end{figure}

%\begin{enumerate}
%\item 
1. \emph{Coulomb-blockade phenomena:}
When $V$ is varied on a large scale of tens of mV  for fixed $\Vg$
  [\Fig{fig:generic-IV}], the $I$-$V$ curves have a typical
  ``Coulomb-staircase form'' characteristic of SETs: zero current at
  low $|V|$ (the ``Coulomb blockade'' regime), sloping steps equally
  spaced in $V$, and step thresholds sensitive to $\Vg$.  This proves
  that the tunnel current flows only through {\em one\/} grain. The
  maximal width of the flat step of zero current around $|eV|=0$ is
  governed, in order of magnitude, by $\Ec$
  and typically varies between 5 and 50~mV.  

As $V_g$ is varied, the
  $I$-$V$ curves periodically repeat.
The  ``orthodox theory'' for Coulomb blockade
phenomena \cite{averin-likharev,grabert92,schoen98,PR-VDR}
explains this as follows: 
the electrostatic work required to add $\Nex$ excess electrons with
a total charge of $e \Nex$ (with $e <0$) to a grain with initial random
off-set charge $Q_0$, while the time-independent voltages $V_\ssL$, $V_\ssR$
and $V_\ssg$ of the left and right leads and the gate electrode, respectively,
are held fixed, has the form 
\begin{eqnarray}
  \label{eq:Epot}
  \Epot(\Nex) =  e \VD \Nex \, + \,\Ec \Nexsq  \; .
\end{eqnarray}
Here $\VD \equiv \left(Q_0 + \sum_{r = {\rm L,R,g}} C_r V_r
\right)/C $ represents the electrostatic potential on the grain,
and $\Ec \Nexsq$ represents the Coulomb interaction energy of the $\Nex$
excess electrons due to their mutual repulsion.
Since, for given $\Vg$,  the system  adjusts 
$\Nex$ such as to minimize $\Epot (\Nex)$, the  $I$-$V$ characteristics
are $\Vg$-periodic, with period   $e/C_\ssg$. \vspace*{2mm}

\begin{figure}[t]
  \centerline{\epsfig{figure=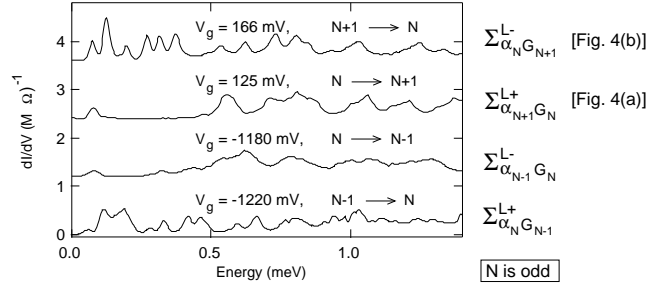,%
width=0.65\linewidth}}
\caption[Excitation Spectra of a Superconducting
Grain with an Even or Odd Number of Electrons]{ Excitation spectra for the
  same sample \cite{rbt97} as in \Fig{fig:generic-IV}, measured at $T=50$~mK
  and $H=0.05$~T (to drive the Al leads normal), for four different
  $\Vg$-values, corresponding to different values for the grain's average
  electron number (from top to bottom: $N+1,N,N,N-1$).  The curves are
  artificially offset on the vertical axis and each is labeled by the
  associated bottleneck tunneling rate $\Sigma^{r \pm}_{\alpha_N \G_{N'}}$
(from initial state $|\G\rangle_{N'}$ to final state
  $|\alpha\rangle_N$, two of which are illustrated schematically in
  \Fig{fig:Vg-V}), the bottleneck barrier being $r=\ssL$ in this case.
  Plotted is $\ddr I/\ddr V$ vs.\ energy, where the latter is given by the
  voltage-to-energy conversion factor $|eV| (C_\ssR + C_\ssg/2)/C = 0.73 |e
  V|$, which reflects the voltage drop across barrier L (for a derivation of
  this factor, see Sec.~2.3 of Ref.~\cite{PR-VDR}).  The sizeable
  spectroscopic gap between the first two peaks in the middle two curves, and
  its absence in the top and bottom curves, reflects the pairbreaking energy
  cost in the excitation spectrum of a superconducting grain with an {\em
    even\/} number of electrons, and implies that $N$ is odd.  }
\label{fig:sc-spectra(h=0)}
\end{figure}
\begin{figure}[t]
\centerline{
  \epsfig{figure=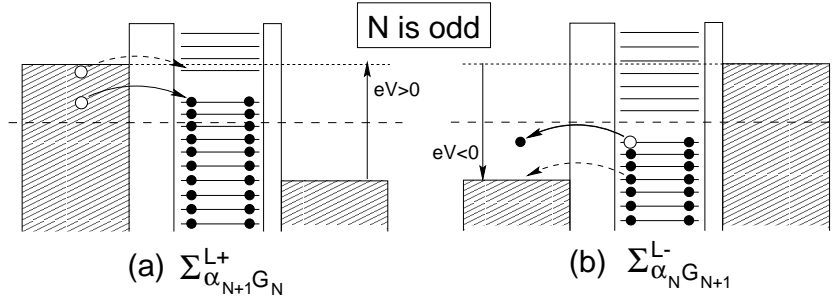,width=0.85\linewidth}}
  \caption[Cartoon of bottleneck tunneling processes]{
    Schematic  depiction of bottleneck tunneling 
     processes governing the excitation spectra 
    of Fig.~\protect\ref{fig:sc-spectra(h=0)},
    for $N$ being odd, with rates: 
    (a)     $\Sigma^{L+}_{\alpha_{N+1} G_N}$ and
    (b) $\Sigma^{L-}_{\alpha_N G_{N+1}}$, corresponding
to two different choices of $\Vg$ just below or above
the degeneracy point at which the $N$- and $(N+1)$-electron
ground states are degenerate. The long-dashed line indicates the
   equilibrium, $V=0$ chemical potential of the L and R leads. 
   Solid (dashed) arrows depict bottleneck tunneling
    transitions into the lowest- (highest) energy {\em final\/} states
    accessible for the chosen value of $V$, and filled circles
    represent the electron configuration of the lowest-energy {\em
      final\/} state.   (For a more detailed discussion
  of such diagrams, in particular how they change with
   $\Vg$, see Sec.~2 of Ref.~\protect\cite{PR-VDR}.)
}
\label{fig:Vg-V}
\label{fig:explainV-Vg}
\end{figure}

\begin{figure}[t]
\centerline{\epsfig{figure=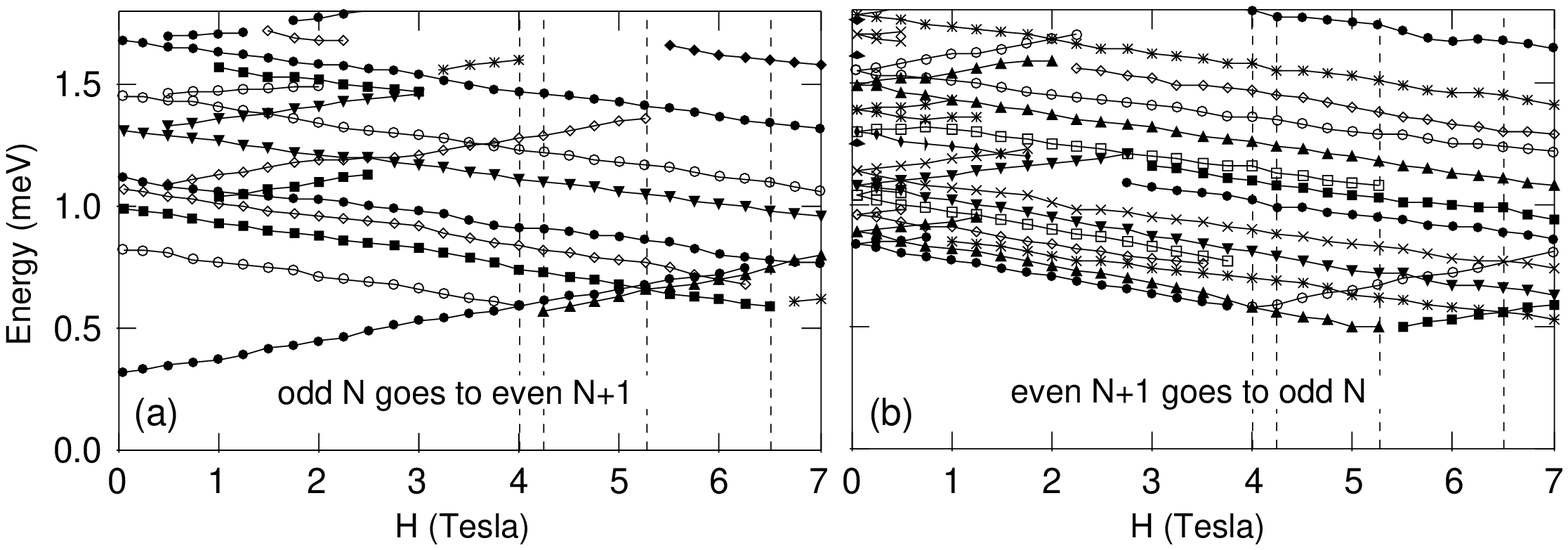,width=\linewidth}}
  \caption[Experimental excitation spectra for superconducting
grains]%
{ Magnetic field dependence \protect\cite{rbt97} of excitation spectra
  such as those of \Fig{fig:sc-spectra(h=0)} and
  taken for the same grain, at (a) $\Vg \approx 110$~mV and (b) $\Vg
  \simeq180$~mV [corresponding to the second and topmost curves of 
  \Fig{fig:sc-spectra(h=0)}, respectively]. Each line 
  represents a distinct conductance peak in the $\ddr I / \ddr V$ curve
  and traces how its energy changes with $H$. 
  Upward-moving peaks are broader and less distinct than
  downward-moving ones (for reasons poorly understood) and can be
  followed only for a limited range of $H$ before they are lost in the
  background.  The distances between lines directly give the grain's
  (a) fixed-($N+1$) and (b) fixed-$N$ excitation spectrum, where $N$
  is odd; the corresponding bottleneck processes are schematically
  illustrated in \Figs{fig:Vg-V}(a) and (b), respectively. The vertical dashed
  lines indicate the first four level-crossing fields $H_{s,s'}$
  (defined in \Eq{eq:hcrit} and assigned by comparison with
  \Fig{fig:spectra}, see \Sec{sec:tunneling-spectra}),
  namely $H_{0,1} \!=\!4$T, $H_{1/2,3/2}\!=\!4.25$T,
  $H_{1,2}\!=\!5.25$T and $H_{3/2,5/2}\!=\!6.5$T with uncertainty $\pm
  0.13$T (half the $H$-resolution of 0.25T). }
\label{fig:sc-magneticfield} 
\label{fig:experimental-spectra}
\end{figure}

%\item 
2. \emph{Fine-structure due to level discreteness:} When $V$ is varied
on the much smaller scale of a few mV around the threshold of the
Coulomb blockade regime and the temperature is sufficiently low ($T
\ll d$), the $I$-$V$ curves have a step-like substructure, and
correspondingly, the differential conductance $({\rm d}I / {\rm d} V)$
curves contain a series of fine peaks, see \Fig{fig:sc-spectra(h=0)}.
As first pointed out by Averin and Korotkov \cite{averin90}, such
small steps in the $I$-$V$ curve are expected to occur whenever the
voltage drop across one of the tunnel junctions (say $r={\rm L,R}$)
equals the threshold energy at which the rate for tunneling across
that junction into or out of one of the grain's {\em discrete energy
  eigenstates\/} becomes nonzero, since this opens up another channel
for carrying current across that junction.  Such tunneling processes
are ilustrated schematically in \Fig{fig:explainV-Vg}. 

More formally,
let $\Sigma^{r \pm}_{\alpha_N \alpha'_{N'}}$ denote the rate for
the tunneling transition $|\alpha'\rangle_{N'} \to
|\alpha\rangle_N$ between two grain eigenstates (with electron number
$N'$ and $N = N' \pm 1$, and eigenenergies $\E^{N'}_{\alpha'}$,
$ \E^N_\alpha$, respectively,), induced by transferring an electron
across barrier $r$ onto (upper sign) or off (lower sign) the grain. A
golden rule calculation (see Sec.~2.3.3 of \Ref{PR-VDR} for details)
then shows that these rates have the form
  \begin{equation}
    \label{eq:goldenrule}
    \Sigma^{r \pm}_{\alpha \alpha'} = 
   f\bigl( \E^N_\alpha - \E^{N'}_{\alpha'} \mp
    e (V_r - V_\ssD) \bigr) \Gamma^{r \pm}_{\alpha \alpha'} \; ,
  \end{equation}
where $f(E) = 1/(e^{E/\kB T} + 1)$ is the Fermi function,
$V_r - V_\ssD$ is the voltage drop (electrostatic potential
difference) between lead $r$ and the grain, and $\Gamma^{r
  \pm}_{\alpha \alpha'}$ are 
transition probabilities. With increasing transport
voltage $V = V_\ssL - V_\ssR$, a current step thus occurs each
time a ``bottleneck rate'' $ \Sigma^{r \pm}_{\alpha \alpha'}$
(associated with a rate-limiting tunneling process)
is switched from ``off'' (exponentially small) to ``on''
(of order $\Gamma^{r \pm}_{\alpha \alpha'}$),
\ie\ each time $e (V_r - V_\ssD)$ passes through a threshold at which
one of the inequalities 
\begin{equation}
  \label{eq:Vrthresholds}
  \label{eq:current-steps}
\pm e (V_r - V_\ssD) \ge \E^N_\alpha - \E^{N'}_{\alpha'}
\end{equation}
becomes true. The tunneling spectrum yields particularly
useful information if the initial 
 state for all bottleneck tunneling processes is always the ground state, 
$|\alpha'\rangle_{N'} = |\G\rangle_{N'}$
(whose electron number $N'$ is determined by $\Vg$). This will
be the case if the following conditions are met:
(i) for the given (fixed) value of gate voltage $\Vg$,
all bottleneck processes involve the \emph{same}\footnote{If 
bottleneck rates for tunneling across \emph{both} barriers are comparable,
the probability of finding $N'$ and $N$ electrons on
the grain will be comparable, and 
the analysis is considerably more complicated; 
see Sec.~2.3.4 of \protect\cite{PR-VDR} for details.} 
barrier; 
(ii) the temperature is sufficiently low ($T \ll d$);
and (iii) nonequilibrium effects are negligible
(requiring relaxation rates on the grain to be much
greater than tunneling rates). Under these conditions,
which are satisfied by the data of \Fig{fig:sc-spectra(h=0)} (and
  \Fig{fig:sc-magneticfield} below), the distances between the current
  steps or conductance peaks   directly reflect the 
so-called \emph{fixed-$N$ excitation spectrum} of the grain,
\ie\ the set of energy differences 
  \begin{equation}
    \label{eq:fixed-N-spectrum}
    \delta \E_{\alpha \bar \alpha}^N = \E^N_{\alpha} - \E^N_{\bar \alpha}
  \end{equation}
between the eigenenergies  $ \E^N_{\alpha}$
 of all those $N$-electron eigenstates $| \alpha \rangle_N$ that
  are accessible final states from the 
initial ground state  $|\G\rangle_{N'}$,
via a tunneling processes onto the grain if $N= N'+1$,
or off the grain if $N = N' - 1$. 
\vspace*{2mm}
%\end{enumerate}

For devices having a gate, 
two  very interesting options exist: Firstly,  by
tuning $\Vg$ such that the Coulomb blockade regime is large or small,
so that the $V$-threshold at which current begins to flow is large or
small, {\em nonequilibrium effects can be maximized or minimized},
respectively, depending on whether one chooses to study them or not.
In this review, we consider only the near-equilibrium situation.
Strong nonequilibrium was studied in Refs.~\cite{rbt97}-\cite{agam98}
and is reviewed in Sec.~6 of \Ref{PR-VDR}. Secondly, by
tuning $\Vg$ by an amount large enough ($\simeq \Ec/e$) to change $N'$
by one unit, the {\em influence on the spectrum of the parity\/} of
the number of electrons on the grain can be studied.

Parity effects occur, for instance,
in the magnetic-field dependence of the fixed-$N$ excitation spectrum,
which  can be
obtained by simply tracing the motion of the conductance peak positions as a
magnetic field is turned on (at fixed $\Vg$).  This is shown in
\Fig{fig:sc-magneticfield} below, which nicely illustrates the breaking of
Kramers degeneracy by the applied field: For $H=0$, the grain will have
time-reversal symmetry.  For an even-$N$ grain, the many-electron wave
function for the ground state will be a spin singlet, in order that the
orbital energy be minimized.  In contrast, the ground state of an odd-$N$
grain for $H=0$ necessarily is two-fold degenerate, by Kramers' theorem,
forming a Kramers doublet.  When $H$ is turned on, this doublet is
Zeeman-split by $\pm \half \muB \, g \, H$.  Therefore, for an {\em
  even\/}-$N$ grain at small $H$, the lowest-lying tunneling excitations
correspond to transitions from the even-grain ground state singlet to the
odd-grain ground state doublet, \ie\ to two states split by $H$, so that the
lowest-$V$ conductance peak will exhibit Zeeman splitting in an applied
field [\Fig{fig:sc-magneticfield}(b)]. 
 On the other hand, for an {\em odd\/}-$N$ grain with $T \ll \muB \, g
\, H/ \kB$, the odd-grain ground state will be the lower-energy state of the
Kramers doublet; the lowest-lying tunneling excitation will thus consist only
of a single transition from this odd-grain ground state to the even-grain
ground state singlet, so that the lowest-$V$ conductance peak will {\em
  not\/} split into two as a function of $H$
[\Fig{fig:sc-magneticfield}(a)]. It follows
that in  \Fig{fig:sc-magneticfield}, $N$ is odd.

\section{A gap in the excitation spectrum}
\label{sec:gap-in-spectrum}
\label{sec:sc-introduction}
\label{sec:sc-measuredgap}

The spectra shown in \Figs{fig:sc-spectra(h=0)} and
\ref{fig:sc-magneticfield}, which are typical for RBT's results for largish Al
grains ($r \! \gtrsim \!5$ nm), reveal a very striking feature: if the number
of electrons on the grain in the final state after the bottleneck tunneling
process is even (middle two curves of \Fig{fig:sc-spectra(h=0)}), the
excitation spectra display a spectroscopic gap between the first two
conductance peaks that is significantly larger than the mean spacing between
subsequent peaks, whereas no such gap occurs if the final-state electron
number is odd (top and bottom curves of \Fig{fig:sc-spectra(h=0)}).  In other
words, even-$N$ excitation spectra (number parity $p =0$) are gapped, whereas
odd-$N$ excitation spectra ($p = 1$) are not.  This is even more clearly
apparent when comparing \Figs{fig:sc-magneticfield}(a) and
\ref{fig:sc-magneticfield}(b), which give the magnetic-field ($H$) dependence
of an even-$N$ and odd-$N$ excitation spectrum, respectively.  However, in
their smallest grains ($r \lesssim 3$ nm) no such clear parity-dependent
spectroscopic gap could be discerned.

BRT interpreted these observations as evidence for {\em
  superconducting pairing correlations} in their larger grains, using
notions from the BCS theory of superconductivity: in an even grain,
all \emph{excited} states involve at least one broken Cooper pair, \ie\ two
BCS quasiparticles, and hence lie at least $2 \Delta$ above the
fully-paired BCS ground state; in contrast, in an odd grain {\em
  all}\/ states have at least one unpaired electron, \ie\ at least one
quasiparticle, and hence no significant gap exists between ground- and
excited states.  \Fig{fig:explainV-Vg} is a cartoon illustration of
this interpretation, if one imagines two electrons drawn on the same
energy level to represent a ``Cooper pair'' (making this cartoon
precise will be one of the main goals of review): in
\Fig{fig:explainV-Vg}(a)  the final
electron number is even, and all final excited states (reached via
dashed arrows) have one less ``Cooper pair'' than the final ground
state (reached via the solid arrow); in contrast, in
\Figs{fig:explainV-Vg}(b) the final
electron number is odd, and the final ground and excited states have
the same number of ``Cooper pairs''.

The approximately linear $H$-dependence of the excitation spectra in
\Fig{fig:sc-magneticfield} was attributed by RBT to the Zeeman energy
shifts of discrete levels in a magnetic field
(its effect on orbital motion is neglible, as will be discussed in
\Sec{sec:orbitalmagnetism}). The fact that the lowest
state in \Figs{fig:sc-magneticfield}(a) or (b) does not or does
display Zeeman splitting, respectively, implies that $N$ is odd,
as mentioned above.  The
reduction of the spectroscopic gap in \Fig{fig:sc-magneticfield}(a)
therefore is purely due to Zeeman energy shifts and has nothing to do
with the reduction of the BCS gap parameter due to pair-breaking that
occurs in bulk samples in a magnetic field \cite{tinkham-book}.  A
detailed discussion of the spectra's magnetic field dependence will be
given in \Secs{sec:CC-transition} and 
\ref{sec:tunneling-spectra-prb97}.

For completeness, it should be remarked that a spectral gap in
ultrasmall superconducting grains was observed as long ago as 1968 by
Giaever and Zeller \cite{GiaeverZeller-68,ZellerGiaever-69}, who
studied tunneling through granular thin films containing electrically
insulated Sn grains. They found gaps for grain sizes right down to the
critical size estimated by Anderson (radii of 2.5~nm in this case),
but were unable to prove that smaller particles are always ``normal''.
RBT's experiments are similar in spirit to this pioneering work, but
their ability to focus on \emph{individual} grains makes a much more
detailed study possible.

\section{A discrete BCS model for ultrasmall grains with pairing correlations}
\label{sec:model}

In this section we construct a model for an isolated ultrasmall grain
with pairing correlations, using phenomenological arguments valid for
the regime $d \lesssim \tilde \Delta$. The model,
which we shall call ``\dbcsm'', allows a rather
detailed qualitative understanding of the measurements of RBT
\cite{braun97,braun99} and hence is ``phenomenologically successful''.
For $d\gg \tilde \Delta$ it is unrealistically simple, however, and
should rather be viewed as a toy model for studying how pairing
correlations change as the grain is made smaller and smaller.

\subsection{A simple reduced BCS interaction plus a Zeeman term}
\label{sec:Hamiltonian}

Following the philosophy of the ``orthodox theory'' for Coulomb blockade
phenomena \cite{averin-likharev,PR-VDR}, we assume that the only effect of the
Coulomb interaction is to contribute an amount $\Ec \Nexsq$ [cf.\ 
\Eq{eq:Epot}] to the eigenenergy of each eigenstate of the grain. Since the
charging energy is huge (5 to 50 meV) in ultrasmall grains, this term strongly
suppresses number fluctuations around the optimal value of $\Nex$, so that, to
an excellent approximation, all energy eigenstates will also be number
eigenstates.\footnote{ An exception occurs \emph{at} a so-called degeneracy
  point, where $ \Epot(\Nex) = \Epot(\Nex+1)$; the resulting complications
  will not be considered here.}  Since $\Ec \Nexsq$ is constant within every
fixed-$N$ Hilbert space, we shall henceforth ignore it, with the understanding
that the model we are about to construct should always be solved in a
fixed-$N$ Hilbert space (and that the use of grand-canonical approaches below,
after having dropped $ \Ec \Nexsq$, simply serves as a first approximation to
the desired canonical solution).

The only symmetry expected to hold in realistic, irregularly-shaped
ultrasmall grains at zero magnetic field is time-reversal symmetry. We
therefore adopt a single-particle basis of pairs of time-reversed
states $|j \pm \rangle$, enumerated by a discrete index $j$. Their
discrete energies $\varepsilon_j$ are assumed to already incorporate
the effects of impurity scattering and the average of
electron-electron interactions, etc.  As simplest conceivable model
that incorporates pairing interactions and a Zeeman coupling to a
magnetic field, we adopt a Hamiltonian $\op H = \op H_0 + \op H_\red$
of the following reduced BCS form:
\begin{equation} 
  \label{eq:hamiltonian}
  \label{eq:hamilton-1}
 \op  H_0 = \sum_{j, \sigma= \pm} (\varepsilon_j -  \mu  - \sigma h) 
    c^\dagger_{j\sigma}c^\ds_{j\sigma} \; , \qquad
  \op H_\red = 
    - \lambda d \sum_{ij}
    c^\dagger_{i+}c^\dagger_{i-}c^\ds_{j-}c^\ds_{j+} \; .
\end{equation}
Here $ - \sigma h \equiv \sigma \frac12 \mu_Bg H$ is the Zeeman energy
of a spin $\sigma$ electron in a magnetic field $H$, and we shall take
$h>0$ below.  Models of this kind had previously been studied by
Strongin \etalia\ \cite{Strongin-70}, M\"uhlschlegel \etalia\ 
\cite{Muehlschlegel-72,muehlschlegel94} and Kawataba
\cite{kawabata-80,kawabata-81}. The first application to RBT's grains
for $h=0$ was by von Delft \etalia\ \cite{vondelft96} and for $h \neq
0$ by Braun \etalia\ \cite{vondelft96,braun97,braun99}. 

Due to level repulsion the $\varepsilon_j$'s will, to first
approximation, be uniformly spaced. Unless otherwise specified, we
shall for simplicity always (except in \Sec{sec:sc-level-statistics})
take a completely uniform spectrum with level spacing $d$.
Fluctuations in the level spacings have been studied with methods of
random matrix theory \cite{smith96}, with qualitatively similar
results (see \Sec{sec:sc-level-statistics}).  For a system with a
total of $N$ electrons, where the \emph{electron number parity} $p
\equiv N \mbox{mod} 2$ is equal to $0$ for even $N$ and $1$ for odd
$N$, we use the label $j=0$ for the lowest-lying non-doubly-occupied
level (with occupation number $p$) in the $T=0$ Fermi sea, which we
shall denote by $|\F_N\rangle$. We choose the Fermi energy at $\eF
\equiv 0$ write\footnote{ This convention differs slightly from that
  used in \cite{vondelft96,braun97,braun99}, namely $\varepsilon_j = j
  d + \varepsilon_0$. The latter is a little less convenient,
  resulting, \eg, in a $p$-dependent chemical potential for the
  variational BCS ground states discussed below, $\mu_p^\BCS =
  \varepsilon_0 + (p-1)d/2$, whereas (\ref{eq:ejs}) results simply in
  $\mu_p^\BCS = 0$. \label{f:mu-value}}
\begin{equation}
  \label{eq:ejs}
\varepsilon_j = j d + (1-p)d/2 \; ,
\end{equation}
thereby taking the doubly-occupied and empty levels of $|\F_N\rangle$
to lie symmetrically above and below $\eF$
(see \eg\ \Fig{fig:alpha-states} below).  The parameter $\mu$ 
in \Eq{eq:hamiltonian} is, in \gc\ theories, 
the chemical potential, whose
value${}^{\ref{f:mu-value}}$ determines the average particle number.
For canonical theories, which make no reference to a chemical
potential, $\mu$ is not needed and can be dropped (\ie\ set equal to
0).

%Due to level repulsion the $\varepsilon_j$'s will, to first
%approximation, be uniformly spaced. Unless otherwise specified, we shall
%for simplicity take a completely uniform spectrum with level spacing
%$d$, $\varepsilon_j = j d + \varepsilon_0$. Fluctuations in the level
%spacings have been studied with methods of random matrix theory
%\cite{smith96}, with qualitatively similar results (see
%\Sec{sec:sc-level-statistics}).  For a system with a total of $N$
%electrons, where the \emph{electron number parity} $p \equiv N
%\mbox{mod} 2$ is equal to $0$ for even $N$ and $1$ for odd $N$, we use
%the label $j=0$ for the first level whose occupation is not 2 but $p$
%in the $T=0$ Fermi sea, which we shall denote by $|\F_N\rangle$.  The
%parameter $\mu$ is, in \gc\ theories, the chemical potential.  For
%canonical theories, which make no reference to a chemical potential,
%$\mu$ is not needed and can be dropped, or else viewed simply as a
%reference energy (whose actual value is irrelevant) for the
%single-particle energies $\varepsilon_j$.

The pairing interaction is of the reduced BCS form, in that it scatters a pair
of electrons from one pair of time-reversed states into another.\footnote{Note
  that the use of a \emph{reduced} BCS interaction means that couplings
  between non-time-reversed pairs of states are neglected.  A theoretical
  motivation, based on random matrix theory, for this reduced form may be
  found in \Ref{agam98}, or in Sec.~6.1.3 of Ref.~\protect\cite{PR-VDR}.
  Experimental evidence for the sufficiency of the reduced form is discussed in
  Refs.~\protect\cite{braun97,braun99} and, in most detail, in Sec.~4.7 of
  Ref.~\protect\cite{PR-VDR}.}  It is taken to include only states whose
energy separation from the Fermi energy lies within the cutoff given by the
Debye frequency: $|\varepsilon_j | < \omegaD$.  The pair-coupling constant in
\Eq{eq:hamilton-1} is written as $\lambda d$, where $\lambda$ is a
dimensionless parameter independent of the grain's volume, to make it explicit
that both $\hat H_0$ and $\hat H_\red$ make extensive ($\propto \Vol$)
contributions to the ground state energy (since the number of terms in each
sum $\sum_j$ in \Eq{eq:hamilton-1} scales with $N$, and $d \propto \eF /N$).
The ``bulk gap'' of the model, obtained by solving the standard BCS gap
equation [\Eq{eq:gap-bulk}] at $T=0$ in the bulk limit, thus is
\begin{equation}
  \label{eq:lambda-definition}
  \tilde \Delta =\omegaD/\sinh(1/\lambda) \; .
\end{equation}
To be precise, by ``bulk limit'' we shall always mean $d/ \tilde
\Delta \to 0$ and $N \to \infty$ while the product $N  d $ is kept
fixed, and use $d \sum_j \to \int \! {\ddr} \varepsilon_j$.

An applied magnetic field will completely penetrate an ultrasmall
grain, since its radius (typically $r \lesssim 5$nm) is much smaller
than the penetration length of 50 nm for bulk Al. The Zeeman term in
Eq.~(\ref{eq:hamiltonian}) models the fact that RBT's measured tunnel
spectra of \Fig{fig:experimental-spectra} evolve approximately
linearly as a function of magnetic field, with $g$ factors
between\footnote{\label{f:wrong-g} Claims of smaller $g$ factors made
  in \Ref{rbt96a} are wrong, the result of confusing different orbital
  states as Zeeman-split spin states.  This was made clear in
  \Ref{rbt97}, where it was observed that upward-trending Zeeman
  states can have significantly smaller amplitude than
  downward-trending states, making them difficult to observe.}  $1.95$
and $2$ (determined from the differences between measured slopes of
up- and downward-moving lines).  Deviations from $g = 2$ probably
result from spin-orbit scattering, known to be small but nonzero in
thin Al films \cite{Meservey-70,Meservey-94}, but neglected below
(where $g=2$ is used).

Intuitively speaking, it is clear that the \dbcsm\ introduced above
contains all ingredients necessary to make contact with the spectra of
\Fig{fig:sc-magneticfield}: it is formulated in terms of discrete
levels, it contains a pairing interaction which is known, from bulk
BCS theory \cite{BCS-57,tinkham-book}, to cause a gap in the
excitation spectrum, and it contains a Zeeman term that will cause
eigenenergies to linearly depend on an applied magnetic
field.  Indeed, we shall see in \Sec{sec:tunneling-spectra-prb97}
that it can be used to obtain a rather detailed qualitative
understanding of the spectra of \Fig{fig:sc-magneticfield}.

\subsection{Why orbital diamagnetism is negligible in ultrasmall grains}
\label{sec:orbitalmagnetism}

Of course, a magnetic field in principle also couples to the orbital
motion of the conduction electrons -- in bulk samples,
this is the origin of the Meissner effect. 
 Orbital effects in spherical and
cylindrical superconductors whose dimensions are smaller than the
penetration depth were first considered by Larkin \cite{Larkin65}.
However, in grains as small as those of RBT, orbital diagmagnetic
effects are negligible \cite{salinas99},
just as for thin films in a parallel magnetic field
\cite{Meservey-70,Meservey-94}. The reason is as follows:

Let $H_\orb$ denote the
field scale above which orbital diamagnetism becomes important.
If, in a random-matrix description of the grain's
spectrum,   $H_\orb$ is formally associated
with the field at which the crossover between
the symplectic and unitary ensembles,
driven by the orbital effects of the magnetic field,
is complete, it is found  \cite{Kravtsov92} that
\begin{eqnarray}
  \label{eq:Horbital}
  H_\orb \approx { \Phi_0 \over r^2 \sqrt{ \ETh / d} } \; ,
\end{eqnarray}
where $\ETh$ is the Thouless energy and 
$\Phi_0 $ ($= hc/2e = 2067 {\rm T.nm}^2$) is the flux quantum.
An intuitive understanding for the
origin of this result can be obtained by the
following argument \cite{glazman-priv}: 
Associate $H_\orb$ with the field  at which the orbital splitting of the
eigenenergies of two time-reversed states $|j \pm \rangle$ becomes comparable
to the mean level spacing.  
Denoting the angular momentum of these states by
$\pm \langle \hat l_z \rangle_j \hbar$, the orbital diamagnetic contribution
to their eigenenergies is $\pm \langle \hat l_z \rangle_j \muB H$, hence 
$ H_\orb \approx d/(2 \langle \hat l_z \rangle_j \muB)$. 
 Now, the angular momentum
of an electron traversing the ``closed trajectory'' corresponding to a
discrete quantum level can be estimated as $\langle \hat l_z \rangle \hbar
\approx m (A_\typ d/\hbar)$, where the bracketed factor is the typical
(directed) area $A_\typ$ covered by its trajectory divided by the period
$\hbar/ d$ of its motion.  The number of bounces off the grain's boundaries
during this time is roughly $\ETh/d$, since
the  $\hbar / \ETh$ is the time to cross the grain once.  Hence
the directed area is $A_\typ \approx r^2 \sqrt{\ETh /d} $, where the square
root accounts for the fact that the direction of motion after each bounce is
random \cite{matveev00}.  Collecting the various estimates 
results in $H_\orb \approx \Phi_0/(\pi r^2 \sqrt{ \ETh / d} )$,
which, up to a factor of $\pi$, agrees with \Eq{eq:Horbital}.

Now, assuming ballistic electron motion in the grain, the Thouless energy has
the form $\ETh \approx \hbar \vF / (a 2 r)$, where $a$ is a geometrical
constant of order unity.  Using $d$ from \Eq{eq:d-estimate},
we see that  $H_\orb$ grows like 
$r^{-3}$ with decreasing grain size.
Taking 
%the ballistic estimate of
%\Eqs{eq:g-dimensionlessconductance} and (\ref{eq:define-Thouless})
%for $\gdc$, with 
$a = 3$ (as in  \Ref{davidovich99}), we find 
from \Eq{eq:Horbital} that
hemispherical Al grains with radii of (say) $r \approx 3$ or 5~nm
have  $H_\orb \approx 19$ or 7~T, respectively.
If larger values are used for $a$, as  would be
appropriate for more pancake-shaped grains
\cite{agam97a}, $H_\orb$ would be even larger.

We may thus conclude that orbital diagmagnetic effects only begin
to play a role for largish grains
($\gtrsim 5$~nm), and then only for the highest fields (of
7~T) studied by RBT.
  Indeed, some larger grains do show slight
deviations from $H$-linearity \cite{rbt96a} for large fields, which
probably reflect the onset of such orbital effects;
% (according to \cite{Bahcall-priv}, these give 
%corrections to the
%eigenenergies of the order of $\hbar v_\ssF r^3(H/\Phi_0)^2$); 
however, these are much smaller than Zeeman effects in the grains of present
interest, and will be neglected here.  Thus, the \dbcsm\ assumes that Pauli
paramagnetism due to the Zeeman energy completely dominates orbital
diamagnetism, similarly to the case of thin films in parallel magnetic fields
\cite{Meservey-70,Meservey-94}.

\subsection{Choice of numerical values for model parameters}

When doing numerical calculations for the \dbcsm, some choices must be
made for the numerical values of its parameters (though slight changes
in their values will not change the results qualitatively).  We shall
follow the choices made by Braun \etalia\ \cite{braun99}, since these
led to  reasonable agreement between experimental and theoretical
excitation spectra. For the Debye frequency they used the textbook
value \cite{ashcroft-mermin} for Al of $\omegaD=34$meV.  Making an
appropriate choice for the ``bulk gap'' $\tilde \Delta$ is less
straightforward, since its experimental value for systems of reduced
dimensionality often differs from that of a truly bulk system,
presumably due to (poorly-understood) changes in the phonon spectrum
and the effective electron-phonon coupling.  For example, for thin Al
films \cite{Strongin-70,Garland-68} it is known that $\tilde
\Delta_{\rm thin \, film} \simeq 0.38$~meV, which is about twice as
large as the gap of a truly bulk system, $\tilde \Delta_\bulk =
0.18$~meV.  (This increase in $\tilde \Delta$ is not universal,
though; \eg, for Nb $\tilde \Delta$ is smaller in thin films than in
the bulk.)  Since ultrasmall grains are in many ways analogous to thin
films in a parallel magnetic field [see \Sec{sec:CC-transition}],
%only smaller, we
%shall assume that the further dimensional reduction increases
%$\tilde\Delta$ by another factor of about $2$, 
Braun \etalia\ adopted the thin-film value for grains too, \ie\ used
$\tilde\Delta \simeq 0.38$meV.  These choices imply that the
dimensionless pair-coupling constant $\lambda= [\sinh^{-1}(\omegaD/
\tilde \Delta)]^{-1} $ [cf.\Eq{eq:lambda-definition}] has the value
$\lambda = 0.194$.  (In \Sec{sec:tunneling-spectra-prb97} we shall
see, \emph{a posteriori}, that the choices $\tilde \Delta = 0.34$ and
$\lambda = 0.189$ would have been slightly more appropriate.)
Finally, for those numerical calculations that are explicitly cut-off
dependent, Braun \etalia\ smeared the cutoff of the BCS interaction
over two single-electron levels; this smooths out small
discontinuities that would otherwise occur in $d$-dependent quantities
each time the energy $|\varepsilon_j|$ of some large-$|j|$ level moves
beyond the cutoff $\omegaD$ when $d$ is increased.

Note that the above way of choosing $\lambda$ lumps into a single
phenomenological constant all the poorly-understood effects of reduced
dimensionality \cite{Strongin-70} on the phonons that mediate the
attractive electron-electron interaction. %%??%% give a citation...
Studying these effects in detail would be interesting in its own
right, but would require systematic investigations with grains of
well-controlled shapes and sizes.  For the case of RBT's
irregularly-shaped grains, using a phenomenological coupling constant
seems the best one can do. Note, though, that the precise value of
$\lambda$ is not very important as long as all energies are measured
in units of $\tilde \Delta$ (as we shall do for all numerical
calculations), since most of the $\lambda$-dependence is thereby
 normalized away.  Therefore, the slight difference between the
$\lambda$-values proposed above and those used in
\cite{braun98,braun-thesis,sierra99} (namely 0.224) hardly matters.

\subsection{Some general properties of the eigenstates
-- the blocking effect}
\label{sec:generalproperties}

The eigenstates of the \dbcsm\ of \Eq{eq:hamiltonian} have some
simple but general properties that are worth stating at the outset.

Firstly, every eigenstate of $\hat H$ will also be an eigenstate of
the number operator $\hat N = \sum_{j \sigma} c^\dagger_{j \sigma}
c^\ds_{j \sigma}$, since $[\hat H, \hat N] = 0$.

Secondly, since the interaction only involves levels within the cutoff
energy $\omegaD$ of $\eF$, the dynamics of those lying outside this
range is trivial. We shall thus ignore them henceforth and focus only
on the remaining set of \emph{interacting} levels, denoting this set
by $\I$. 

\begin{figure}
  \centerline{\epsfig{figure=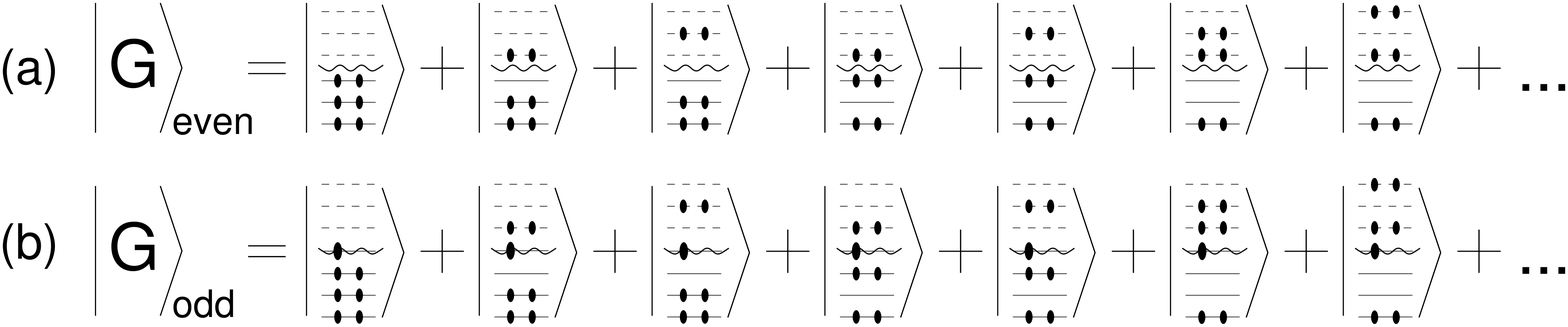,%
width=0.95\linewidth}}
  \caption[Cartoon of the exact even and odd ground states of a 
  reduced BCS Hamiltonian]{A cartoon depiction of the exact ground
    state for a reduced BCS Hamiltonian, for $N$ being even (a) or odd
    (b): they are coherent superpositions of eigenstates of $\hat H_0$
    (whose respective amplitudes are not depicted) that all have the
    same $N$; the leftmost of these is (a) the even or (b) the odd Fermi
    ground state $|\F_N \rangle$, whose Fermi energy is indicated
    by the wavy line.  }
    \label{fig:exactgroundstate}
  \end{figure}

Thirdly, {\em singly-occupied}\/ levels do not participate in the
pair scattering described by $\hat H$: ``unpaired'' electrons in such
levels are not scattered to other levels, hence the labels of
singly-occupied levels are good quantum numbers.  Moreover, every
unpaired electron Pauli-blocks the scattering of other pairs into its
own singly-occupied level, \ie\ it restricts the phase space available
to pair scattering and thereby weakens the amount of pairing
correlations, as we shall see in detail later.  This was called the
\emph{``blocking effect''} by Soloviev \cite{Soloviev-61}, who
discussed it extensively in the early 1960's in the context of nuclear
physics.  The eigenstates $|\alpha \rangle$ and corresponding
eigenenergies $\E_\alpha$ of $\hat H$ thus have the following general
forms:
\begin{eqnarray}
|\alpha \rangle &= & |\Psi_n,\B\rangle = 
\prod_{i \in \B} c_{i \sigma_i}^\dagger |\Psi_n\rangle ,
  \label{eq:generaleigenstate} 
\\
  \label{eq:generaleigenstate-2} 
 |\Psi_n \rangle & = & \sum_{j_1, \dots, j_n}^\U
\psi (j_1, \dots , j_n) \prod_{\nu=1}^n b_{j_\nu}^\dagger  
|\Vac \rangle
\; , 
\\
\label{eq:generaleigenenergy}
\E_{\alpha} &=& \E_n + \E_\B (h) \, , 
\qquad   \E_\B  (h) = \sum_{i \in \B} 
(\varepsilon_i - \mu - \sigma_i h) \, .
\end{eqnarray}
This describes $N = 2n + b$ electrons, $b$ of which are unpaired and
sit in a set $\B$ of singly-occupied, blocked levels, making a
contribution $\E_\B (h)$ to the total eigenenergy.
%$\E_n + \E_B$ of $|n,b\rangle$,
The remaining $n$ pairs of electrons, created by the pair operators
$b_j^\dagger = c^\dagger_{j +} c^\dagger_{j -}$, are distributed among
the remaining set $\U= \I \backslash \B$ of \emph{unblocked} levels,
with wavefunction $\psi (j_1, \dots , j_n)$ ($\sum_j^\U \equiv
\sum_{j \in \I\backslash \B}$ denotes a sum over all {\em unblocked\/}
levels in $\I$).  The corresponding state $|\Psi_n \rangle$ is an
eigenstate of the pair number operator and a Hamiltonian $\hat H_\U$
involving only pair operators:
\begin{eqnarray}
  \label{eq:eigenpsi}
  &&  \sum_j^\U b_j^\dagger b^\ds_j |\Psi_n\rangle
  = n |\Psi_n\rangle 
   , \qquad    \hat H_\U |\Psi_n\rangle = \E_n |\Psi_n\rangle \; , 
\\
 && \hat H_\U = \sum_{ij}^\U \left[ 2 (\varepsilon_j - \mu) \delta_{ij} -
 \; \lambda \, d \right]  b_i^\dagger b^\ds_j \; .
 \label{1}
 \end{eqnarray}
 Each eigenstate $|\Psi_n ,\B\rangle$ may be visualized as a coherent
 superposition of eigenstates of $\hat H_0$ that all lie in the same fixed-$N$
 Hilbert space, and in all of which each pair of unblocked $(j \in \U$),
 time-reversed levels $|j \pm\rangle $ is either doubly occupied or empty.
 This is illustrated in \Figs{fig:exactgroundstate}(a) and (b), which
 schematically depict the exact ground states for even and odd $N$,
 respectively.  The odd ground state has a single blocked level, at the Fermi
 energy, containing an unpaired electron. The latter somewhat weakens pairing
 correlations relative to the even ground state and hence leads to parity
 effects, which will be extensively discussed in later sections.
  
  A useful measure for the amount of energy gained by $|\alpha
  \rangle$ via its correlations is its ``condensation energy''
  relative to the uncorrelated state $|\alpha \rangle_0$,
\begin{equation}
  \label{eq:define-condensation-energy}
  E^\cond_\alpha = \E_\alpha -
 {}_0 \langle \alpha | \hat H | \alpha \rangle_0 , \qquad
\mbox{where} \quad | \alpha \rangle_0 = \prod_{i \in \B} c^\dagger_{i
\, \sigma_i} |\U  \rangle_0 \, ,
\end{equation}
and $|\U \rangle_0$ is the ``Fermi ground state'' in $\U$, for which
the $n$ pairs occupy the $n$ lowest-lying levels in $\U$.
 
 Note that $\hat H_\U$ is $h$-independent, since the total Zeeman
 energy of any \emph{pair} of electrons is zero. Hence the full
 $h$-dependence of the eigenenergies resides in the rather trivial
 contribution $\E_\B(h)$ of the blocked levels, which is a very
 important and useful simplification.
 
 Diagonalizing $\hat H_\U$ would be trivial if the $b$'s were true
 bosons.  However, they are not, and in the subspace spanned by the
 set $\U$ of all non-singly-occupied levels instead satisfy the
 ``hard-core boson'' relations,
\begin{eqnarray}
\label{hard-core-boson-1}
  b^{\dagger 2}_j = 0, \qquad
\label{hard-core-boson-2}
\mbox{[} b^\ds_j, b^\dagger_{j^{\prime}}\mbox{]} =
\delta_{j j^{\prime}} (1 - 2 b^\dagger_j b^\ds_j), \qquad 
\label{hard-core-boson-3}
\mbox{[} b^\dagger_{j} b^\ds_{j},  
b^\dagger_{j'} \mbox{]} &=& \delta_{j j'} b^\dagger_j \; ,
\end{eqnarray}
which reflect the Pauli principle for the fermions from which the
$b$'s are constructed. In particular, $b^{\dagger 2}_j=0$ implies that
only those terms in (\ref{eq:generaleigenstate-2}) are non-zero for
which the indices $j_1, \dots j_n$ are all distinct.

The task of finding the eigenstates $|\Psi_n \rangle$ is thus highly
non-trivial. Nevertheless, an exact solution does exist. Unbeknownst
to most of the condensed-matter physics community, it was found and
studied extensively by Richardson in the 1960's and will be presented
in \Sec{sec:sc-richardson}. Throughout the present part I,
however, we shall use more well-known
approaches based on the variational wavefunctions introduced
by BCS \cite{BCS-57}, and that had been used to study the \dbcsm\ 
before Richardson's solution was revived towards the end of 1998.

\section{Canonical characterization of pairing correlations}
\label{sec:canonical-pair-mixing}
\label{sec:meaningofsc}
\label{chap:fixedN}

Since the \dbcsm\ has the standard reduced BCS form, the most natural
first step toward understanding its $T=0$ properties is to use
BCS-like variational wavefunctions (or equivalently Bogoljubov's
mean-field approach), and indeed this will be done in
\Secs{chap:generalBCS} to \ref{sec:BCS-parity}.  However, the
limitations of such an approach should be realized at the outset: the
spectra measured by RBT are excitation spectra for a grain with a {\em
  fixed electron number\/} $N$, and hence should be calculated for a
grain with definite electron number $N$ (\ie\ completely isolated from
the rest of the world, \eg\ by infinitely thick oxide barriers).  In
contrast, the variational wavefunctions of BCS [\Eq{eq:BCSground}
below] do not have the fixed-$N$ form [\Eq{eq:generaleigenstate-2}]
which any true eigenstate should have, but instead are formulated in a
{\em grand-canonical\/} (\gc) framework (as is the Bogoljubov
mean-field approach to which they are intimately related).

When considering a truly isolated superconductor such as a perfectly
insulated grain (another example would be a superconductor levitating
in a magnetic field due to the Meissner effect), one therefore needs
to address the following question, which will be the main theme of the
present section: \emph{how is one to incorporate the fixed-$N$
  condition into BCS theory, and how important is it to do so?}  This
issue is well understood and was discussed at length in the early days
of BCS theory (Rickayzen's book gives a beautiful discussion
\cite{rickayzen-book}), in particular in its application to pairing
correlations in nuclei \cite[p.~439]{RingSchuck-80} (see also the
general remarks in \cite{Lipkin-60}). Nevertheless, for pedagogical
reasons the arguments are well worth recapitulating in the present
context.
% Readers familiar with the
% relevant arguments may prefer to skip this section.

We shall first remind the reader that the use of a \gc\ framework is
only a matter of convenience, since the essence of the pairing
correlations that lie at the heart of BCS theory is by no means
inherently \gc\ and can easily be formulated in canonical language
\cite{vondelft96}. We then show how standard BCS theory fits into this
scheme, point out that the differences between results obtained using
\gc\ and canonical wavefunctions are negligible for $d \ll \tilde
\Delta$, and conclude that for the purposes of gaining a
phenomenological understanding of the experimental data, standard
grand-canonical BCS theory should be sufficient.  Nevertheless, the
fundamental question of how to improve on this theory, in order to
achieve a truly canonical description and to properly treat
fluctuation effects, which become important for $d \gtrsim \tilde
\Delta$ \cite{muehlschlegel62,Janko-94,matveev97}, is interesting and
important in its own right and will be addressed at length in
\Sec{sec:sc-canonical}.

For simplicity, throughout the present subsection
\ref{sec:canonical-pair-mixing} we shall consider only the even ground
state in the thermodynamic limit (in which even-odd differences are
negligible), so that $\U=\I$ and blocking effects need not be worried
about.

\subsection{The grand-canonical BCS wavefunction}

Conventional BCS theory describes the pairing correlations induced by
an attractive pairing interaction such as $\hat H_\red$ of
\Eq{eq:hamiltonian} within a \gc\ ensemble, formulated on a Fock space
of states in which the total particle number $N$ is not fixed.  This
is illustrated by BCS's famous variational ground state Ansatz
\begin{eqnarray}
  \label{eq:BCSground}
    |\BCS \rangle = \prod_j 
    (u_j + \eer^{\ii\phi_j} 
    v_j b^\dagger_j )\,|\Vac\rangle \; , 
    \qquad \mbox{with} \quad u_j^2 + v_j^2 = 1,
\end{eqnarray}            
where the variational parameters $u_j$ and $v_j$ are real and $\phi_j$
is a phase (which, it turns out, must be $j$-independent, for reasons
discussed below).  $|\BCS \rangle$ is not an eigenstate of $\hat N$
and its particle number is fixed only on the average by the condition
$ \langle \hat N \rangle_\BCS = N$, which determines the \gc\ chemical
potential $\mu$.  Likewise, the commonly used \gc\ definition
\begin{equation}
  \label{eq:BCS-gap}
  \Delta_{\MF} \equiv \lambda \, d \sum_j \langle b_j
\rangle_\BCS = \lambda \, d  \sum_{j} u_j v_j \eer^{\ii\phi_j}
 \; 
\end{equation}
for the superconducting pairing parameter only makes sense in a \gc\ 
ensemble, since $\langle b_j \rangle$ would trivially give
zero when evaluated in a canonical ensemble, formulated on a strictly
fixed-$N$ Hilbert space of states.  (We shall use the term ``pairing
parameter'' instead of ``order parameter'', since the latter carries
the connotation of a phase transition, which would require the
thermodynamic limit $N\to \infty$, which is not applicable for
ultrasmall grains).

\subsection{Canonically meaningful definition
for the pairing parameter}
\label{sec:meaningfulDelta}

A theory of strictly fixed-$N$ superconductivity must evidently entail
modifications of conventional BCS theory. However, these are only of
technical, not of conceptual nature, since the essence of the pairing
correlations discovered by BCS can easily be formulated in a
canonically meaningful way, including a definition for the pairing
parameter.  We shall now attempt to explain, in intuitive,
non-technical terms, how this may be done (our discussion is indebted
to that of Rickayzen \cite{rickayzen-book}).  Readers with a
preference for rigor may consult \Secs{sec:sc-richardson} to
\ref{sec:bulk-few-n-differences} for a corroboration, using
Richardson's exact solution, of the arguments presented below.

Let $|\G \rangle$ be the exact even ground state of the system,
depicted in \Fig{fig:exactgroundstate}(a). As explained in
\Sec{sec:generalproperties}, it is a coherent superposition of
eigenstates of $\hat H_0$ that all have the same $N$ and in all of which
each pair of time-reversed levels $|j \pm\rangle $ is either doubly
occupied or empty.  Due to this coherent superposition, $|\G \rangle$
entails strong pairing correlations, whose essential properties may be
understood by investigating how they modify the correlators
\begin{eqnarray}
  \label{eq:defineu2v2}
\label{eq:C_ij}
 C_{ij}  \equiv  \langle 
    b_i^\dagger b^\ds_j \rangle \, , \qquad 
 \bar v_j^2 \equiv C_{jj} = \langle b_j^\dagger b^\ds_{j}
 \rangle \, ,  \qquad 
  \bar u_j^2  \equiv \; \langle  b^\ds_{j} b_j^\dagger \rangle \,  , 
\qquad 
\end{eqnarray}
relative to the form these take on for the Fermi ground state
$|\F_N\rangle$:
\begin{eqnarray}
(C_{ij})_\F  = \delta_{ij} (\bar v_j^2)_\F \, , \qquad
(\bar v_j^2)_\F
= \theta (  -
  \varepsilon_j) \, ,  \qquad
 (\bar u_j^2)_\F  = \theta ( \varepsilon_j  ) \, .
\end{eqnarray}
$C_{ij} (= C_{ji}^\ast)$
is the matrix element for the interaction to be able to scatter a pair
of electrons from level $j$ to $i$, and 
$\bar v_j^2$ and $\bar u_j^2$ are the probabilities to find level $j$
doubly occupied or empty, respectively.
%the product $(\bar u_j \bar v_j)_\F $ equals zero for all $j$.
The pairing correlations in $|\G\rangle$ must be such that $\hat
H_\red$ lowers the ground state energy below that of the uncorrelated
Fermi sea $|\F_N\rangle$ by an amount that is extensive ($\propto N
\propto d^{-1}$)
in the thermodynamic limit.  Clearly, this requires that $ \langle \op
H_\red \rangle_\G - \langle \op H_\red \rangle_\F $ is negative and
extensive, \ie\ that
\begin{eqnarray}
  \label{eq:exptation-Hred}
 \lambda \, d  \sum_{ij}  
\mbox{[} C_{ij} - (C_{ij})_\F \mbox{]}  \simeq 
 \lambda \, d  \sum_i \sum_{j < i} 2  \re ( C_{ij} ) 
\, \propto \, N \;  \quad \mbox{(and positive}).
\end{eqnarray}
In the second expression we neglected the diagonal terms, since their
number is so small (only $\propto N $) that $\lambda d \sum_j [\bar
v_j^2 - (\bar v_j^2)_\F ] $ is at best of order unity in the
thermodynamic limit.  For \Eq{eq:exptation-Hred} to hold, $|\G\rangle$
must have two properties:
\begin{enumerate}
\item[(i)] the number of $C_{ij}$'s that differ significantly from
  zero (\ie\ are of order unity) should scale like $N^2$, \ie\ one
  power of $N$ per index \cite[p.~167]{rickayzen-book};
\item[(ii)] most or all of the $C_{ij}$ for $i<j$ should have the same
  phase, since a sum over random phases would average out to zero.
\end{enumerate}
Since a suitable pairing parameter should vanish in the thermodynamic
limit unless both these conditions hold, the definition
\begin{eqnarray}
  \label{eq:canonical-order-parameter}
  \Delta^2_{\can} \equiv  (\lambda \, d )^2 
\sum_{ij} (C_{ij} - 
\langle c^\dag_{i+} c^\ds_{j+} \rangle
 \langle c^\dag_{i-} c^\ds_{j-} \rangle )
\end{eqnarray}
(or its square root) suggests itself, where the subscript emphasizes
that (\ref{eq:canonical-order-parameter}) is meaningful in a canonical
ensemble too, and we subtracted\footnote{This subtraction was
  suggested to us by Moshe Schechter, who pointed out that then
  \Eq{eq:canonical-order-parameter} has a natural generalization to
  position space: it is the spatial average,
  $  \Delta^2_\can  \equiv  (\lambda \, d)^2
 \int \! \! \ddr \vec r_1  \ddr \vec r_2 \, {\cal F}(\vec r_1,
 \vec r_2)$, 
of the two-point function
\begin{eqnarray}
\nonumber 
{\cal F}(\vec r_1, \vec r_2) \equiv
 \langle \psi^\dag_+ (\vec r_1) \psi_-^\dag (\vec r_1) 
          \psi_-^\ds (\vec r_2) \psi_+^\ds (\vec r_2) \rangle 
  \;  - \; \langle \psi^\dag_+ (\vec r_1) \psi_+^\ds (\vec r_2) \rangle
     \langle \psi^\dag_- (\vec r_1) \psi_-^\ds (\vec r_2) \rangle
 \,    \label{eq:Delta_can-realspace}
\end{eqnarray}
(with $\psi_\sigma (\vec r) \equiv \Vol^{-1/2} \sum_{\vec k} \eer^{\ii
  \vec k \cdot \vec r} c_{\vec k \sigma}$), which evidently measures
the amplitude for the propagation of \emph{pairs} as opposed to
uncorrelated electrons.  Other definitions for a canonically
meaningful pairing parameter have been suggested
\protect\cite{vondelft96,braun98,braun99}, such as $\lambda d \sum_j
\bar u_j \bar v_j$ or $ \lambda d \sum_j [ \langle b^\dagger_{j}
b^\ds_{j}\rangle - \langle c^\dagger_{j+}c^\ds_{j+} \rangle \langle
c^\dagger_{j-}c^\ds_{j-} \rangle ]^{1/2} \; $, but these focus only on
requirement (i) and fail to incorporate requirement (ii).  A quantity
very similar to Eq.~\protect\ref{eq:canonical-order-parameter} was recently
proposed in Eq.~(55) of Ref.~\protect\cite{Rossignoli-99a}, 
namely $(\lambda d)^2
\sum_{ij} \mbox{[} C_{ij} - (C_{ij})_{\lambda=0} \mbox{]}$. }
the ``normal-state
contribution to  $C_{ij}$.''  If (i) and (ii) hold, $\Delta_\can$ will
take on a finite value; its relation to a gap in the spectrum will
become clear below.  In the bulk limit,
$\Delta_\can$ can be shown [see \Sec{sec:bulk-few-n-differences}]
to reduce to the ``bulk pairing parameter'' $\tilde \Delta$
of \Eq{eq:lambda-definition}.

\subsection{Redistribution of occupation probability
across $\eF$}

\begin{figure}[t]
   \centerline{\epsfig{figure=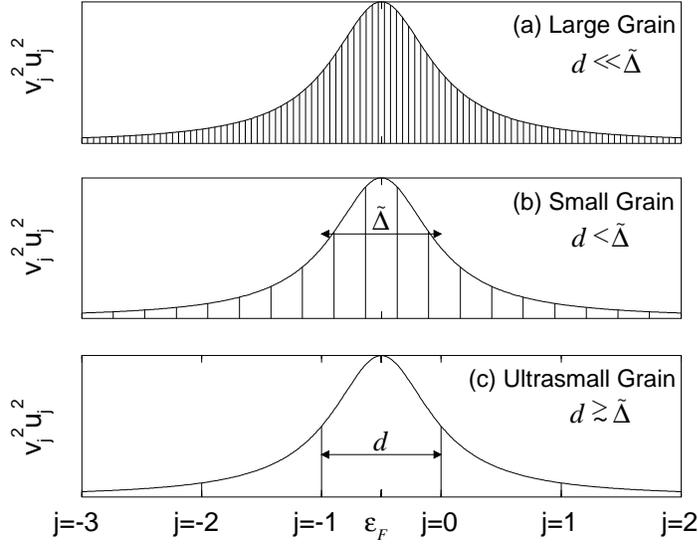,%
width=0.7\linewidth}}
  \caption[Cartoon of breakdown of superconductivity]{A cartoon
    depiction of why ``superconductivity breaks down'' when the sample
    becomes sufficiently small. Vertical lines are drawn at each
    single-particle energy $\varepsilon_j$, spaced with a mean level
    spacing $d$ corresponding to (a) a ``large'' grain
    ($d\ll\tilde\Delta$); (b) a ``small'' grain ($d \simeq 0.25 \tilde
    \Delta$); (c) an ``ultrasmall'' grain ($d \simeq \tilde \Delta$).
    The lines' height represents the function $u_j^2 v_j^2 = \frac14
    \tilde \Delta^2/(\varepsilon_j^2 + \tilde \Delta^2)$ of standard
    bulk BCS theory, to illustrate the energy range (of width $\tilde
    \Delta$ around $\eF$) within which pairing correlations are
    strongest.  Loosely speaking, the number of single-particle levels
    in this regime, namely $ \tilde \Delta /d$, corresponds to ``the
    number of Cooper pairs'' of the system.  Evidently, this number
    becomes less than one when $d \gtrsim \tilde \Delta$ as in (c), so
    that it then no longer makes sense to call the system
    ``superconducting'' [cf.\ \Sec{sec:qual-disc}].}
    \label{fig:v2u2-prb97}
  \end{figure}

Now, property (i) can be realized if all $C_{ij}$ in a finite
($d$-independent) range of $\varepsilon_i$'s and $\varepsilon_j$'s
around the Fermi energy  $\eF$ differ significantly from
zero; the width of this range will evidently determine the magnitude
of $\Delta_\can$ (provided (ii) also holds), which conversely can be
viewed as a measure of this width. But a nonzero $C_{ij}$ evidently
requires \emph{both} $b^\dagger_i b^\ds_j | \G \rangle \neq 0$,
implying $(\bar v_{j})_\G \neq 0$ and $(\bar u_i)_\G \neq 0$,
\emph{and also} $\langle \G | b^\dagger_i b^\ds_j \neq 0$, implying
$(\bar v_i)_\G \neq 0$ and $(\bar u_{j})_\G \neq 0$.  The product
$(\bar u_j \bar v_j)_\G$ must thus be different from zero [in contrast
to $(\bar u_j \bar v_j)_\F = 0$] for all $\varepsilon_j$ within a
finite range around $\eF$ (cf.\ \Fig{fig:v2u2-prb97}).  This can be
achieved by smearing out the sharp steps of the $\theta$-functions of
$(\bar v_j)_\F $ and $(\bar u_j)_\F$, so that $(\bar v_j)_\G$ [or
$(\bar u_j)_\G$] is nonzero also for a finite range of $\varepsilon_j$
above [or below] $\eF$.  In other words, for $|\G\rangle$ some
occupation probability must be redistributed (relative to
$|\F_N\rangle$) from below to above $\eF$, as illustrated in \Fig
{fig:exactgroundstate}.  This redistribution, which was called
pair-mixing in \cite{vondelft96,braun99}, frees up phase space for
pair scattering and so achieves a gain in interaction energy (provided
(ii) also holds) that more than compensates for the kinetic energy
cost incurred thereby.

  Furthermore, note that properties (i) and (ii) imply, even without
  detailed calculations, that the spectrum will be gapped.  Consider,
  for example, a ``blocking excitation'' that disrupts pairing
  correlations by having $|j +\rangle$ definitely occupied and $|j
  -\rangle$ definitely empty; since pair-scattering involving level
  $j$ is blocked, the energy cost is \label{p:blocking}
\begin{eqnarray}
&& (\varepsilon_j - \mu) - [
(\varepsilon_j - \mu)  2\langle 
b^\dagger_j  b^\ds_j \rangle) - \lambda \, d  
\sum_{i (\neq j)} 
\langle b^\dagger_i b_j^\ds + b^\dagger_j b^\ds_i  \rangle ]
\\ & &
\label{eq:excitationenergy}
= (\varepsilon_j - \mu) (1- 2 \bar v_j^2)
 +  \lambda \, d  
\sum_{i (\neq j)} ( C_{ij} + C_{ji} ) \, ,
\end{eqnarray}
in which the restriction on the sum reflects the blocking of
scattering events involving level $j$.  Since the first term of
(\ref{eq:excitationenergy}) is positive definite (particle-hole
symmetry ensures that $(\half - \bar v_j^2) \gol 0$ if $\varepsilon_j
- \mu \gol 0$) and the second of order $\Delta_\can$, the excitation
energy will be \emph{finite} even for $d\to 0$, implying the existence
of a gap of order $\Delta_\can$. Similarly, ``phase-breaking
excitations'' that violate the fixed-phase condition (ii) are gapped
too: for example, if $(C_{i j})_{\rm excited} = - (C_{ij})_{\rm
  ground}$ for a given $j$ and all $i (\neq j)$, the energy cost is $
- \lambda d \sum_{i(\neq j)} [
(C^{\phantom{\ast}}_{ij} + C^\ast_{j i})_{\rm excited} 
- (C^{\phantom{\ast}}_{ij} + C_{j i})_{\rm ground} ]
$, which is at least of order $ 2 \Delta_\can$.

We see, therefore, that the essence of pairing correlations can
readily be formulated in a canonical framework: (i) a redistribution
of occupation probability across $\eF$ occurs, such that each level
$j$ in a finite range around $\eF$ has a finite probability of both
being doubly occupied or empty, and (ii) any two components of the
ground state wavefunction that differ only by the exchange of a pair
of electrons between two levels $i$ and $j$ have the same phase.

Since pairing correlations with these properties are the
microscopic property at the heart of all manifestations of
``superconductivity'', it seems reasonable to call a sample
``superconducting'' as long as it exhibits pairing correlations with
measurable consequences.  And by this criterion the gap observed in
the even grains of RBT certainly qualifies.

\subsection{Gauge symmetry breaking}

Note that property (ii) will be preserved under the gauge
transformation $c_{j\sigma} \to \eer^{\ii\phi^\prime_j} c_{j \sigma}$,
\ie\ $C_{ij} \to \eer^{-2 \ii ( \phi^\prime_i -\phi^\prime_j)} C_{ij}$,
only if all $\phi^\prime_j $ are equal, say $\phi^\prime_j =
\phi^\prime$.  Property (ii), and likewise the pairing parameter
$\Delta_\can$, therefore (a) are not gauge invariant ``locally'' in
$j$-space, but (b) are gauge invariant globally. These are obvious
consequences of the facts that (a) a \emph{correlated fixed-$N$} state
consists of a \emph{phase-coherent} superposition of many different
components, and hence cannot be invariant under arbitrary changes of
the phases of individual components; and that (b) all of these
components contain the \emph{same} number of electrons $N$ and hence
under a global gauge transformation all pick up the \emph{same} phase
factor $\eer^{\ii N \phi'}$. Obviously, global gauge symmetry can
therefore never be broken in a canonical ensemble.  In contrast, the
breaking of global gauge symmetry by the \gc\ pairing parameter
$\Delta_\MF$ of Eq.~(\ref{eq:BCS-gap}), which transforms as
$\Delta_\MF \to \eer^{2 \ii \phi^\prime} \Delta_\MF$, is an inevitable
consequence or artefact of its \gc\ definition
\cite[p.~142]{rickayzen-book}.

\subsection{Making contact with standard BCS theory}
\label{sec:contact-with-BCS}

One of the breakthrough achievements of BCS was, of course, to propose
a simple variational ground state which has precisely the properties
(i) and (ii) described above: when evaluating the correlators of
\Eq{eq:defineu2v2} using $|\BCS\rangle$ of \Eq{eq:BCSground}, one
finds
\begin{eqnarray}
  \label{eq:uvC-BCS}
 (\bar u_j)_\BCS^2 = u_j^2,  \qquad (\bar v_j)_\BCS^2 = v_j^2,
\qquad (C_{ij})_\BCS = u_i v_i u_j v_j \eer^{- \ii(\phi_i - \phi_j)}\; ,
\end{eqnarray}
and  also $(\Delta^2_\can)_\BCS = |\Delta_\MF|^2$.  The definite-phase
requirement (ii) can thus be implemented by choosing all the phases
$\phi_j$ to be the same, say $\phi_j = \phi$ for all $j$, thereby
breaking local gauge invariance (usually one simply takes $\phi = 0$);
and requirement (i) is fulfilled automatically when minimizing the
expectation value $\langle \hat H \rangle_\BCS$ w.r.t.\ $u_j$ and
$v_j$, since this does yield smeared-out step functions,
namely  \cite{BCS-57,tinkham-book}
%\begin{equation}
%\label{vj}
%v_j^2 = \frac{1}{2} \left(1 - {\varepsilon_j - \mu \over
%\sqrt{(\varepsilon_j - \mu)^2 
%+ |\Delta_\MF|^2}} \right) .
%\end{equation} 
\begin{equation}
\label{vj-bulk}
v_j^2 = \half  \left[1 - (\varepsilon_j - \mu)/E_j \right] \, ,
\qquad E_j \equiv \sqrt{(\varepsilon_j - \mu)^2 
+ |\Delta_\MF|^2} \; .
\end{equation} 
Here we neglected terms that vanish for $d \to 0$, and 
 $\Delta_\MF$ is determined by the famous gap equation
(for $T=0$), 
\begin{eqnarray}  \label{eq:gap-bulk}
  \frac1\lambda & = & 
d \sum_{|\varepsilon_j| < \omegaD} \frac1{2 E_j} \; . 
\end{eqnarray}
The BCS wavefunction instructively illustrates some of
the general properties discussed above.
Firstly, the product $u_j^2v_j^2$, shown in \Fig{fig:v2u2-prb97}, has a
bell-shaped form with a well-developed peak around $\eF$ of width
$\simeq |\Delta_{\MF}|$, illustrating that pairing correlations are
strongest within a region of width $|\Delta_{\MF}|$ around the Fermi
surface.  Secondly,  the energy
of a blocking excitation  [\Eq{eq:excitationenergy}]
reduces to $(\varepsilon_j - \mu) (1 - 2 v_j^2) + 2 u_j v_j |\Delta_{\MF}| =
E_j$, which is just the well-known energy of the 
Bogoljubov quasiparticle state $\gamma^\dag_{j +} | \BCS \rangle$,
where 
\begin{eqnarray}
  \label{eq:Bogoljubov}
\gamma_{j \sigma} = u_j c_{j \sigma} - \sigma v_j e^{\ii \phi}
  c^\dagger_{j-\sigma} \; . 
\end{eqnarray}
%$\gamma^\dagger_{j+} |\BCS\rangle$, 
%$where $\gamma_{j \sigma} = u_j c_{j \sigma} - \sigma v_j e^{\ii \phi}
%  c^\dagger_{j-\sigma}$.  
Thirdly, an example of a phase-breaking excitation is 
\begin{eqnarray}
  \label{eq:phase-breaking-excitation}
\gamma^\dag_{j +} 
\gamma^\dag_{j -} | \BCS \rangle
 = (- v_{j} e^{- \ii \phi} + u_{j} 
b^\dag_{j})
\prod_{i (\neq {j})} 
(u_i + v_j e^{\ii \phi} b^\dag_i) | \Vac \rangle
 \; ,   
\end{eqnarray}
which has $(C_{i j})_{\rm excited}
 = - u_i v_i u_{j} v_{j}$
and energy $2 E_{j}$.

It should be appreciated, however, that BCS chose a
\emph{grand-canonical} construction purely for calculational
convenience (as is made clear on p.~1180 of their original paper
\cite{BCS-57}): the trick of using a factorized form of
\emph{commuting} products in (\ref{eq:BCSground}), at the cost of
$N$-indefiniteness, makes it brilliantly easy to determine the
variational parameters $u_j$ and $v_j$. In fact, BCS proposed
themselves to use the projection of $|\BCS \rangle$ to fixed $N$ as
the actual variational ground state, namely \cite{rickayzen-book}
\begin{eqnarray}
  \label{eq:BCSground-N}
    | \PBCS \rangle & \equiv & \int_0^{2 \pi} \!\! \ddr
    \phi\, \eer^{- \ii \phi N} \! \prod_j 
    (u_j \! + \! \eer^{2 \ii \phi}
    v_j b^\dagger_j )\,| \Vac \rangle
\\
  \label{eq:BCSground-N-2}
 & = & {1 \over (N/2)!} \Bigl( \prod_j u_j \Bigr)
\Bigl( \sum_j {v_j \over u_j} b^\dagger_j
\Bigr)^{N/2} | \Vac \rangle \; 
\end{eqnarray}      
(PBCS for \underline{P}rojected BCS), which \emph{is} of the general
form of \Eq{eq:generaleigenstate-2}.  In the bulk limit ($d/\tilde
\Delta \ll 1$), however, it is completely adequate to use $| \BCS
\rangle$: firstly, the relative error which its factorized form
causes, by taking the occupation amplitude of level $j$ to be
independent of that of level $i$, scales like $1/N$
\cite[pp.~150,163]{rickayzen-book}; and secondly, the fluctuations in
its particle number, $(\Delta N^2)_\BCS \equiv \langle N^2
\rangle_\BCS - N^2 = \sum_j (2 u_j v_j)^2$, are equal to $\pi \tilde
\Delta / d$ in the bulk limit, in which the relative fluctuations
$(\Delta N^2)_\BCS / N^2 \propto d \tilde \Delta / \eF^2 $ therefore
vanish.  Thus, bulk results obtained from $ | \PBCS \rangle$ or $ |
\BCS \rangle$ are essentially identical.  In fact, Braun
\cite{braun98,braun-thesis} checked by explicit calculation that the
functions $(\bar v_j^2)_\G$, $(\bar v_j^2)_\PBCS$ and $v_j^2$ are
practically indistinguishable even for $d/ \tilde \Delta$ as large as
0.5 [see \Sec{sec:bulk-few-n-differences}]. Significant differences
\emph{do} develop between them once $d / \tilde \Delta$ increases past
0.5, however, as will be discussed in \Sec{sec:bulk-few-n-differences}.

To end this section, note that \Fig{fig:v2u2-prb97} offers a very
simple intuitive picture for why pairing correlations weaken with
increasing level spacing until, in Anderson's words \cite{anderson59},
``superconductivity is no longer is possible'' when $d \gtrsim \tilde
\Delta$: an increase in level spacing implies a decrease in the number
of levels within $\tilde \Delta$ of $\eF$ for which $u^2_j v^2_j$
differs significantly from zero, \ie\ a decrease in the number of
pairs with significant pairing correlations.  This number, namely
$\tilde \Delta/d$, can roughly speaking be viewed as the ``number of
Cooper pairs'' of the system, and when it becomes less than one, as in
\Fig{fig:v2u2-prb97}(c), it no longer makes sense to call the system
``superconducting''. However, this should not be taken to imply that
pairing correlations cease altogether in this regime; remnants of them
do persist, in the form of fluctuations, up to arbitrarily large
$d/\tilde \Delta$, as will be discussed in detail in
\Sec{sec:bulk-few-n-differences}.

\section{Generalized variational BCS approach}
\label{chap:generalBCS}
\label{sec:generalBCS}
\label{excited}

%%??%%
% Ideas for shortening:
% c) Give the numerical results first, then discuss
%    them qualitatively, as in PRL
% d) Perhaps put some of the technicalities in an Appendix
%    (which could perhaps include the general formulas
%    for Ansatz, gap equation, and assumptions made 
%    when solving everything.       
 
In the next several sections we review the generalized variational BCS
approach used by Braun \etalia\ \cite{braun98,braun99,braun-thesis} to
describe the paramagnetic breakdown of superconductivity in nm-scale
grains in a magnetic field.  This theory produces theoretical
excitation spectra that are in good qualitative agreement with the
measurements of BRT shown in \Fig{fig:sc-magneticfield} and thereby
yields the most direct confirmation available of the relevance to
experiment of the \dbcsm.  Moreover, it sheds considerable light on
how ``superconductivity breaks down'' (more precisely, how pairing
correlations weaken) with increasing $d$ and $h$: As mentioned in the
previous paragraph, in grains with $d \simeq \tilde \Delta$ (bulk
gap), near the lower size limit \cite{anderson59} of observable
superconductivity, the number of free-electron states with strong
pairing correlations (those within $\tilde \Delta $ of
$\eF$) is of order one. Thus, even in grains in which a
spectral gap can still be observed, pairing correlations are expected
to become so weak that they might be destroyed by the presence of a
single unpaired electron \cite{vondelft96}.  This can be probed
directly by turning on a magnetic field, since its Zeeman energy
favors paramagnetic states with nonzero total spin.

The theory reviewed below exploits analogies to thin films in a
parallel magnetic field \cite{Meservey-70,Meservey-94}, but explicitly
takes account of the discreteness of the grain's spectrum.  Since in
RBT's experiments the temperature $T= 50\mbox{mK}$ is much smaller
than all other energy scales ($d, \tilde \Delta$), we shall neglect
finite-temperature effects and set $T=0$.  In \Secs{sec:BCS-ansatz}
the eigenenergies $\E_\alpha$ of the grain's lowest-lying eigenstates
$|\alpha \rangle$ are calculated approximately using a
\emph{generalized \gc\ variational BCS approach} that goes beyond
standard mean-field theory by using a different pairing parameter
$\Delta_\alpha$ for each $|\alpha \rangle$.  The $\E_\alpha$ are then
used to discuss various observable quantities, such as $h$-dependent
excitation spectra (\Sec{sec:tunneling-spectra-prb97})
%direct
%experimental evidence for the dominance of purely \emph{time-reversed}
%states in the pairing interaction (\Sec{sec:time-reversed}), and
and various parity effects (\Sec{sec:BCS-parity}).

The reasons for deciding to calculate the excitation spectra, despite
their \emph{fixed $N$} nature, within a \emph{grand-canonical}
framework are as follows: Firstly, its simplicity.  Secondly and
perhaps most importantly, the exact eigenenergies have the general
form $\E_\alpha= \E_n + \E_\B (h)$ [\Eq{eq:generaleigenenergy}], in
which \emph{all $h$-dependence resides in the exactly known
  contribution $\E_\B (h)$ from the blocked levels.}  The choice of
approximation scheme therefore only affects $\E_n$, which determines
the $h=0$ properties of the spectrum, such as the size of the
zero-field spectral gap, etc., but not the qualitative features of the
$h$-dependence. In particular, this means that all of the analysis
below could easily be ``made exact'' by simply replacing the \gc\ 
approximations for $\E_n$ by the exact values from Richardson's
solution. However, this is expected to cause only slight quantitative
differences, since, thirdly, canonical calculations (mentioned after
\Eq{eq:BCSground-N-2} and discussed in \Sec{sec:sc-canonical}) yield
very similar results to \gc\ ones as long as $d/\tilde \Delta \lesssim
0.5$, which, by inspection of \Fig{fig:sc-magneticfield}, does seem to
be the case for the grain in question (the analysis of
\Sec{sec:tunneling-spectra-prb97} yields $d / \tilde \Delta \simeq
0.67$).

\subsection{The generalized variational Ansatz}
\label{sec:BCS-ansatz}

\begin{figure}[t]
  \centerline{\epsfig{figure=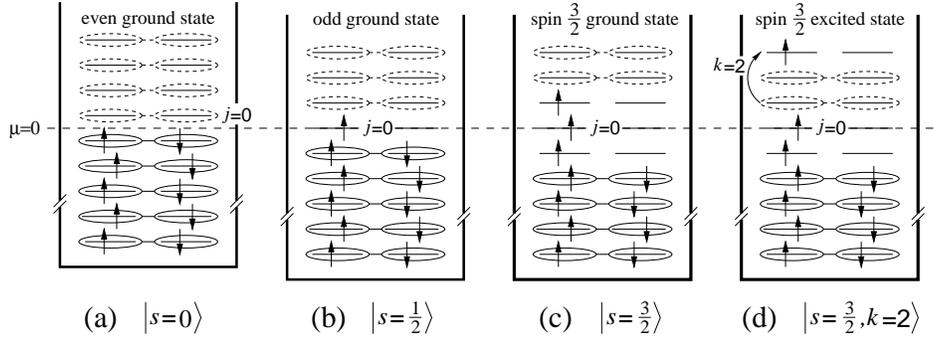,%
width=0.95\linewidth}}
  \caption[Generalized BCS wavefunctions]{Cartoon of 
    four typical variational states, 
    labeled using the notation of \Eq{eq:ansatzs} for (a-c) and
    \Eq{eq:ansatzs12} for (d).  They represent (a) the even ground
    state $|0\rangle$; (b) the odd ground state $|\half\rangle$; (c)
    the spin-${3 \over 2}$ ground state $|{3\over 2}\rangle$; (d) a
    spin-${3 \over 2}$ excited state $|{3\over 2},2\rangle$.  The
    single-particle levels are drawn for $h=0$, with the chemical
    potential half-way between levels 0 and 1 for even systems (a),
    but exactly on level 0 for odd ones (b,c,d).  The ellipses joining
    states on the same level are meant to represent a ``Cooper pair'',
    and signify its being empty or doubly occupied with amplitude
    $(u_j + v_j b_j^\dag)$; solid (dashed)
    ellipses are used for levels that would be completely filled
    (empty) in the absence of pairing correlations.  }
    \label{fig:alpha-states}
\end{figure}

The Zeeman term in the Hamiltonian of \Eq{eq:hamiltonian} favors
states with a nonzero total $z$-component of the total spin, 
$s=\half \sum_{j\sigma} \sigma c^\dagger_{j \sigma}
c^\ds_{j \sigma}$ 
(henceforth simply called ``spin''). Increasing $h$ will thus
eventually lead to a series of ground state changes to states with
successively larger spins.  In general, therefore, we are interested
in pair-correlated states with nonzero spin, and in particular in
their eigenenergies.  Following Braun \etalia\ 
\cite{braun97,braun99,braun-thesis}, we now show how this can be
calculated variationally, using the following general BCS Ansatz for a
state $|s,\bbalpha \rangle$ with $N=2n + 2s$
electrons and a definite total spin $s \ge 0$
(first introduced by Soloviev for application in nuclei
\cite{Soloviev-61}):
\begin{eqnarray}
  \label{eq:ansatz}
    |s, \bbalpha\rangle = \prod_{i \in \B} c^\dagger_{i+} 
                \prod^\U_j (u^{(s,\bbalpha)}_j + v^{(s,\bbalpha)}_j
       b^\dagger_j )\,|\Vac\rangle.
\end{eqnarray}                                               
If the spin is nonzero, it is built up by placing $2s$ unpaired
spin-up electrons in a set $\B$ of $b=2s$ single-particle levels
[cf.\ \Eq{eq:generaleigenstate}]
while the remaining single-particle levels have BCS-like amplitudes to
be either empty $(u^{(s,\bbalpha)}_j)$ or doubly occupied by a pair
$(v^{(s,\bbalpha)}_j)$, with
$(u^{(s,\bbalpha)}_j)^2+(v^{(s,\bbalpha)}_j)^2=1$.  The subscript $\U$
over products (and over sums below) indicates exclusion of the singly
occupied levels in $\B$, for which $u^{(s,\bbalpha)}$,
$v^{(s,\bbalpha)}$ are not defined. The product
$\prod_j^\U$ thus constitutes a \gc\ approximation
to the state $|\Psi_n\rangle$ of \Eq{eq:generaleigenstate-2}.

More specifically, in a given spin-$s$ sector
of Hilbert space the following two types of
specializations of \Eq{eq:ansatz} were studied in detail
($p = 2s\,{\rm mod}\,2$):
\begin{eqnarray}
  \label{eq:ansatzs}
    |s\rangle &=& 
    \prod_{i=-s + p/2}^{s-1+p/2}  c^\dagger_{i +} 
                \prod^\U_j (u^{s}_j + v^{s}_j
       b^\dagger_j)\,|\Vac\rangle.
\\
  \label{eq:ansatzs12}
    |s,k \rangle &= & 
    c^\dagger_{( s-1 +  p/2 + k ) +} c_{( s-1 +  p/2) +} 
     |s\rangle \, . 
\end{eqnarray}
$|s\rangle$ is the spin-$s$ state with the lowest energy,
\ie\ the ``variational spin-$s$ ground state'', obtained by placing
the $2s$ unpaired electrons as close as possible to $\eF$
[\Fig{fig:alpha-states}(b,c)], in order to minimize the kinetic
energy cost of having more spin ups than downs. $|s,k\rangle$ is a
particular type of excited spin-$s$ state, obtained from $|s\rangle$
by moving one electron from its topmost occupied level ($s-1+p/2$)
upwards by $k$ units of $d$ into a higher level ($ s-1+p/2 + k$).
These constructions are
illustrated in \Fig{fig:alpha-states}, of which 
 (a) and (b) represent the
variational ground states of a grain with an even or odd number of
electrons, respectively.
%, namely
%\begin{eqnarray}
%  \label{eq:p-ansatz}
%  \begin{array}{rcccl}
%    |\BCS\rangle_{\rm even} & = & & \displaystyle
%    \prod_j &\!\! (u^{\rm even}_j + v^{\rm even}_j
%    c^\dagger_{j+}c^\dagger_{j-})\,|\Vac\rangle, \, \\
%    |\BCS\rangle_{\rm odd\,} & = & c^\dagger_0 & \displaystyle
%    \prod_{j\neq 0} & \!\!
%    (u^{\rm odd}_j + v^{\rm odd}_j 
%    c^\dagger_{j+}c^\dagger_{j-})\,|\Vac\rangle \; , 
%  \end{array}
%\end{eqnarray}            
%These two states illustrate an essential difference between a system
%of even and odd number parity \cite{vondelft96}:
%in the even but not in the odd case all electrons can pair-correlate,
%implying that pair-correlations are weaker in the odd case, as we
%shall see below.  Similarly, by comparing
%Figs.~\ref{fig:alpha-states}(a-c) it is intuitively clear that the
%larger the spin $s$, the less electrons can pair-correlate.

The orthogonality of the wavefunctions, 
 $\langle s, \bbalpha | s', \bbalpha'\rangle
= \delta_{ss'}\delta_{\bbalpha\bbalpha'}$, implies
that the variational
parameters $v_j^{(s,\bbalpha)}$ and $u_j^{(s,\bbalpha)}$ must be found
\emph{anew} for each $(s, \bbalpha)$ (hence the superscript),
by minimizing the variational ``eigenenergies'' 
\begin{eqnarray}
  \label{eq:Esalpha}
\lefteqn{  
\E^\GC_{s,\bbalpha} (h,d)  \equiv  
    \langle s,\bbalpha | \hat H | s,\bbalpha \rangle }
\\   &=& 
  -2sh + \sum_{i \in \B} (\varepsilon_{i} - \mu)
  + \sum^\U_j \left[2 (\varepsilon_j - \mu) (v^{(s,\bbalpha)}_j)^2 
 + \lambda d (v^{(s,\bbalpha)}_j)^4 \right] 
\\
 & &
  - \lambda d\Big(\sum^\U_{j}
  u^{(s,\bbalpha)}_jv^{(s,\bbalpha)}_j\Big)^2 , \nonumber
\end{eqnarray}
which we use as approximations to the exact  eigenenergies
$\E^\ex_{s,\bbalpha}(h,d)$. The 
$v_j^4$ term is not extensive and hence neglected in
the bulk case where only effects proportional to the system volume are
of interest.  Here it is retained, since in ultrasmall systems it is
non-negligible (but not dominant either) \cite{braun97,braun99}.
Solving the variational conditions 
%\begin{eqnarray}
%  \label{eq:extremal}
$ 
 \frac{\partial \E^\GC_{s,\bbalpha}}{\partial v^{(s,\bbalpha)}_j} = 0
$
%\end{eqnarray}
in standard BCS fashion yields 
\begin{equation}
\label{vj}
(v_j^{(s,\bbalpha)})^2 = \half ( 1 - \xi_j /
 [\xi_j^2 + \Delta_{s,\bbalpha}^2]^{1/2}) \; , \qquad
\xi_j \equiv \varepsilon_j-\mu - \lambda d
(v_j^{(s,\bbalpha)})^2 \, ,
\end{equation} 
where  the ``pairing parameter'' 
$\Delta_{s,\bbalpha}$ is defined by the relation
\begin{eqnarray}
  \label{eq:gap1}
  \Delta_{s,\bbalpha} & \equiv & \lambda d \sum^\U_j u_j^{(s,\bbalpha)}
  v_j^{(s,\bbalpha)} \; ,\qquad\mbox{or}  \qquad 
  \label{eq:gap}
  \frac1\lambda  =  
d\sum^\U_j \frac1{2 \sqrt{\xi_j^2+\Delta_{s,\bbalpha}^2}} \; ,
\end{eqnarray}
which in the limit $d/\tilde \Delta \to 0$ reduces to the standard
bulk $T=0$ gap equation.  Note that it is $h$-independent, because it
involves only unblocked levels $j\in \U$, which are populated by pairs
with zero total Zeeman energy. Note also that in \Eq{vj} the $ \lambda
d (v_j^{(s,\bbalpha)})^2$ shift in $\xi_j$, usually neglected because
it simply renormalizes the bare energies, is retained, since for large
$d$ it somewhat increases the effective level spacing near $\eF$ (and
its neglect turns out to produce a significant upward shift in the
$\E^\GC_{s,\bbalpha} (h,d)$'s, which one is trying to minimize).

The chemical potential $\mu$ is fixed by requiring that
\begin{eqnarray}
  \label{eq:mu}
   2n + 2s =  \langle s,\bbalpha | \hat N | s,\bbalpha \rangle
  = 2s + 2\sum_j^\U (v_j^{(s,\bbalpha)})^2.
\end{eqnarray} 
In contrast to conventional BCS theory, the pairing parameter
$\Delta_{s,\bbalpha}$ can in general not be interpreted as an energy
gap and is \emph{not} an observable. It should be viewed simply as a
mathematical auxiliary quantity which was introduced to conveniently
solve the variational conditions. 
However, by parameterizing  $v_j^{(s,\bbalpha)}$ and $u_j^{(s,\bbalpha)}$,
$\Delta_{s,\bbalpha}$ does serve as a measure of the pairing
correlations present in $|s,\bbalpha\rangle$: for
vanishing $\Delta_{s,\bbalpha}$ the latter reduces to an uncorrelated
paramagnetic state $|s,\bbalpha\rangle_0$ with spin $s$
and energy $\E^0_{s,\bbalpha}$, namely
\begin{eqnarray}
  \label{eq:param}
  |s,\bbalpha\rangle_0 \equiv 
\prod_{i \in \B} c_{i +}^\dagger 
                          \prod_{j< 0}^\U b^\dagger_j |0\rangle \; ,
\qquad \mbox{with} \quad \E^0_{s,\bbalpha} \equiv  
{}_0\langle s,\bbalpha| \hat H |s,\bbalpha\rangle_0 \; ,
\end{eqnarray}
and the condensation energy $E_{s,
  \bbalpha}^\cond \equiv \E_{s,\bbalpha}^\GC-\E^0_{s,\bbalpha}$ of
$|s,\bbalpha\rangle$ reduces to zero.

\subsection{General numerical solution -- illustration
of the blocking effect}
\label{sec:generalnumerics}
\label{sec:qual-disc}

\begin{figure}[t]
  \centerline{\epsfig{figure=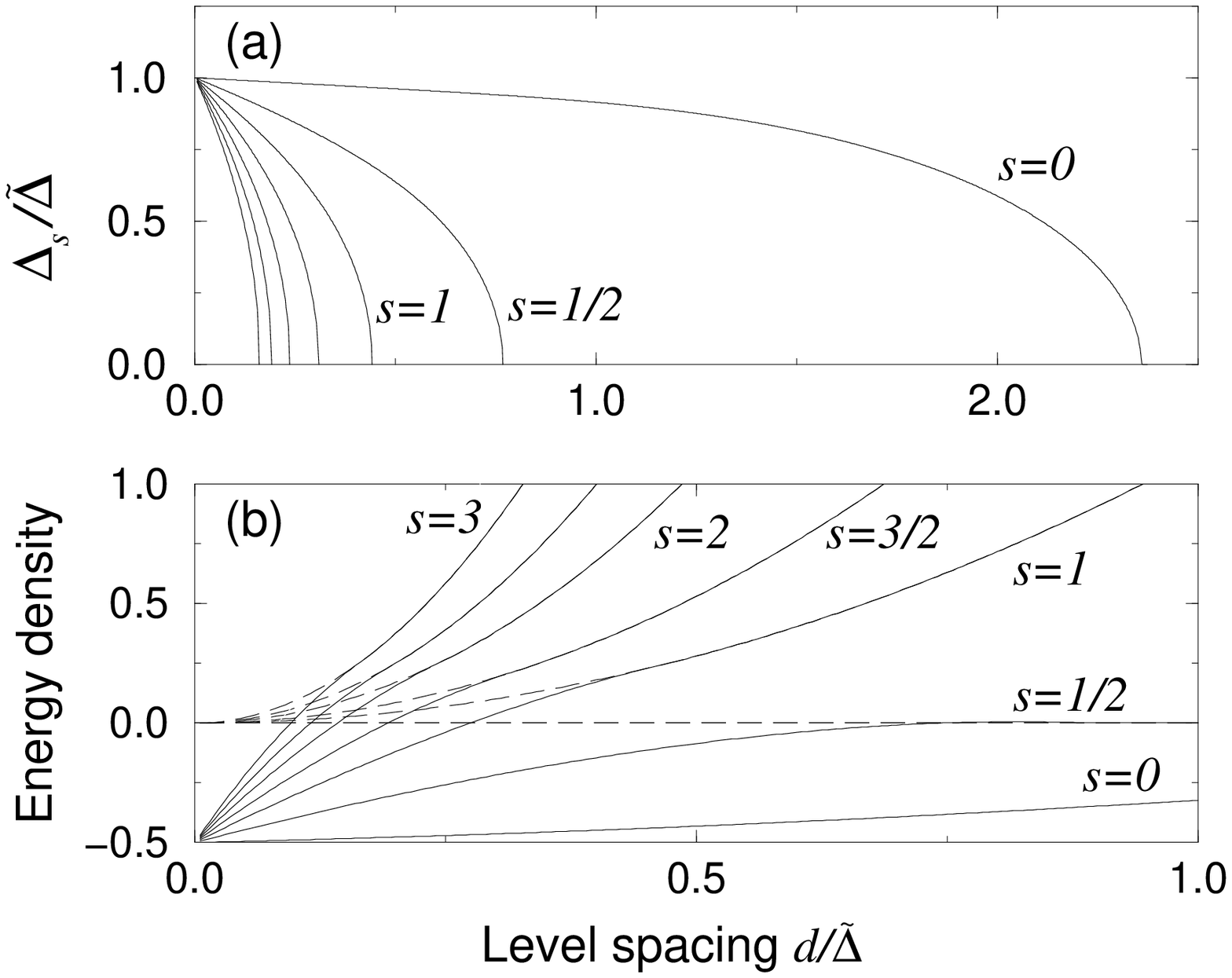,%
height=0.40\linewidth}
\epsfig{figure=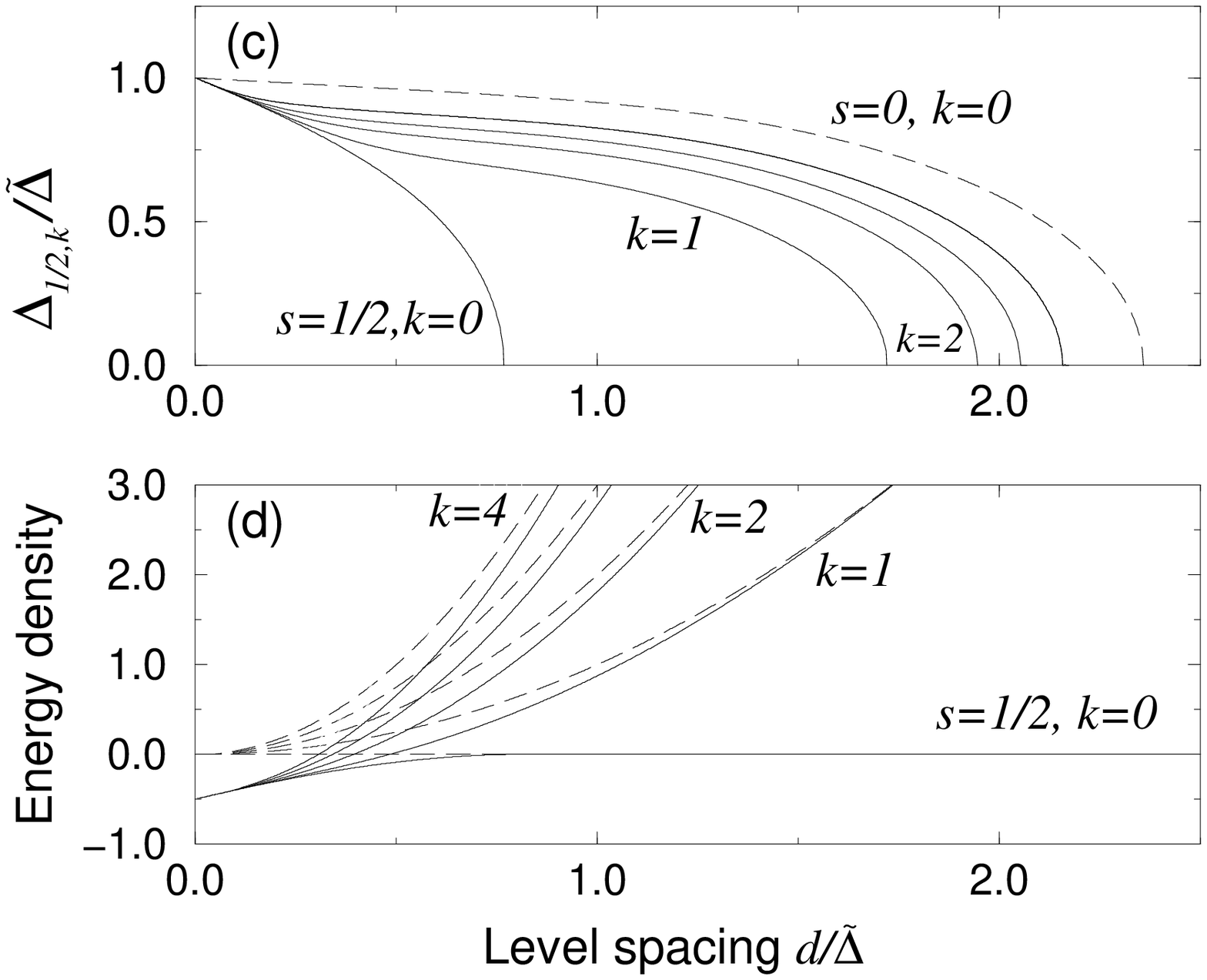,%
height=0.40\linewidth}}
   \caption[Behavior of spin-$s$ ground states and spin-$\half$
  excited states]{Properties of (a,b) spin-$s$ ground states
    $|s\rangle = |s,0\rangle$ [Eq.~(\protect\ref{eq:ansatzs})] and
    (c,d) spin-$\half$ excited states $|\half, k \rangle$ (for $k = 0,
    \dots, 4$) [Eq.~(\ref{eq:ansatzs12})], as functions of $d/\tilde
    \Delta$ (\ie\ decreasing grain size), calculated for $\lambda =
    0.194$.  (a) The pairing parameters $\Delta_s(d)/ \tilde \Delta$,
    which vanish at critical level spacings $d^\GC_{s}$ of 2.36, 0.77,
    0.44, 0.31,$\ldots$ for $s = 0, 1/2, 1, 3/2, \dots$, respectively.
    (c) The pairing parameters $\Delta_{1/2, k}$, together with their
    $k \to \infty$ limit, which equals $\Delta_0$ (dashed line).
    (b,d) show relative energy \emph{densities} (since normalized by
    $d/\tilde \Delta^2 \propto \Vol^{-1}$) at $h=0$ for both
    correlated (solid) and uncorrelated (dashed) states, the latter
    obtained by setting $\Delta_{s, \B} = 0$ in the former.  (b) shows
    $(\E^\GC_s - {\E}^0_{p/2})d / \tilde \Delta^2$ (solid) and
    $({\E}^0_s - {\E}^0_{p/2}) d / \tilde \Delta^2$ (dashed), the
    energy differences of $|s\rangle$ and $|s\rangle_0$ relative to
    the uncorrelated spin-$p/2$ Fermi sea $|p/2\rangle_0$.  (d) shows
    $(\E^\GC_{1/2,k} - {\E}^0_{1/2}) d / \tilde \Delta^2$ (solid) and
    $(\E^0_{1/2,k} - {\E}^0_{1/2}) d / \tilde \Delta^2$ (dashed), the
    energy differences of $|\frac12,k\rangle$ and
    $|\frac12,k\rangle_0$ relative to the uncorrelated spin-$\half$
    ground state $|\frac12,0\rangle_0 = |\frac12\rangle_0$.  Solid and
    dashed lines meet at the critical level spacing $d^\BCS_{s,k}$ at
    which $\Delta_{s , k}$ becomes 0 and the condensation energy
    $E^\cond_{s,k} = {\cal E}^\BCS_{s,k} - {\cal E}_{s,k}^0$
    vanishes.}
    \label{fig:pairing-parameter}
\end{figure}

The simultaneous solution of \Eqs{vj}, (\ref{eq:gap}) and
(\ref{eq:mu}) is a straightforward numerical exercise which Braun and
von Delft performed \cite{braun97,braun99}, for the sake of
``numerical consistency'', without further approximations. (Analytical
solutions can be found only in the limits $d\ll\tilde\Delta$ and
$d\gg\Delta_{s,B}$, see App.~A of \cite{braun99}.) The numerical
results are summarized in \Fig{fig:pairing-parameter}, which shows the
pairing parameters $\Delta_{s,\bbalpha}$
[\Figs{fig:pairing-parameter}(a,c)] and energies
$\E^{\BCS}_{s,\bbalpha}$ [\Figs{fig:pairing-parameter}(b,d), solid
lines] of some selected variational states $|s,\bbalpha \rangle$, as
well as the energies $\E^0_{s,\bbalpha}$ of the corresponding
uncorrelated states $|s,\bbalpha \rangle_0$
[\Figs{fig:pairing-parameter}(b,d), dashed lines]; both
$\E^{\BCS}_{s,\bbalpha}$ and $\E^0_{s,\bbalpha}$ are plotted relative
to the energy $\E^0_{p/2}$ of the uncorrelated spin-$p$ Fermi sea
$|p/2\rangle$.  The results have a number of salient features:

(i) In the bulk limit $d/\tilde \Delta \to 0$, all of the pairing
parameters $\Delta_{s,\B}$ reduce to $\tilde \Delta$, as expected, and
the energy differences $\E^{\BCS}_{s,\B} - \E^0_{p/2}$ between the
correlated states $|s,\B\rangle$ and the uncorrelated Fermi sea $|p/2
\rangle$ reduce to $- \half \tilde \Delta^2 /d = - \half \N(\eF)
\tilde \Delta^2 $, which is the standard bulk result for the
condensation energy.

(ii) Each $\Delta_{s,\bbalpha}$ in \Figs{fig:pairing-parameter}(a,c)
decreases with increasing $d$.  This reflects the fact that with
increasing $d$, the number of pair-correlated states within $\tilde
\Delta$ of $\eF$ decreases [cf.\ \Fig{fig:v2u2-prb97} and the last
paragraph of \Sec{sec:contact-with-BCS}], so that the amount of
pairing correlations, for which $\Delta_{s,\bbalpha}$ is a measure,
decreases too.

(iii) Each $\Delta_{s, \B}$ vanishes abruptly at a critical level
spacing $ d^\GC_{s,\bbalpha}$ (whose precise numerical value depends
sensitively on model assumptions such as the value of $\lambda$ and
the use of uniformly-spaced levels \cite{smith96}).  For $d >
d^\GC_{s,\bbalpha}$ no pairing correlations exist at this level of
approximation, so that that the condensation energy
$E^\cond_{s,\bbalpha} $ (difference between solid and dashed lines)
vanishes and the solid and dashed lines in
\Figs{fig:pairing-parameter}(b,d) meet.

(iv) In \Figs{fig:pairing-parameter}(a), the pairing parameters
$\Delta_s$ for the spin-$s$ ground states decrease rapidly with
increasing $s$ at fixed $d$ (and $d^\GC_{s} < d^\GC_{s'}$ if $s>s'$).
[This is a generalization of a parity effect discussed by von Delft
\etalia\ \cite{vondelft96}, who studied only ground state pairing
correlations and found that these are weaker in odd $(s=1/2)$ grains
than in even $(s=0)$ grains, $\Delta_{1/2} < \Delta_0$, cf.\ 
\Sec{sec:BCS-parity}.]  This tendency is a direct consequence of the
\emph{blocking effect} described in \Sec{sec:generalproperties} and
is independent of model details: larger $s$ means more unpaired
electrons, more terms missing from the sum $\sum_j^\U$, less
correlated pairs and hence smaller $\Delta_{s , \bbalpha}$.

(v) As $d$ increases the blocking effect described in (iv) becomes
stronger, \ie\ the difference between the various $\Delta_{s}$ for
different $s$ becomes more pronounced, since then the relative weight
of each term missing in the sum $\sum_j^\U$ increases.
The blocking effect is most dramatic in the
regime $d/\tilde\Delta\in[0.77, 2.36]$ in which $\Delta_{0} \neq 0 $
but $\Delta_{s \neq 0 } = 0$. This is a regime of ``minimal
superconductivity'' \cite{braun97,braun99}, in the sense that all
pairing correlations that still exist in the \emph{even} 
variational ground state $|0\rangle$ 
(since $\Delta_0 \neq 0$) are completely destroyed by the addition of
a single electron or the flipping of a single spin (since $\Delta_{s
  \neq 0} = 0$).

(vi) Considering the spin-$\half$ excited states $|\half, k \rangle$ of
\Figs{fig:pairing-parameter}(b,d), one finds that the larger $k$, the
longer the pairing correlations survive with increasing $d$: the
critical spacings $d^\GC_{1/2,k} $ increase with $k$, approaching the
value $d^\GC_{0}$ of the spin-0 case as $k \to \infty$;
correspondingly, the larger $k$, the larger the $d$-value at which the
condensation energies $E^\cond_{1/2,k}$ [differences between solid and
dashed lines in \Fig{fig:pairing-parameter}(d)] vanish. The intuitive
reason why the amount of pairing correlations in an excited
$|s,k\rangle$ increases with $k$ is of course quite simple: the
further the unpaired electron sits from the Fermi surface where
pairing correlations are strongest, the less it disrupts the latter
(since $u_k v_k$ becomes very small for large $k$, see
\Fig{fig:v2u2-prb97}).  In fact, the state $|\frac12,k \to \infty
\rangle$ will have just about the same amount of pairing correlations
as the even ground state $|0 \rangle$ ($\Delta_{1/2,k \to \infty}
\simeq \Delta_0$).

(vii) Similar effects hold for excited states in other spin sectors
 (not shown): The higher the excitation, the larger the
pairing parameter $\Delta_{s,\bbalpha}$. However, the concomittant
gain in correlation energy is always less than the
kinetic-energy cost of having an
unpaired electron far from $\eF$. 

(viii) The strong dependence of $\Delta_{s,\bbalpha}$ on $s$ and $d$
for $d \gtrsim \tilde \Delta$ illustrates why in this regime a
conventional mean-field treatment is no longer sufficient: \emph{the
  system cannot be characterized by a {\bf single} pairing parameter,
  since the amount of pairing correlations vary from state to state,
  each of which is characterized by its own pairing parameter.}
Instead, the present variational approach is, roughly speaking,
equivalent to a doing a \emph{separate} mean-field calculation for
each new choice $\U$ of unblocked levels within the Fock space spanned
by them (\ie\ replacing $b_j \to \{ b_j - \langle b_j \rangle \} +
b_j$ and neglecting terms quadratic in the fluctuations $\{b_j -
\langle b_j \rangle \}$).  Indeed, the behavior of $\Delta_{
  s,\bbalpha} (d)$ near $d^\GC_{s,\bbalpha}$ has the standard
mean-field form $\sqrt{1-d/d^\GC_{s,\bbalpha}}$, as can be shown
analytically \cite[App.~A]{braun99}.

To summarize: pairing correlations decrease with increasing $d$ and
$s$ and decreasing $k$. These features survive also in more accurate
canonical calculations. This is not the case, however, for the abrupt
vanishing of $\Delta_{s, \B}$ at $d^\BCS_{s, \B}$, which signals the
breakdown of the \gc\ approach once $d$ becomes of order $\tilde
\Delta$: canonical methods show that, regardless how large $d$
becomes, some remnants of pairing correlations survive and the pairing
parameters $(\Delta_{s,\B})_\can$ do not vanish
[\Sec{sec:sc-canonical}], in accordance with the rule of thumb that
``in a \emph{finite system} no abrupt phase transition can occur
between a zero and nonzero order parameter.''

\section{Softening of the $H$-induced 
transition to a paramagnetic state}
\label{sec:CC-transition}
\label{sec:paramagnetic-breakdown}

Since  states with nonzero spin are favored
by the Zeeman energy but have smaller correlation energy due
to the blocking effect, a competition arises between Zeeman energy and
correlation energy.  The manifestations of the blocking effect can
thus be probed by turning on a magnetic field; if it becomes large
enough to enforce a large spin, excessive blocking will destroy all
pairing correlations.

The situation is analogous to ultra-thin films in a parallel magnetic
field \cite{Meservey-70,Meservey-94}, where orbital diamagnetism is
negligible for geometrical reasons and \emph{superconductivity is
  destroyed at sufficiently large $h$ by Pauli paramagnetism.}  This
occurs via a first order transition to a paramagnetic state, as
predicted by Clogston and Chandrasekhar (CC)
\cite{Clogston-62,Chandrasekhar-62} by the following argument (for
bulk systems): A pure Pauli paramagnet chooses its spin $s$ 
such that the
sum of the kinetic and Zeeman energies, $s^2 /\N(\eF) - 2
h s$, is minimized, and hence has spin $s = h \N(\eF)$ and ground
state energy $-h^2 \N (\eF)$.  When this energy drops below the bulk
correlation energy $-\frac12\tilde\Delta^2 \N(\eF)$ of the
superconducting ground state, which happens at the critical field
$h_{\CC} =\tilde\Delta/\sqrt{2}$, a transition will occur from the
superconducting to the paramagnetic ground state. The transition is
first-order, since the change in spin, from 0 to $ s_{\CC} = h_{\CC}
\N(\eF) = \tilde \Delta / (d \sqrt 2)$, is macroscopically large
($\N(\eF) = 1/d \propto {\rm Vol}$).

\begin{figure}[t]
  \centerline{\epsfig{figure=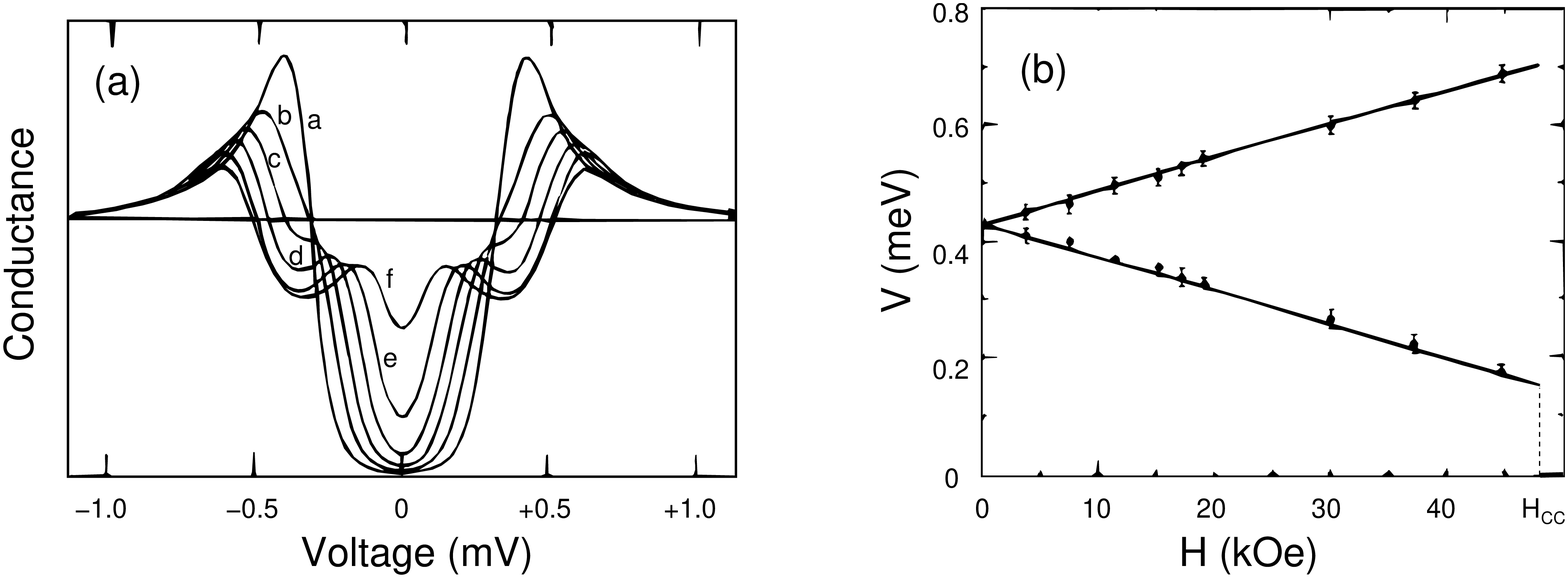,%
width=0.95\linewidth}}
    \caption[Superconductivity in thin Al films]{Thin 
      films in a magnetic field (Figs.~11 and 12 of
      \protect\cite{Meservey-70}). (a) Tunneling conductance from a
      normal metal through a tunnel barrier into a thin
      superconducting Al film, as function of voltage, for several
      magnetic fields labeled in increasing order ``a'' to ``f''. The
      conductance reflects the BCS quasiparticle density of states,
      whose single peak (for a given sign of $V$) for $H=0$ splits
      into two separate peaks for $H\neq0$, corresponding to the
      Zeeman energy difference between quasiparticles with spin up and
      down. (b) Voltage corresponding to the maxima of spin-up and
      spin-down density of states as a function of magnetic field. At
      the critical field $H_\CC$ superconductivity is destroyed and the
      tunneling threshold drops abruptly to zero.}
    \label{fig:MT}
\end{figure}

This transition has been directly observed by Meservey and Tedrow
\cite{Meservey-70,Meservey-94} in ultra-thin (5nm) superconducting Al
films ($\tilde\Delta = 0.38$meV), whose density of states
[\Fig{fig:MT}(a)] they measured via the tunnel conductance through an
oxide layer between a normal metal and the film.  They found that in a
magnetic field the BCS quasiparticle peak splits up into two subpeaks,
separated in energy by $2 \muB H$ [\Fig{fig:MT}(b)], which simply
reflects the Zeeman splitting of quasiparticles states\footnote{Recall
  that the BCS quasiparticles $\gamma^\dag_{j,\sigma} = u_j
  c^\dag_{j,\sigma} - \sigma v_j c^\ds_{j,-\sigma}$ have well-defined
  spins.} with spin up or down (and $g=2$). Remarkably, the tunneling
threshold abruptly dropped to zero at a field of 4.7~T
[\Fig{fig:MT}(b)], which they associated with the field $H_\CC$ at
which the phase transition from the superconducting to the
paramagnetic ground state occurs.  Indeed, \Fig{fig:MT}(b)
demonstrates clearly that the transition to the normal state is first
order: \emph{the mean of the spin-up and spin-down spectral gaps, \ie\ 
  the pairing parameter $\tilde\Delta$, is constant until the critical
  field $H_{\CC}$ is reached, at which it abruptly drops to zero.}

For the case of isolated ultrasmall grains, the above picture of the
transition needs to be rethought in two respects due to the
discreteness of the electronic spectrum: Firstly, the spin must be
treated as a discrete (instead of continuous) variable, whose changes
with increasing $h$ can only take on (parity-conserving) integer
values.  Secondly, one needs to consider more carefully the
possibility of $h$-induced transitions to nonzero spin states that are
still \emph{pair-correlated} (instead of being purely paramagnetic),
such as the variational states $ |s, \bbalpha\rangle $ discussed
above.  (In the bulk case, it is obvious that such states play no
role: the lowest pair-correlated state with nonzero spin obtainable
from the ground state by spin flips is a two-quasiparticle state,
costing energy $2\tilde \Delta - 2h$; when $h$ is increased from 0,
the paramagnetic transition at $h_{\CC} = \tilde \Delta/\sqrt 2$ thus
occurs before a transition to this state, which would require $h =
\tilde \Delta$, can occur.)

\begin{figure}[t]
  \centerline{\epsfig{figure=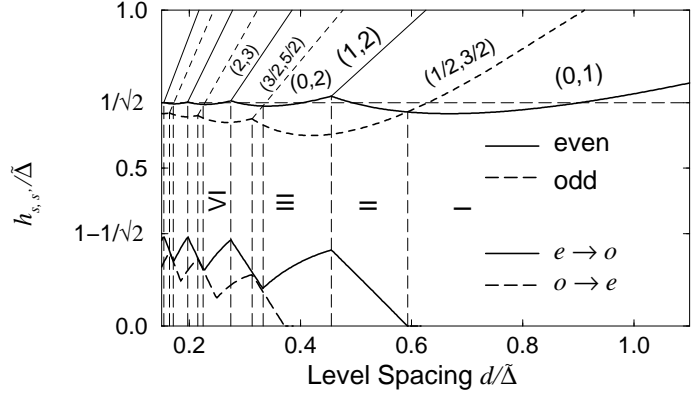,%
width=0.65\linewidth}}
  \caption[$d$-dependence of the 
  level-crossing fields $h_{s,s'} (d)$.]{$d$-dependence of the
    level-crossing fields $h_{s,s'} (d)/ \tilde \Delta$
    [Eq.~(\ref{eq:hcrit})] at which $\E^\BCS_{s'}$ drops below
    $\E^\BCS_s$ with increasing $h$. Only those level crossing fields
    are shown that belong to the cascade of (\emph{fixed}-$N$) ground
    state changes (GSCs) $(s_0,s_1)$; $(s_1,s_2)$; \dots that occur as
    $h$ increases from 0 at given $d$.  Solid (dashed) lines are used
    for even (odd) grains with integer (half-integer) spins, and some
    are labeled by the associated GSC $(s,s')$. (In contrast, in
    \Fig{fig:spectra} the $N$-\emph{changing} tunneling transitions in
    are labeled by $|s_i\rangle \to |s_f\rangle$.)  The size $|\Delta
    \E_{s_1,f'} - \Delta \E_{s_0,f}|$ of the first jump (occuring at
    the level-crossing field $h_{\CC}(p,d) = h_{s_0,s_1}$) in the
    lowest line of the tunneling spectra of \Fig{fig:spectra} is shown
    by the lowest two (jagged) curves (solid for $e\to o$ and dashed
    for $o \to e$ tunneling spectra), which both approach the CC value
    $1 - 1/\sqrt2$ as $d \to 0$.}
      \label{fig:h-crit}
\end{figure}

Quite generally, the effect of increasing $h$ from 0
can be analyzed as follows: At given $d$ and $h$, the grain's ground
state is the lowest-energy state among all possible spin-$s$ ground
states $|s \rangle$ having the correct parity $p = 2s\,{\rm mod}\,2$.
Since $\E_{s}(h,d) = \E_{s}(0,d) - 2hs$, level crossings occur with
increasing $h$, with $ \E_{s'}$ dropping below $\E_s$ at the
\emph{level crossing field}
\begin{eqnarray}
  \label{eq:hcrit}
  h_{s,s'} (d) = \frac{\E_{s'}(0,d)-\E_s(0,d)}{2(s'-s)}. 
\end{eqnarray}
Therefore, as $h$ is slowly turned on from zero with initial ground
state $|s_0 = p/2\rangle$, a cascade of successive ground-state
changes (GSCs) to new ground states $|s_1 \rangle$, $|s_2 \rangle$,
\dots will occur at the fields $h_{s_0, s_1}$, $h_{s_1, s_2}$, \dots
Let us denote this cascade by $(s_0,s_1); (s_1,s_2);\ldots$; for each
of its GSCs the corresponding level-crossing fields $h_{s,s'} (d)$ is
shown in \Fig{fig:h-crit}.  Generalizing CC's critical field to
nonzero $d$, let us denote the (parity-dependent) field at which the
\emph{first} transition $(s_0, s_1)$ occurs by $h_{\CC} (d,p) \equiv
h_{s_0, s_1} (d) $, which simply is the lower envelope of the
level-crossing fields $h_{s_0, s_1}$ in \Fig{fig:h-crit} (shown as
bold solid and dashed lines for $s_0=0$ and $s_0=\frac12$,
respectively).  In the limit $d\to0$ it is numerically found
to reduce to the Clogston-Chandrasekhar value, \ie\ $h_{\CC}(0,p) =
\tilde\Delta/\sqrt{2}$, as expected. 

In general, the order in which the GSCs occur with increasing $h$
within a cascade (\ie\ the order of $h_{s,s'}$ lines encountered when
moving vertically upward in \Fig{fig:h-crit}) depends sensitively on
$d$ and an infinite number of distinct regimes (cascades) I, II, III,
\dots can be distinguished: Starting at large $d$ we find the typical
normal-grain behavior $(0,1); (1,2); (2,3); \ldots$ for even grains
and $(\frac12,\frac32); (\frac32,\frac52); \ldots$ for odd grains,
with $h_{0,1} < $ (or $>$) $ h_{\frac12,\frac32}$ in regimes I (or
II).  In regimes III and IV of somewhat smaller $d$, the order of GSCs
is $(0,2); (2,3); \ldots$ and $(\frac12,\frac32); (\frac32,\frac52);
\ldots$, etc, \ie\ the spin $s_1$ attained after the first GSC
$(s_0,s_1)$ has increased to 2 in the even case.  This illustrates a
general trend: \emph{the spin $s_1 (d)$ after the first transition
  increases with decreasing $d$ and becomes macroscopically large in
  the $d\to0$ limit}, where $s_1 = h_{\CC} / d = \tilde \Delta / (d
\sqrt 2)$, as explained in recounting CC's argument above.

Furthermore, it turns out that $\Delta_{s_1} (d)= 0$ and therefore
$\E^\GC_{s_1}=0$ for \emph{all} $d$, implying that after the first GSC
the new ground state $|s_1\rangle$ is \emph{always} (not only in CC's
bulk limit) a purely paramagnetic state, \ie\ without any pairing
correlations in the \gc\ framework (canonical calculations would yield
some weak remnant pairing correlations in the form of fluctuations).
In this regard, CC's picture of the transition remains valid
throughout as $d$ is increased: at $h_{\CC} (d,p)$, a transition
occurs from the superconducting ground state to a paramagnetic,
\emph{uncorrelated} state $|s_1 \rangle_0$, the transition being
first-order in the sense that $\Delta_{s_1} (d)= 0$; however,
\emph{the first-order transition is ``softened'' with increasing $d$,
  in the sense that the size of the spin change, $s_1 - s_0$,
  decreases from being macroscopically large in the bulk to being
  equal 1 at $d \gg \tilde \Delta$ (regimes I and II)}.

To conclude this section, we mention that the above analysis of the
paramagnetic breakdown of superconductivity has recently been
generalized to finite temperatures
%%??%% by Canosa and Rossignoli 
\cite{Rossignoli-00}, using the so-called static path approximation
[explained in \Sec{sec:SPA}] to treat fluctuation effects properly. 
%%??%% It is found that the latter increasingly smooth out the
%superconducting-to-paramagnetic transition as the particle size
%diminishes.

\section{Excitation spectrum in a magnetic field}
\label{sec:tunneling-spectra-prb97}
\label{sec:tunneling-spectra}
\label{sec:magfield}

In this section we compare the theoretical tunneling spectra for a
grain coupled to leads, calculated as functions of $h$ and $d$
\cite{braun97,braun99,braun-thesis}, and compare these to RBT's
measurements of \Fig{fig:sc-magneticfield}.

The form of the tunneling spectrum depends in a distinct way on the
specific choice of level spacing $d$ and on the electron number parity
$p$ of the final states $|f\rangle$ of the bottleneck tunneling
processes $|i\rangle \to |f \rangle$ (or $|\alpha' \rangle \to |\alpha
\rangle$ in the notation of \Sec{subsec:ultrasmallset}).
However, for the uniformly spaced $\varepsilon_j$-levels used here,
particle-hole symmetry ensures that there is no difference between
electron addition or removal spectra $|i_{N \mp 1} \rangle \to |f_N
\rangle$.  To calculate the spectrum \emph{for given $d$ and $p$},
Braun \etalia\ \cite{braun97,braun99,braun-thesis} proceeded as
follows: they first analyzed at each magnetic field $h$ which
tunneling processes $|i \rangle \to |f \rangle$ are possible, then
calculated the corresponding tunneling energy thresholds $\Delta
\E_{if} (h) \equiv \E_f(h) - \E_i(h)$ [cf.\ \Eq{eq:Vrthresholds}] and
plotted $\Delta \E_{if} (h) - \Delta\E_{{\rm min}} (0)$ as functions
of $h$ for various combinations of $|i\rangle$ and $|f\rangle$, each
of which gives a line in the spectrum.  Since the selection rule $s_f
- s_i = \pm 1/2$ holds, only slopes of $\pm 1$ can occur.  The reason
for subtracting $\Delta\E_{{\rm min}} (0)$, the $h=0$ threshold energy
cost for the \emph{first} (lowest-lying) transition, is that in
experiment, this energy depends on $V_g$ and hence yields no
significant information (see also Sec.~2.4.6 of \Ref{PR-VDR}).
Neglecting
nonequilibrium effects \cite{rbt97,agam97a,agam97b,agam98} (which were
minimized in the present experiment by tuning $\Vg$),
the initial state is always taken to be
the ground state of a given spin-$s$ sector.  The appropriate
$s_i(h,d)$ must be determined from Fig.~\ref{fig:h-crit}.

\begin{figure}[t]
  \centerline{\epsfig{figure=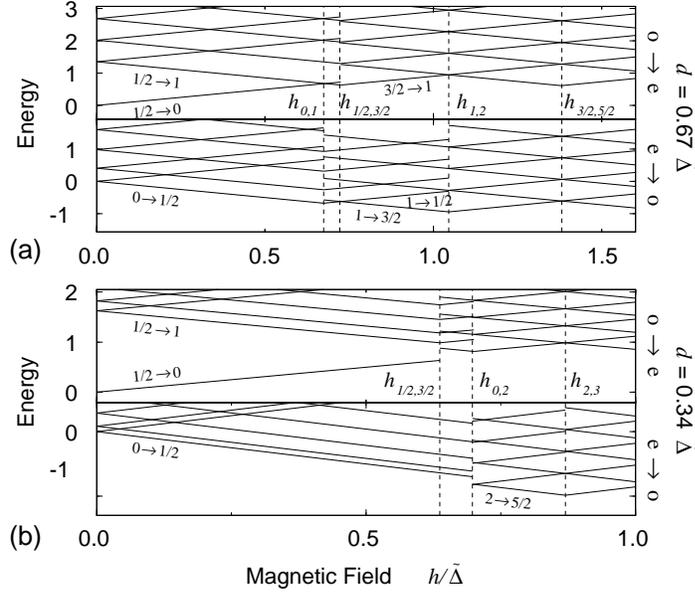,%
width=0.7\linewidth}}
  \caption[Theoretical tunneling spectra]{The theoretical 
    odd-to-even and even-to-odd tunneling spectra $(\Delta \E_{if} -
    \Delta\E_{{\rm min}} (0))/ \tilde \Delta $ predicted for an
    ultrasmall superconducting grain as a function of magnetic field
    $h$, to be compared with the experimental spectra of
 \Fig{fig:sc-magneticfield}, 
for two different level spacings: (a) $d=0.67\tilde\Delta$
    and (b) $d=0.34\tilde\Delta$ (corresponding to regimes I and III
    of Fig.~\protect\ref{fig:h-crit}, respectively).  Some lines are
    labeled by the corresponding $s_i\to s'_i$ tunneling transitions.
    Not all possible higher lines (corresponding to excited final
    states $|s,j\rangle$) are shown.  Vertical dashed lines indicate
    those level-crossing fields $h_{s,s'}$ [\Eq{eq:hcrit}]
    at which kinks or jumps occur, with $h_{0,1}<h_{1/2,3/2}
    < h_{1,2} < h_{3/2,5/2}$ in (a) and 
    $h_{1/2,3/2}<h_{0,2} < h_{2,3}$ in (b). }
      \label{fig:spectra}
%%??%% make bounding box thicker than main lines.
\end{figure}

\Fig{fig:spectra} shows four typical examples of such theoretical
tunneling spectra, with some lines labeled by the corresponding $|i
\rangle \to |f \rangle$ transitions.  Whenever $h$ passes through one
of the level-crossing fields $h_{s_i, s_{i'}}$ of \Eq{eq:hcrit},
the grain experiences a ground state change $(s_i, s_{i'})$, at which
the set of allowed tunneling transitions changes from $|s_{i} \rangle
\to \{ |s_{f} \rangle \}$ to $|s_{i'} \rangle \to \{ |s_{f'} \rangle
\}$. Therefore, at $h_{ s_i, s_{i'}}$ one set of lines in the
tunneling spectrum ends and another begins, producing kinks or
discontinuities.  A \emph{kink} occurs if one of the new final states
coincides with one of the old ones, $|f'\rangle = |f\rangle$, meaning
that it can be reached from both $|s_i \rangle $ and $|s_{i'} \rangle$
[\ie\ $s_f - s_i = - (s_f - s_{i'})$], in which case $\Delta \E_{if}
(h)$ and $\Delta\E_{if'} (h)$ have slopes of opposite sign.  However,
for most lines this is not the case, so that at $h_{ s_i, s_i'}$ the
line $|s_{i}\rangle \to |f \rangle$ simply ends while new lines
$|s_{i'} \rangle \to |f' \rangle$ begin.  This results in
\emph{discontinuities} (or ``jumps'') in the spectrum at $h_{ s_i, s_i'}$ of
size $(\Delta\E_{i'f'} - \Delta \E_{if})(h_{ s_i, s_i'})$, unless by
chance some other final state $|f '\rangle $ happens to exist for
which this difference equals zero.

Since the order in which the GSCs $(s_i, s_{i'})$ occur as functions
of increasing $h$ depend on $d$ and $p$, as indicated by the distinct
regimes I, II, III, \ldots in \Fig{fig:h-crit}, one finds a
distinct kind of tunneling spectrum for each regime, differing from
the others in the positions of its jumps and kinks.  In regime~I,
where the order of occurrence of GSCs with increasing $h$ is $(0,1);
(\frac12,\frac32); (1,2); (\frac32,\frac52);\ldots$, there are no
discontinuities in the evolution of the lowest line [see
\Fig{fig:spectra}(a)].  For example, for the $e\to o$ spectrum,
the lowest $|0 \rangle \to |1/2\rangle$ line changes
\emph{continuously} to $|1\rangle \to |1/2\rangle$ at $h_{0,1}$, since
$|s_f - s'_i | = 1/2$.  However, in all other regimes the first change
in ground state spin (at $h_{0, s_1}$ from 0 to $s_1$) is $ > 1$,
implying a \emph{jump} (though possibly small) in all $e\to o$ lines,
as illustrated by \Fig{fig:spectra}(b).

The jump's magnitude for the tunneling thresholds, \ie\ the
\emph{lowest} $e\to o$ and $o\to e$ lines, is shown as function of $d$
in the lower part of \Fig{fig:h-crit}.  It starts at $d=0$ from the CC
value $\tilde \Delta (1 - 1/\sqrt2)$ measured for thin Al films
\cite{Meservey-70,Meservey-94}, and with increasing $d$ decreases to 0
(non-monotonically, due to the discrete spectrum).  This
\emph{decrease of the size of the jump in the tunneling threshold}
reflects the fact, discussed in \Sec{sec:CC-transition}, that the
change in spin at the first ground state change $(s_0,s_1)$ decreases
with increasing $d$ (as $s_1 \!-\! s_0 \sim h_{\CC}/d$), and signals
the softening of the first-order superconducting-to-paramagnetic
transition.
 
The fact that the measured tunneling thresholds in
\Fig{fig:experimental-spectra} show no jumps at all, which might at
first seem surprising when contrasted to the threshold jumps seen at
$h_{\CC}$ in \Fig{fig:MT} for thin films in a parallel field
\cite{Meservey-70,Meservey-94}, can therefore naturally be explained
\cite{braun97,braun99} by assuming the grain to lie in the ``minimal
superconductivity'' regime I of \Fig{fig:h-crit} (where the jump size
predicted in \Fig{fig:h-crit} is zero).  Indeed, \emph{the overall
  evolution (\ie\ order and position of kinks, etc.)  of the lowest
  lines of \Fig{fig:experimental-spectra} qualitatively agrees with
  those of a regime I tunneling spectrum, \Fig{fig:spectra}(a)}. This
important result rather convincingly establishes the phenomenological
success of the \dbcsm. It also allows one to deduce the following
values for the level-crossing fields $H_{s_i,s'_i}$ (indicated by
vertical dashed lines in Figs.~\ref{fig:experimental-spectra}
and~\ref{fig:spectra}): $H_{0,1} = 4$T, $H_{1/2,3/2} = 4.25$T,
$H_{1,2} = 5.25$T and $H_{3/2,5/2} = 6.5$T. \label{HSS} As
corresponding uncertainties we take $\Delta H_{s_i,s'_i} = 0.13$T,
which is half the $H$ resolution of 0.25T used in experiment.

By combining the above $H_{s_i,s'_i}$ values with \Fig{fig:h-crit},
some of the grain's less-well-known parameters can be determined
somewhat more precisely: 
\begin{itemize}
\item[(i)]
To estimate the grain's ``bulk $H_{\CC}$'',
note that since $H_{1/2,3/2} / H_{0,1} \simeq 1.06$, this grain lies
just to the right of the boundary between regions II and I in
\Fig{fig:h-crit} where $d/ \tilde \Delta \simeq 0.63$,
at which we
have $h_{0,1} / h_{\CC} \simeq 0.95$, so that $H_{\CC} = H_{0,1}/ 0.95
\simeq 4.2$~T.  This is quite close to the value $H_{\CC} \simeq 4.7$~T
found experimentally \cite{Meservey-70,Meservey-94} in thin films in a
parallel field, confirming our expectation that these correspond to
the ``bulk limit'' of ultrasmall grains as far as paramagnetism is
concerned.
\item[(ii)] The grain's corresponding bulk gap is $\tilde \Delta =
  \sqrt 2 \mu_B H_{\CC} \simeq 0.34$~meV, implying a coupling constant
  of $\lambda = 0.189$ [by \Eq{eq:lambda-definition}].  \emph{A
    posteriori}, these values can be regarded as being more
  appropriate for the present grain than the choices $\tilde \Delta =
  0.38$~meV and $\lambda = 0.194$ made in \Sec{sec:generalnumerics},
  though the differences are evidently not significant (12\% for
  $\tilde \Delta$ and 3\% for $\lambda$).
\item[(iii)] 
The mean level spacing implied
by $d/ \tilde \Delta \simeq 0.63$ is $d \simeq 0.21$~meV.
\label{p:estimate-d}
The crude volume-based value $d \simeq 0.45$~meV cited in
the caption of \Fig{fig:generic-IV} thus seems to have been an
overestimate.  It would be useful if this determination of $d$
could be checked via an independent accurate experimental
determination of $d$ directly from the spacing of lines in the
excitation spectrum. Regrettably, this is not possible: the measured
levels are shifted together by pairing interactions, implying that
their spacing does not reflect the mean \emph{independent}-electron
level spacing $d$. Nevertheless, note that the
measured spacing of $0.05$~meV 
between the lowest two states
of the  odd grain agrees quite well with the crude
BCS estimate $\sqrt{\tilde \Delta^2 + d^2} - \tilde \Delta$
[cf.\ \Eq{vj-bulk}], which gives $0.06$~meV when evaluated for
$d=0.21$~meV and $\tilde \Delta = 0.34$~meV. 
%%??%% What about using Richardson here?
\end{itemize}

The higher lines plotted in \Fig{fig:spectra} correspond to
transitions into spin-$s_f$ state of the form $|s_f, k \rangle$ [cf.\ 
\Eq{eq:ansatzs12} and \Fig{fig:alpha-states}(d)] (for simplicity
these were the only ones considered in
\cite{braun97,braun99,braun-thesis}, though in general others are
expected to occur too). The jumps in these lines, \eg\ in
\Fig{fig:spectra}(a) at $h_{1,2}$, occur whenever the two final
excited states $|s_f, k_f \rangle$ and $|s_{f'}, k_{f'} \rangle$
before and after the GSC at $h_{s_i, s'_i}$ have different correlation
energies.  (Recall that the correlation energy of an excited state
$|s_f, \bbalpha_f \rangle$ can be nonzero even if that of the
corresponding ground state $|s_f \rangle$ is zero, since the former's
unpaired electrons are further away from $\eF$, so that
$\Delta_{s_f,\bbalpha_f}>\Delta_{s_f}$, see point (vi) of
\Sec{excited}.)  Experimentally, these jumps have not been observed.
This may be because up-moving resonances lose amplitude and are
difficult to follow \cite{rbt97} with increasing $h$, or because the
widths of the excited resonances ($\simeq 0.13\tilde\Delta$) limit
energy resolution \cite{agam97a,agam97b,agam98}.

For somewhat larger grains, the present theory predicts jumps even in
the lowest line, as illustrated in \Fig{fig:spectra}(b).  It remains
to be investigated, though, whether orbital effects, which rapidly
increase with the grain size, would not smooth out such jumps.

To conclude this section, we emphasize once again that more than
qualitative agreement between theory and experiment can not be
expected, since both the model and our variational treatment thereof
are very crude: the model neglects, for instance, fluctuations in
level spacing and in pair-coupling constants, and the \gc\ wave
functions become unreliable for $d /\tilde \Delta \gtrsim 0.5$.
Furthermore, we neglected nonequilibrium effects in the tunneling
process and assumed equal tunneling matrix elements for all processes.
In reality, though, random variations of tunneling matrix elements
could suppress some tunneling processes which would otherwise be
expected theoretically.

\section{Measurable consequences of the 
 blocking effect: parity effects}
\label{sec:BCS-parity}

This section is devoted to various measurable manifestations of
the blocking effect, in the form of parity effects,
\ie\ differences between a grain with an even or
odd number of electrons.

\subsection{Bulk consequences of blocking}
\label{sec:two-qp}

The most obvious measurable manifestation of the blocking effect is 
the very existence of a spectral gap: ``breaking a pair'' and
placing the two newly unpaired electrons in two singly-occupied levels
costs a significant amount of correlation energy, because the unpaired
electrons loose pairing energy themselves and also 
disrupt the pairing correlations of the other pairs. This,
of course, is already present in standard bulk mean-field BCS theory
via the energy cost of at least $2 \tilde \Delta$ involved in creating
two quasiparticles, and is one of the hallmarks of superconductivity.

In the context of ultrasmall grains, let 
us denote the \emph{pair-breaking energies}
for an even (odd) grain, \ie\ the  minimum
energy cost per electron for breaking a pair by flipping a
single spin at $h=0$, by $\Omega_e$ ($\Omega_o$):
\begin{equation}
  \label{eq:spectralgaps}
      \Omega_e \equiv \half (\E_1 - \E_0)_{h=0}
%= h_{0,1} 
, \qquad   
\Omega_o \equiv \half (\E_{3/2} - \E_{1/2})_{h=0} \, . 
% = h_{1/2,3/2}.
\end{equation}
%[$h_{0,1}$ and $h_{1/2,3/2}$ are level crossing fields, see
%\Eq{eq:hcrit}.] 
The even pair-breaking gap $\Omega_e$ is of course strikingly visible
in RBT's $h=0$ spectra as a large spectral gap for even grains [cf.\ 
\Figs{fig:sc-spectra(h=0)} and \ref{fig:sc-magneticfield}; the latter
gives $\Omega_e = 0.26$~meV].  Its presence is direct evidence for the
existence of pairing correlations in the grain, which in that sense
can still be called ``superconducting''. 

In contrast, the odd pair-breaking gap $\Omega_o$ can not be obtained
from $h=0$ spectra, since in an odd grain the lowest excitation does
not involve breaking a pair, but simply exciting the unpaired
electron, which does not require a correlation-induced gap to be
overcome.  To measure $\Omega_o$, a finite field is needed: by
\Eq{eq:hcrit}, $ \Omega_e = h_{0,1}$ and $\Omega_o = h_{1/2,3/2}$,
hence both spin-flip gaps are equal to level-crossing fields that can
be deduced from $h \neq 0$ data, as explained in
\Sec{sec:tunneling-spectra-prb97}.  For \Fig{fig:sc-magneticfield}
this yields $ \Omega_e = 0.23\pm 0.01$~meV and $\Omega_o = 0.24 \pm
0.01$~meV [a result further discussed in \Sec{sec:parity-pairbreaking}].
The reason that the $\Omega_e$-value determined in this way
is somewhat {\em smaller}\/ than the above-mentioned 0.26~meV
determined at $h=0$ is presumably that the experimental
spectral lines are not perfectly linear in $h$ (having a small
$h^2$-contribution due to orbital diamagnetism,
 which should cause the spectroscopic gap to
decrease faster with $h$ than in our model).

Another consequence of the blocking effect is that the condensation energies
$E^\cond_{p/2} = \E_{p/2} - \E_{p/2}^0$ for an even and odd grain differ: the
unpaired electron of an odd grain weakens its pairing correlations relative to
an even grain, so that $E^\cond_{1/2}$ is less negative than $E^\cond_0$.  In
the bulk limit their difference approaches $ E^\cond_{1/2} - E^\cond_0 \to
\tilde \Delta$, the energy of a single quasiparticle.  For large mesoscopic
islands (with $d/\tilde \Delta \ll 1$) this energy difference has indeed been
directly observed: it causes a change from $e$- to $2e$-periodicity in the
gate-voltage dependence of Coulomb oscillations
\cite{hanna91,Tuominen-92,Tuominen-93,Tinkham-95,Saclay,eiles93}.  For
ultrasmall grains, however, ground state energy differences are currently not
directly measurable, due to experimental difficulties\footnote{In brief: when
  $\Vg$ is changed over a sufficiently large range to see Coulomb oscillations
  ($\sim e/C_\ssg \simeq 1 $~V), sudden rigid shifts in the background off-set
  charge $Q_0$ are encountered at random values of $\Vg$, which spoil the
  $2e$-periodicity which would have been expected otherwise.
\label{f:offset}} explained in
detail in Sec.~2.4.6 of \Ref{PR-VDR}.

The parity effects discussed above survive in the bulk limit. Let us
now turn to parity effects that result from even-odd differences in
the \emph{$d$-dependence} of various quantities.

\subsection{Parity-dependent pairing parameters}

As is evident from
\Fig{fig:pairing-parameter}(a,b), not only the condensation energies
$E_{p/2}$ are parity dependent; as soon as one leaves the bulk regime,
the pairing parameters $\Delta_{p/2}$ become parity-dependent too,
with $\Delta_{0} > \Delta_{1/2}$.  In the context of ultrasmall grains
this was first emphasized by von Delft \etalia\ 
\cite{vondelft96}, but it had been anticipated before by Janko, Smith
and Ambegaokar \cite{Janko-94} and Golubev and Zaikin \cite{Golubev-94},
who had studied the first correction to the bulk limit, finding
$\Delta_0 - \Delta_{1/2} = d/2$ to leading order in $d/\tilde
\Delta$; and this result, in turn, had already been published by
Soloviev in the nuclear physics literature as long ago as 1961
\cite{Soloviev-61}.

The \gc\ results of \Fig{fig:pairing-parameter}(a), in particular the
fact that the critical level spacing $d^\BCS_{p/2}$ at which
$\Delta_{p/2}$ vanishes is smaller for odd than even grains
($d_{1/2}^\BCS < d_{0}^\BCS$), suggest that ``pairing correlations
break down sooner in odd than even grains'' \cite{vondelft96}.
However, it should be remembered that the vanishing of $\Delta_{p/2}$
signals the breakdown of the \gc\ approach. A more accurate statement,
that is born out by the canonical calculations reviewed in
\Sec{sec:sc-canonical}, is that the inequality $\E_{1/2} > \E_0$
persists for arbitrarily large $d$ (see \Fig{fig:exact-gse} in
\Sec{comparison}), \ie\ pairing correlations are always weaker for odd
than even grains, although they never vanish altogether in either.

\subsection{Matveev-Larkin parity parameter}
\label{sec:ML-parameter}

\begin{figure}[t]
  \centerline{\epsfig{figure=%
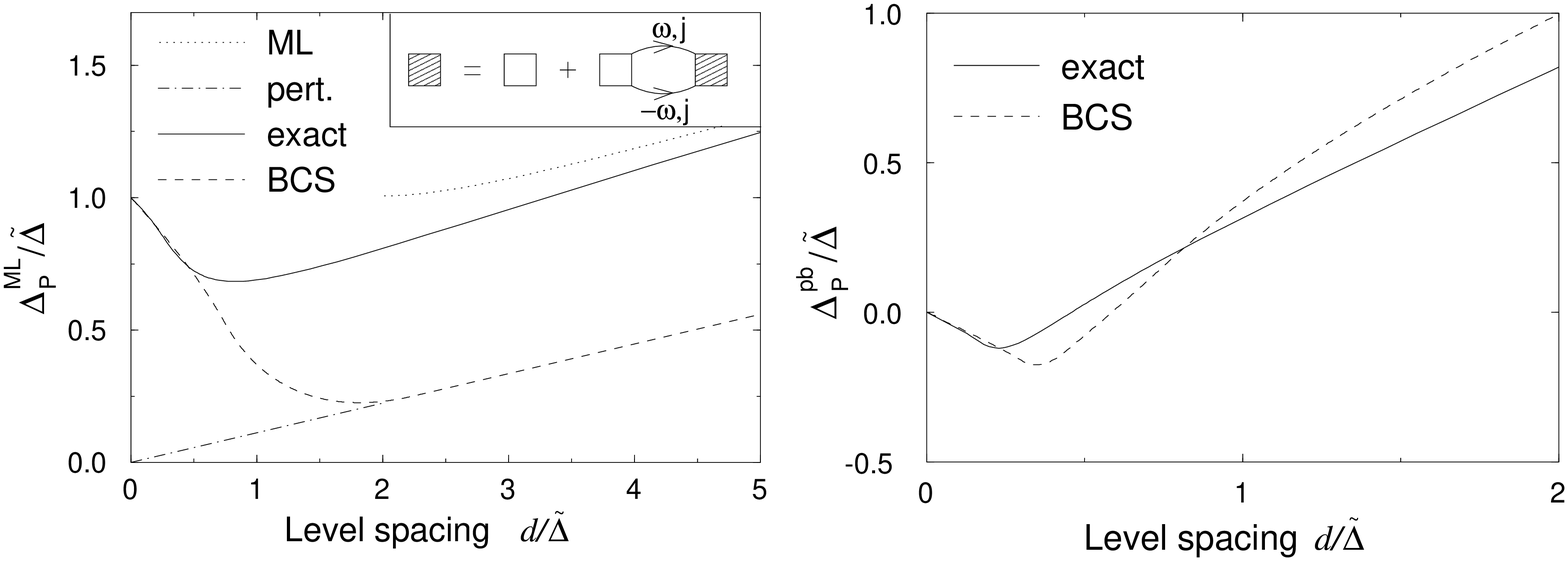,height=0.34\linewidth}}
  \caption[Two parity parameters]{The 
    parity parameters (a) $\Delta^\ML_\PP$ of Matveev-Larkin
    [\Eq{eq:def-even-odd-effects}] and (b) $\Delta^\pb_\PP$ for the
    pair-breaking energies [\Eq {eq:Delta-pairbreaking}], as functions
    of $d/\tilde \Delta$, calculated using the \gc\ variational BCS
    approach of \Sec{sec:BCS-ansatz} (dashed lines), and Richardson's
    exact solution of \Sec{sec:richardson} (solid lines).  In (a), we
    also show the perturbative result for the uncorrelated Fermi sea,
    $(\Delta_\PP^\ML)_\pert= \frac12 \lambda d$ (straight dash-dotted
    line); and the renormalized result $(\Delta_\PP^{\ML})_\ren \simeq
    d/[2\ln(ad/\tilde\Delta)]$ of \Eq{MLren}, in its range of validity
    $d / \tilde \Delta \gg 1$ (dotted line).  The parameter $a=1.35$
    is chosen to ensure quantitative agreement with the exact result in
    the limit $d/\tilde \Delta \gg 1$.  For a summary of the results
    of various other canonical calculations of $\Delta_\PP^{\ML}$, see
    Fig.~\protect\ref{fig:exact-gse}. The inset of (a) shows the Dyson
    equation used to calculated the renormalized coupling $\tilde
    \lambda$ in \Eq{eq:lambda-Dyson}.}
  \label{fig:spectral-gap}
  \label{fig:Matveev-parity}
\end{figure}
To capture the difference between correlations in even and odd grains
in terms of measurable quantities (which $\Delta_{p/2}$ are not),
Matveev and Larkin \cite{matveev97} proposed the parameter (sometimes
called ``pairing energy'' in nuclear physics \cite{richardson65a})
\begin{eqnarray}
  \label{eq:def-even-odd-effects}
  \Delta^\ML_\PP & \equiv & 
\E_{1/2}^{N+1} - \half (\E_0^{N}  + \E_0^{N+2} )
  \qquad \mbox{(where $N$ is even)} \, ,
\end{eqnarray}
\ie\ the difference between the ground state energy of an odd grain
and the mean of the ground state energies of the even grains obtained
by removing or adding one electron.  \Fig{fig:Matveev-parity}(a) shows
its behavior as function of $d/ \tilde \Delta$.  In the bulk limit we
have $\E_0^{N} \simeq \E_0^{N+2} $ and $\Delta^\ML_\PP \simeq \tilde
\Delta$, which is simply the energy cost for having an unpaired
electron on the odd grain.  With increasing $d/\tilde \Delta$, this
energy cost decreases since pairing correlations get weaker, hence
$\Delta_\PP^\ML$ initially decreases.  It begins to increase again for
$d \gtrsim \tilde \Delta$, since then pairing correlations are so weak
that the behavior of $\Delta_\PP^\ML$ is governed by the
``self-energy'' of the one extra pair in $\E_0^{N+2}$ relative to
$\E_0^N$. For example, in the \gc\ variational BCS result for
$\Delta^\ML_\PP$, namely
\begin{equation}
  \label{eq:ML-BCS}
(\Delta^{\ML}_\PP)_\BCS = \E^\BCS_{1/2}
- \E^\BCS_0 + \lambda d/2 \, ,   
\end{equation}
it is this self-energy which produces the $ \lambda d/2$ contribution.

A more careful calculation for the regime $d \gg \tilde \Delta$ was
performed by Matveev and Larkin \cite{matveev97}, who considered the
renormalization of $\lambda$ due to ``pairing fluctuations'' about the
uncorrelated Fermi ground state $|p/2 \rangle_0$. Summing up the
leading logarithmic vertex corrections \cite{AGD63} [see inset of
\Fig{fig:Matveev-parity}(a)], they obtained a renormalized coupling
$\tilde \lambda$ given, with logarithmic accuracy, by
\begin{eqnarray}
  \label{eq:lambda-Dyson}
  \tilde \lambda & = & \lambda + \lambda d \left[ \sum_j^U \int
   { \ddr \omega \over 2 \pi} 
   { 1 \over [ \ii \omega - (\varepsilon_j -    \mu) ] }
   { 1 \over [ - \ii \omega - (\varepsilon_j -    \mu) ] } \right] \tilde
   \lambda\\
  \label{eq:renormalized-lambda}
\tilde \lambda & = & {\lambda \over 1 - \lambda d \sum_j^U  {1 \over
2 | \varepsilon_j - \mu|}}  \; \simeq \; 
{ \lambda \over 1 - \lambda \log (\omegaD/d)} \; .
\end{eqnarray}
This result evidently is valid only if $d \gg \omegaD \eer^{-1/\lambda}
\simeq \tilde \Delta/2$ (which, incidentally, is another way of seeing
that $d \simeq \tilde \Delta$ defines the crossover between the
fluctuation-dominated and bulk regimes).  Matveev and Larkin concluded
that
\begin{equation}
  \label{MLren}
  (\Delta_\PP^{\ML})_\ren \simeq \tilde \lambda d / 2 =
 d /(2 \log d/ \tilde \Delta)
\,   \qquad \mbox{for} \quad d \gg \tilde \Delta  \;  .
\end{equation}
This logarithmic renormalization is beyond the reach of the \gc\ 
variational BCS method, but was confirmed using exact methods
\cite{mastellone98,braun-thesis,sierra99} (see \Sec{comparison}).  Its
occurrence, in a regime that in \gc\ variational calculations appears
to be ``uncorrelated'', can be regarded as the ``first sign of pairing
correlations'', in particular since, by \Eq{eq:renormalized-lambda},
the interaction strength \emph{increases} upon renormalization only if
the interaction is attractive ($\lambda <0$ would imply $|\tilde
\lambda | < | \lambda |$).  The pairing fluctuations responsible for
this renormalization will be discussed in more detail in
\Secs{comparison} and \ref{sec:bulk-few-n-differences}.

Unfortunately, $\Delta_\PP^{\ML}$ is at present not measurable in
ultrasmall grains, for the same experimental reasons as apply to
$\E_{1/2} - \E_0$ (see footnote~\ref{f:offset} in \Sec{sec:two-qp}).

\subsection{Parity effect for pairbreaking energies}
\label{sec:parity-pairbreaking}

Braun and von Delft \cite{braun98,braun99,braun-thesis} discussed yet
another parity effect, based on 
\begin{equation}
  \label{eq:Delta-pairbreaking}
\Delta^\pb_\PP =
\Omega_o- \Omega_e  \; , 
\end{equation}
the difference between the \emph{pair-breaking energies} of an even
and an odd grain [see \Eq{eq:spectralgaps}].
\Fig{fig:Matveev-parity}(b) shows its behavior as function of $d$.  In
the bulk limit $\Omega_e \simeq \Omega_o \simeq \tilde \Delta$ and
$\Delta^\pb_\PP \simeq 0$.  The most interesting feature of
$\Delta^{\pb}_\PP$ is that it initially becomes negative as $d/\tilde
\Delta$ increases; this occurs because in an odd grain pairing
correlations are weaker and hence breaking a pair costs less energy
than in an even grain.  $\Delta^{\pb}_\PP$ becomes positive again for
$d / \tilde \Delta \gtrsim 0.5$, since then pairing correlations are
so weak that $\Delta^{\pb}_\PP$ is governed by the kinetic energy cost
of flipping a spin, which is $2d$ for an odd grain but only $d$ for an
even grain.

$\Delta^\pb_\PP$ \emph{is} directly measurable in RBT's grains, via
the level-crossing fields $h_{0,1} = \Omega_e$ and $h_{1/2,3/2} =
\Omega_o $ [\Eq{eq:hcrit}].  The measured values $ \Omega_e = 0.23\pm
0.01$~meV and $\Omega_o = 0.24 \pm 0.01$~meV cited in \Sec{sec:two-qp}
give a positive value of $\Delta^\pb_\PP = 0.1$~meV, implying that the
grain under study was too small to fall in the most interesting regime
where $\Delta^\pb_\PP$ is negative.  Braun and von Delft suggested
that the latter should be observable in a somewhat larger grain with
$h_{1/2,3/2} < h_{0,1}$, i.e.\ in Regime~II of Fig.~\ref{fig:h-crit}.
(This suggestion assumes that despite the increased grain size, the
complicating effect of orbital diamagnetism is still non-dominant in
Regime~II.)  \emph{To look for negative $\Delta^\pb_\PP$
  experimentally would thus require good control of the ratio $d /
  \tilde \Delta$, i.e.\ grain size.}  This might be achievable if
recently-reported new fabrication methods, which allow systematic
control of grain sizes by using colloidal chemistry techniques
\cite{Klein-97,sunmurray,Schmid97}, could be applied to Al grains.

%%HABIL%% This indicates matters that need to be 
%% changed slightly for the habilitation thesis or Annalen der Physik

\newpage

{\large {\bf Part II: Crossover from the bulk to
the limit of a few electrons}} \\

%\section{Superconductivity: crossover from the bulk to
%the limit of a few electrons}

\label{sec:crossover}
\label{sec:sc-overview}
\label{sec:sc-canonical}

%In this section we review theoretical developments concerned with the
%qualitative changes incurred by pairing correlations during the
%crossover from the bulk to the limit of few electrons,
%and compare their results to those
%obtained using Richardson's exact solution of
%the pairing model of \Sec{sec:model}.

Part II of this review is devoted to the question: \emph{How do pairing
correlations change when the size of a superconductor is decreased
from the bulk to the limit of only a few electrons?} In particular,
we shall attempt to refine the answer given by Anderson
\cite{anderson59}, namely that superconductivity as we know it breaks
down for $d \gtrsim \tilde \Delta$.

First steps towards a more detailed answer were taken in the early
1970s by Strongin \etalia\ \cite{Strongin-70} and by M\"uhlschlegel
\etalia\ \cite{Muehlschlegel-72}, who calculated the thermodynamic
properties of ensembles small superconducting grains.  Experimental
realizations of such ensembles were, \eg, the granular films studied
by Giaver and Zeller \cite{GiaeverZeller-68,ZellerGiaever-69}.
%Apart from the work of Kawataba \cite{kawataba-80,kawataba-81},
%%??%% who studied...
The interest of theorists was rekindled in 1995 by RBT's success in
probing \emph{individual} superconducting grains. Apart from
motivating the phenomenological theory of Braun \etalia\ reviewed in
part I, these experiments also inspired a
substantial and still growing number of
theoretical studies 
\vphantom{\cite{vondelft96,braun97,braun99,smith96,%
balian-short,balian-long,%
bonsager98,matveev97,Rossignoli-98,Rossignoli-99a,Rossignoli-99b,%
Rossignoli-00,mastellone98,berger98,braun98,dukelsky99a,dukelsky99b,%
braun-vieweg,sierra99,vondelft-ankara99,dukelsky99c,%
tian99,tanaka99,dilorenzo99}}
\cite{vondelft96}-\cite{dilorenzo99} of how
superconducting pairing correlations in such grains are affected by reducing
the grains' size, or equivalently by increasing its mean level spacing $d
\propto {\rm Vol}^{-1}$ until it exceeds the bulk gap $\tilde \Delta$.

In the earliest of these, von Delft {\em et al.\/} studied the \dbcsm\ 
of \Sec{sec:model} within a parity-projected \gc\ BCS approach
\cite{vondelft96} closely related to the variational BCS method of
\Sec{sec:generalBCS}.  Their \gc\ results suggested that pairing
correlations, as measured by the pairing parameter or the condensation
energy, vanish abruptly once $d$ exceeds a critical level spacing
$d^\BCS_{p/2}$ \emph{that depends on the parity ($p= 0$ or $1$) of the
  number of electrons on the grain}, being smaller for odd grains
($d^\BCS_{1/2} \simeq 0.89 \tilde \Delta$) than even grains $(d^\BCS_0
\simeq 3.6 \tilde \Delta$).  Parity effects were also found in a
number of subsequent papers that used parity-projected \gc\ methods to
study the behavior of the BCS mean-field gap parameter $\Delta_\MF$
and related quantities as functions of level spacing
%\cite{braun97,braun99,smith96,balian-short,balian-long,%
%bonsager98,matveev97,Rossignoli-98,Rossignoli-99a,Rossignoli-99b,%
%Rossignoli-00},
\cite{braun97}-\cite{Rossignoli-00},
temperature and magnetic field.  All these parity effects are
consequences of the blocking effect (cf.\ 
\Sec{sec:generalproperties}): for odd grains, the unpaired electron
somewhat disrupts the pairing correlations of the remaining paired
ones, by reducing the phase space available for pair scattering.

A series of more sophisticated canonical approaches
%\cite{mastellone98,berger98,braun98,%
%dukelsky99a,dukelsky99b,braun-vieweg,sierra99,%
%vondelft-ankara99,dukelsky99c}
\cite{mastellone98}-\cite{dukelsky99c}
(summarized in \Sec{comparison}) {\em confirmed the parity dependence
  of pairing correlations,\/} but established that the abrupt
vanishing of pairing correlations at $d^\BCS_{p/2}$ is an artifact of
\gc\ treatments: \emph{pairing correlations do persist, in the form of
  so-called fluctuations, to arbitrarily large level spacings}
  \cite{matveev97}, and the crossover between the bulk
superconducting (SC) regime $(d \ll \tilde \Delta)$ and the
fluctuation-dominated (FD) regime $(d \gg \tilde \Delta)$ is
completely smooth 
%\cite{dukelsky99a,dukelsky99b,braun-vieweg,sierra99,%
%vondelft-ankara99}.  
\cite{dukelsky99a}-\cite{vondelft-ankara99}.  
Nevertheless, these two regimes are
qualitatively very different
%\cite{braun98,dukelsky99a,dukelsky99b,braun-vieweg,%
%sierra99,vondelft-ankara99}: the
\cite{braun98}-\cite{vondelft-ankara99}: the
condensation energy, \eg, is an extensive function of volume in the
former and almost intensive in the latter, and pairing correlations
are quite strongly localized around the Fermi energy $\eF$, or more
spread out in energy, respectively. Very recently, Di Lorenzo \etalia\ 
\cite{dilorenzo99} suggested that the remnant pairing correlations in
the FD regime might be detectable via susceptibility measurements.

Toward the end of 1998 and after the appearance of most of these
works, R.W.  Richardson pointed out \cite{richardson-private-98} to
their various authors that the discrete BCS Hamiltonian on which they
are based actually has an exact solution, discovered by him in 1963
\cite{richardson63a} (and independently by Gaudin in 1968
\cite{gaudin}). Richardson published his solution in the context of
nuclear physics in a series of papers between 1963 and
1977 
%\cite{richardson63a,richardson63b,richardson64,%
%  richardson65a,richardson65b,richardson66,%
%  richardson66-b,richardson67,richardson77} which seem to have
\cite{richardson63a}-\cite{richardson77} which seem to have
completely escaped the attention of the condensed matter community.
Very recently, the model was also shown to be integrable
\cite{cambiaggio,sierra99b}.  The revival of this remarkably simple
exact solution after such a long and undeserved period of neglect is
perhaps one of the most important consequences of RBT's experimental
breakthrough:
%Richardson's solution allows
%one to elucidate {\em by exact means\/} many important 
%conceptual ingredients
%of the standard BCS theory of superconductivity, and to 
%calculate exactly the
%deviations from it that become important in ultrasmall grains.
%
%Richardson's solution not only allows the exact calculation of 
%essentially all quantities of interest for ultrasmall grains; more
%importantly (since of more general interest),
%
%it also makes possible, and
%this is of general interest and therefore perhaps 
%more important, the
Richardson's solution 
allows the elucidation and illustration by {\em exact\/} means of
many important conceptual ingredients of the standard BCS theory of
superconductivity, such as the nature of pairing correlations, the
importance of phase coherence, the validity of using a mean-field
approximation and a grand-canonical formulation for bulk systems, and
the limitations of the latter approaches for ultrasmall systems.
Moreover, it allows the exact calculation of 
essentially all quantities of interest for ultrasmall grains. 

We shall therefore start part II by discussing the exact solution
[\Sec{sec:richardson}]. We then summarize the other canonical
approaches somewhat more briefly than they perhaps would have deserved
had an exact solution not existed, and compare their results to those
of the exact solution [\Sec{comparison}].  Next we analyze the
qualitative differences between the bulk and FD regimes
[\Sec{sec:bulk-few-n-differences}], then discuss the case of randomly
(as opposed to uniformly) spaced energy levels $\varepsilon_j$
[\Sec{sec:sc-level-statistics}], and finally discuss finite
temperature parity effects [\Sec{sec:finite-T}].
Throughout part II we set $\mu = 0$, since canonical
treatments make no reference to a chemical potential.

\section{Richardson's exact solution}
\label{sec:sc-richardson}
\label{sec:richardson}

In this section we summarize some of the central results of Richardson's
exact solution of the \dbcsm.

\subsection{General eigenstates}
Consider $N= 2n + b$ electrons, $b$ of which are unpaired, as
in \Sec{sec:generalproperties}.  According to the general
discussion there, the nontrivial aspect of solving the model is
finding the eigenenergies $\E_n$ and corresponding eigenstates
$|\Psi_n\rangle$ [\Eq{eq:eigenpsi}] of the pair Hamiltonian [\Eq{1},
in which we set $\mu=0$ below]
\begin{eqnarray}
 && \hat H_\U = \sum_{ij}^\U \left( 2 \varepsilon_j \delta_{ij} -
 \; \lambda d \right)  b_i^\dagger b^\ds_j \; ,  
 \label{pairH}
 \end{eqnarray}
 in the Hilbert space of all states containing exactly $n$ pairs
 $b^\dagger_j = c^\dagger_{j+} c^\dagger_{j-}$ of electrons, where $j$
 runs over the set of all unblocked single-particle levels, $\U = \I
 \backslash \B$ [$I$ is the set of all interacting levels, $B$ the set
 of all blocked levels].  In general, degenerate levels are allowed in
 $\I$, but are to be distinguished by distinct $j$-labels, i.e.\ they
 have $\varepsilon_i = \varepsilon_{j}$ for $i \neq j$.

Richardson showed that the sought-after eigenstates (with
normalization \linebreak
\mbox{$\langle \Psi_n | \Psi_n \rangle = 1$}) and
eigenenergies have the general form
\begin{eqnarray}
  \label{eq:truebosoneigenstates-cc}
  | \Psi_n \rangle = {\cal N}\prod_{\nu=1}^n 
B_{\nu}^\dagger |0\rangle \, , \quad 
\E_n = \sum_{\nu = 1}^n E_{\nu}, 
\quad \mbox{with} \quad 
  B_{\nu}^\dagger = \sum_j^\U 
{b_j^\dagger \over 2 \varepsilon_j - E_{\nu}} \; .
\end{eqnarray} 
Here ${\cal N}$ is a  normalization constant and 
the $n$ parameters $E_{\nu}$ ($\nu = 1, \dots , n$) are a
solution of the set of $n$ coupled algebraic equations
\begin{eqnarray}
  \label{eq:richardson-eigenvalues-cc}
  {1\over \lambda d }  - \sum_j^\U 
{1  \over 2 \varepsilon_j - E_{\nu}}    
+ \sum_{\mu = 1(\neq \nu)}^n {2 \over E_{{\mu}} - E_{{\nu}}} = 0 \; ,
\qquad \mbox{for}\quad \nu = 1, \dots, n \; ,
\end{eqnarray}
which are to be solved (numerically, see
App.~B2 of \Ref{PR-VDR}) subject to the restrictions $E_\mu
\neq E_\nu$ if $\mu \neq \nu$.
Richardson originally derived this remarkably simple result by solving
the Schr\"odinger equation for the wave-function $\psi (j_1, \dots ,
j_n)$ of \Eq{eq:generaleigenstate-2}.  A simpler proof, also due to
Richardson \cite{richardson-private-99}, may be found in
\Ref{vondelft-ankara99} and in App.~B.1 of 
\Ref{PR-VDR}; its strategy is to
verify that $(\hat H_\U - \E_n) |\Psi_n \rangle = 0$ by simply
commuting $\hat H_\U$ past the $B_{\nu}^\dagger$ operators in
(\ref{eq:truebosoneigenstates-cc}).

Below we shall always assume the $\varepsilon_j$'s to be all distinct
(the more general case that degeneracies are present is discussed by
Gaudin \cite{gaudin}).  Then it can be shown explicitly \cite{gaudin}
that (i) the number of distinct solutions of
Eq.~(\ref{eq:richardson-eigenvalues-cc}) is equal to the dimension of
the $n$-pair Hilbert space defined on the set of unblocked levels
$\U$, namely ${N_U \choose n}$, where $N_U$ is the number of unblocked
levels; and (ii) that the corresponding eigenstates
(\ref{eq:truebosoneigenstates-cc}) are mutually orthogonal to each
other, thus forming an eigenbasis for this Hilbert space. This can
easily be understood intuitively, since there exists a simple relation
between the bare pair energies $2\varepsilon_j$ and the solutions of
\Eqs{eq:richardson-eigenvalues-cc}: as $\lambda$ is reduced to 0, it
follows by inspection that each solution $\{ E_{1}, \dots,$ $E_{n}\}$
reduces smoothly to a certain set of $n$ bare pair energies, say $\{2
\varepsilon_{j_1}, \dots, 2\varepsilon_{j_n} \}$; this particular
solution may thus be labeled by the indices $j_1, \dots, j_n$, and the
corresponding eigenstate (\ref{eq:truebosoneigenstates-cc}) written as
$|\Psi_n \rangle \equiv |j_1, \dots j_n \rangle$.  By inspection, its
$\lambda \to 0 $ limit is the state $|j_1, \dots j_n \rangle_{\U,0}
\equiv \prod_{\nu=1}^n b_{j_\nu}^\dagger |0 \rangle $, thus there is a
one-to-one correspondence between the sets of all states $\{ |j_1,
\dots, j_n \rangle_\U \} $ and $\{ | j_1, \dots j_n \rangle_{\U,0}
\}$.  But the latter constitute a complete eigenbasis for the $n$-pair
Hilbert space defined on the set of unblocked levels $\U$, thus the
former do too.

\subsection{Ground state}
\label{sec:richardson-ground-state}

For a given set of blocked levels $B$, the lowest-lying of all states
$|\Psi_n,\B\rangle$ of the form (\ref{eq:generaleigenstate}), say
$|\Psi_n,\B\rangle_{\G}$, is obtained by using that particular
solution $|j_1, \dots j_n \rangle$ for which the total ``pair energy''
${\cal E}_n$ takes its lowest possible value.  The lowest-lying of all
eigenstates with $n$ pairs, $b$ blocked levels and total spin $s =
b/2$, say $|n,s \rangle_\G$ with energy ${\cal E}^\G_s (n)$, is that
$|\Psi_n, \B \rangle_\G$ for which the blocked levels in $\B$ all
contain spin-up electrons and are all as close as possible to $\eF$,
the Fermi energy of the uncorrelated $N$-electron Fermi sea
$|\F_N\rangle$. The $E_{\nu}$ for the ground state $|n,s\rangle_\G$
coincide at $\lambda =0$ with the lowest $n$ energies $2
\varepsilon_j$ ($j = 1, \dots, n$), and smoothly evolve toward
(initially) lower values when $\lambda$ is turned on, a fact that can
be exploited during the numerical solution of
\Eq{eq:richardson-eigenvalues-cc}. As $\lambda$ is increased further,
some of the $E_\nu$'s become complex; however, they always occur in
complex conjugate pairs, so that ${\cal E}_n$ remains real
\cite{richardson66}. For details, see \Ref{richardson66} and 
App.~B.2 of \Ref{PR-VDR}, where some algebraic transformations
are introduced that render the equations less singular and hence
simplify their numerical solution considerably.

\subsection{General comments}
Since the exact solution provides us with wave functions, it is in
principle straightforward to calculate arbitrary correlation functions
of the form $\langle \Psi_n | b^\dagger_{i} b^\dag_j \dots b^\ds_{i'}
b^\ds_{j'} | \Psi_n \rangle$, by simply commuting all $b$'s to the
right of all $b^\dag$'s.  However, due to the hard-core boson
commutation relations (\ref{hard-core-boson-2}) of the $b$'s, the
combinatorics is rather involved. Nevertheless, Richardson succeeded
to derive \cite{richardson65b} explicit results for the normalization
constant ${\cal N}$ of (\ref{eq:truebosoneigenstates-cc}) and the
occupation probabilities $\bar v_j^2$ and correlators $C_{ij}$ of
\Eq{eq:C_ij} (summarized in App.~B.3 of \Ref{PR-VDR}).
The exact result for the $C_{ij}$'s show that they are 
\emph{all} positive, in agreement with
the requirement (ii) formulated in \Sec{sec:meaningfulDelta}.
 It is also
natural to ask whether in the bulk limit ($d \to 0$ at fixed $n \,
d$), the standard BCS results can be extracted from the exact
solution.  Indeed they can, as Richardson showed in
\cite{richardson77} (following unpublished work by Gaudin
\cite{gaudin}), by interpreting the problem of solving the eigenvalue
equations (\ref{eq:richardson-eigenvalues-cc}) for the $E_\nu$ as
a problem in two-dimensional electrostatics (see
App.~B.2 of \Ref{PR-VDR}).  Exploiting this analogy, he showed
that in the bulk limit, Eqs.~(\ref{eq:richardson-eigenvalues-cc})
reduce to the well-known BCS equations determining the gap and
chemical potential at $T=0$ [\Eqs{eq:gap} and (\ref{eq:mu})], and the
ground state condensation energy ${\cal E}^\cond_0(n)$
[\Eq{eq:define-condensation-energy}] to its BCS result, namely $-
\tilde \Delta^2/2d$.

Finally, let us mention that the Cambiaggio, Rivas and Saraceno have
recently shown that the \dbcsm\ is integrable and have constructed
explicit expressions for all its constants of the motion
\cite{cambiaggio}. The latter's relation to Richardson's solution was
clarified by Sierra \cite{sierra99b}, who has also explored possible
connections between the exact solution and conformal field theory.  It
would be an interesting challenge for mathematical physicists to try
to exploit this integrability to calculate finite-temperature
properties exactly --- although these can in principle be obtained
from Richardson's solution by ``simply'' computing the partition
function over all states, this is forbiddingly tedious in practice for
large temperatures, since the eigenenergy of \emph{each} state
requires a separate (non-trivial) numerical calculation.

\section{Comparison of other canonical methods with the
exact solution}
\label{sec:fixedN}
\label{sec:sc-exact-diagonalization}
\label{sec:sc-dmrg}
\label{comparison}

\begin{figure}[t]
\centerline{\epsfig{figure=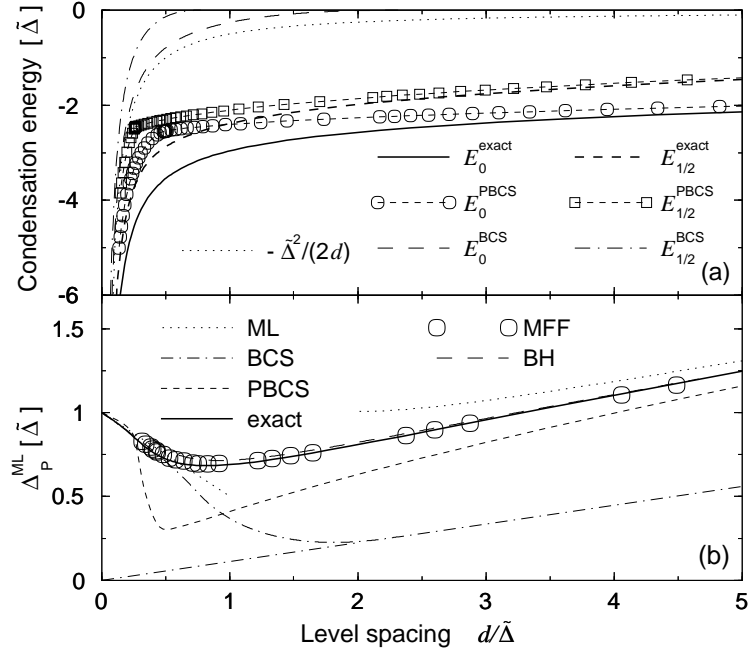,%
width=0.75\linewidth}} 
  \caption[Exact ground state condensation energies]{
    (a) The even and odd $(s=0,1/2)$ condensation energies
    ${E}^\cond_{s}$ of Eq.~(\ref{eq:condensation-again}) [in units of
    $\tilde \Delta$], calculated with BCS, PBCS and exact wave
    functions \cite{sierra99}, as functions of $d/ \tilde \Delta = 2
    \sinh (1/ \lambda) / (2n + 2s)$, for $\lambda = 0.224$.  For
    comparison, the dotted line gives the ``bulk'' result $E_0^{\rm
      bulk} = -\tilde \Delta^2/(2d)$.  (b) Comparison \cite{sierra99}
    of the parity parameters $\Delta^\ML_\PP $
    \protect\cite{matveev97} of Eq.~(\ref{eq:ML-again}) [in units of
    $\tilde \Delta$] obtained by various authors: ML's analytical
    result (dotted lines) [$\tilde \Delta(1-d/2 \tilde \Delta)$ for
    $d\ll \tilde \Delta$, and $d/2\log(a d/\tilde \Delta)$ for $d\gg
    \tilde \Delta$, with $a=1.35$ adjusted to give asymptotic
    agreement with the exact result]; grand-canonical BCS approach
    (dash-dotted line) [the naive perturbative result $\frac12\lambda
    d$ is continued to the origin]; PBCS approach (short-dashed line);
    Richardson's exact solution (solid line); exact
    diagonalization and scaling by MFF (open circles) and BH
    (long-dashed line).  }
  \label{fig:exact-gse}
\end{figure}

In this section we briefly mention the various canonical methods by
which the \dbcsm\ had been investigated prior to the revival of
Richardson's exact solution in 1999. All of these studies used a
half-filled band with fixed width $ 2\omegaD$ of uniformly-spaced
levels [i.e.\ $\varepsilon_j = j \, d + (1-p)d/2$, as in \Eq{eq:ejs}],
 containing $N=2n+b$
electrons.  Then the level spacing is $d= 2 \omegaD /N$ and the bulk
gap is $\tilde \Delta = \omegaD / \sinh (1/\lambda)$. Following
\cite{braun98}, we take $\lambda = 0.224$ throughout this section.  To
judge the quality of the various approaches, we compare in
\Fig{fig:exact-gse} the results which they yield with those from
Richardson's solution, for the even and odd $(s=0,1/2)$ condensation
energies $E_{s}^\cond $ and the Matveev-Larkin parity parameter
$\Delta^\ML_\PP$ [cf.\ \Sec{sec:ML-parameter}]. In the notation of
\Sec{sec:richardson-ground-state}, these are given by
\begin{eqnarray}
  \label{eq:condensation-again}
E^\cond_{s} (n) & = & {\cal E}^\G_{s} (n) -
\langle \F_N| \hat H | \F_N\rangle \; ,
\\
  \label{eq:ML-again}
\Delta^\ML_\PP (n) & = & {\cal E}^\G_{1/2} (n) - 
[ {\cal E}^\G_0(n) + {\cal E}^\G_0 (n+1)]/2 \, .
\end{eqnarray}

Following the initial g.c.\ studies
\cite{vondelft96,braun97,braun99,smith96,matveev97} of the \dbcsm, the
first purely canonical study was that of Mastellone, Falci and Fazio
(MFF) \cite{mastellone98}, who used Lanczos exact diagonalization.
Despite being limited to $n \le 12$, they managed to reach reasonably
small ratios of $d/\tilde\Delta$ by using an ingenious scaling
approach: for a given level spacing $d$, they increased the coupling
constant $\lambda$ to about $0.5$, thereby decreasing $d/\tilde\Delta
= d/\omegaD \sinh(1/\lambda)$ to values as small as 0.5.  This allowed
them to probe, coming from the few-electron side, a remarkably large
part of the crossover to bulk limit. They found, \ia, that the
condensation energies are negative for \emph{all} $d$, showing that
the system can \emph{always} gain energy by allowing pairing
correlations, even for arbitrarily large $d$.

Berger and Halperin (BH) \cite{berger98} showed that almost identical
results can be achieved with less than 6 pairs, thus significantly
reducing the calculational effort involved, by first performing a
``poor man's scaling'' renormalization: they reduce the bandwidth from
$\omegaD \approx n d$ to, say, $\baromegaD \approx \bar n d$
(with $\bar n \le 6$) and incorporate the effect of the removed levels
by using a renormalized coupling constant,
\begin{eqnarray}
  \label{eq:lambda-ren}
  \bar \lambda = \lambda \left[ {1-  \sum_{ \baromegaD
 < |\varepsilon_j | < \omegaD} {\lambda  \over 2  |\varepsilon_j |
}}\right]^{-1}
 \; .
%  = \frac{1}{\log(a_{\bar N,\mu} (2\bar n +1) d/\tilde\Delta)}
\end{eqnarray}
The reduced system is then diagonalized exactly.  Note that the
renormalization of Matveev and Larkin [\Eq{eq:renormalized-lambda}]
corresponds to taking $\baromegaD \simeq d$ in \Eq{eq:lambda-ren},
\ie\ to integrating out the entire band. Also note that the
renormalization prescription of (\ref{eq:lambda-ren}) has the property
that it would leave the bulk gap invariant in the limit $d / \tilde
\Delta \to 0$, for which \Eq{eq:lambda-ren} would imply $\baromegaD
e^{-1/\bar \lambda} \simeq \omegaD e^{-1/\lambda} \simeq \tilde
\Delta$.

To access larger values of $n$, Braun and von Delft \cite{braun98}
used a \emph{fixed-$n$ projected} BCS approach (PBCS), in which
BCS-like variational wavefunctions are projected to fixed particle
number, as in \Eq{eq:BCSground-N}.  The projection integrals occurring
in \Eq{eq:BCSground-N} were evaluated numerically for $n \le 600$,
using tricks developed in the nuclear physics literature by Bayman
\cite{Bayman-60}, Dietrich, Mang and Pradal \cite{Dietrich-64} and Ma
and Rasmussen \cite{Ma-77}, and summarized in part in the book of Ring
and Schuck \cite{RingSchuck-80}.  (A much simpler way of dealing with
the projection, using recursion relations, was recently found by
Dukelsky and Sierra \cite{dukelsky99b}.) The PBCS method gives
condensation energies that (i) are significantly lower than the
grand-canonical ones [see \Fig{fig:exact-gse}], thus the projection much
improves the variational Ansatz, and that (ii) are negative for
\emph{all} $d$, confirming that the abrupt vanishing of the \gc\ 
condensation energies is indeed an artifact of the \gc\ treatment.
The PBCS method is able to fully recover the bulk limit, but the
crossover is not completely smooth and shows a remnant of the \gc\ 
breakdown of pairing correlations: the $d$-dependence of the
condensation energy $({ E}_{s}^\cond)^\PBCS$ changes rather abruptly
[kinks in the short-dashed lines in \Fig{fig:exact-gse}(a)] from being
\emph{extensive} ($\sim 1/d$) to being practically \emph{intensive}
(almost $d$ independent).

It should be mentioned here that a generalization of the PBCS
method to finite temperatures has been worked out by Essebag
and Egido in the context of nuclear physics \cite{Essebag-93}.
The PBCS method has recently also been applied to the attractive
Hubbard model in one dimension by Tanaka and Marsiglio
\cite{tanaka99}, who found even-odd and super-even effects.  The
latter consist of differences between the number of pairs being equal
to $n=2m$ or $2m+1$, and arise if boundary conditions are used that
produce doubly-degenerate levels ($\varepsilon_{\vec k} =
\varepsilon_{- \vec k}$) near the Fermi surface.

Dukelsky and Sierra \cite{dukelsky99a,dukelsky99b} used the density
matrix renormalization group (DMRG) (with $n \le 400$) to achieve
significant improvements over the PBCS results for the \dbcsm, in
particular in the regime of the crossover, which they found to be
\emph{completely smooth}.  In general, the DMRG approach is applicable
to systems that can be divided into two pieces, called block and
environment, which interact via a preferably rather small number of
states. One starts with a small block and environment, computes their
combined density matrix, then enlarges both and recomputes the density
matrix, etc, until a large part of the system has been treated.
Dukelsky and Sierra chose the block and environment to consist,
respectively, of all particle or hole states relative to the Fermi sea
(for a detailed description of the method, see \cite{dukelsky99b}).
Since the pairing correlations involving coherent superpositions of
particle and hole states are peaked in a rather small regime of width
$\tilde \Delta$ around the Fermi energy [compare \Figs{fig:v2u2-prb97}
or \ref{fig:exact-wf}], the ``interaction'' between block and
environment is ``localized'', so that the DMRG can \emph{a priori} be
expected to work rather well for this problem.

Finally, Dukelsky and Schuck \cite{dukelsky99c} showed that a
self-consistent RPA approach, which in principle can be extended to
finite temperatures, describes the FD regime rather well (though
not as well as the DMRG).

To check the quality of the above methods, Braun
\cite{braun-thesis,sierra99} computed $E_{s}^\cond$ (for $s=0,1/2$)
and $\Delta^\ML_\PP$ using Richardson's solution
(Fig.~\ref{fig:exact-gse}).  The exact results
\begin{enumerate}
\item[(i)] quantitatively agree, for $d \to 0$, with the leading $-
  \tilde \Delta^2/2d$ behavior for $E^\cond_{s}$ obtained in the \gc\ 
  BCS approach \cite{vondelft96,braun98,braun99}, which in this sense
  is exact in the bulk limit, corrections being of order $d^0$;
\item[(ii)] confirm that the even ground state energy \emph{always}
lies below the odd one (this had independently been proven
rigoroulsy by Tian and Tang \cite{tian99});
\item[(iii)] confirm that a completely smooth
  \cite{dukelsky99a,dukelsky99b} crossover occurs around the scale $d
  \simeq \tilde \Delta$ at which the g.c.\ BCS approach breaks down;
\item[(iv)] show that the PBCS crossover \cite{braun98} is
qualitatively correct, but not quantitatively, being somewhat too
abrupt; 
\item[(v)] are reproduced remarkably well by the approaches of MFF
\cite{mastellone98} and BH \cite{berger98}; 
\item[(vi)] are fully reproduced by the DMRG of
  \cite{dukelsky99a,dukelsky99b} with a relative error of $< 10^{-4}$
  for $n \le 400$; our figures don't show DMRG curves, since they are
  indistinghuishable from the exact ones and are discussed in detail
  in \cite{dukelsky99a,dukelsky99b}.
\end{enumerate}

The main conclusion we can draw from these comparisons is that the two
approaches based on renormalization group ideas work very well: the
DMRG is essentially exact for this model, but the band-width rescaling
method of BH also gives remarkably (though not quite as) good results
with rather less effort.  In contrast, the PBCS approach is rather
unreliable in the crossover region. To study generalizations of the
\dbcsm, \eg\ using state-dependent couplings of the form $d \sum_{ij}
\lambda_{ij} b^\dagger_i b^\ds_j$, the DMRG would thus be the method
of choice.

\section{Qualitative differences between
the bulk and the few-electron regimes}
\label{sec:bulk-few-n-differences}

\begin{figure}
  \begin{center}
  \epsfig{figure=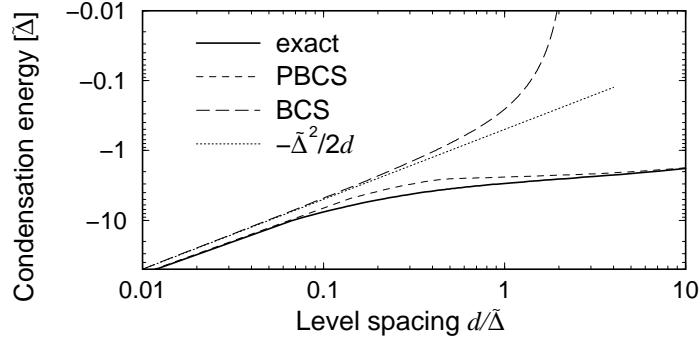,%
width=0.7\linewidth}
%height=0.25\linewidth}
%\quad
%  \epsfig{figure=../Figures/likharev-crossover.eps,%
%height=0.25\linewidth}
    \caption[Condensation energies: log-log plot and
    $\lambda$-dependence]{Log-log plot \cite{braun-thesis} of some of
      the curves of Fig.~\protect\ref{fig:exact-gse}(a) for the even
      condensation energy $E^\cond_0$ [in units of $\tilde \Delta$],
      for $\lambda = 0.224$; its asymptotic $-\tilde\Delta^2/(2d)$
      behavior for $d / \tilde \Delta \to 0$ is shown by the dotted
      line.
%      (b) The coefficients $\alpha (d) $ and $\beta (d) $ of Eq.
%      (\ref{eq:fit}), which characterize the change in the
%      $\lambda$-dependence of $E^\cond_0$ with increasing
%      $d/\tilde\Delta$ \cite{braun-thesis}. 
}
    \label{fig:exact-log-log}
    \label{fig:likharev-crossover}
  \end{center}
\end{figure}

Does the fact that the exact condensation energy $E_{s}^\cond$ is
always negative, even for arbitrarily large $d/\tilde \Delta$, mean
that the system stays ``superconducting'' even if it is arbitrarily
small? The answer is certainly no, since in the fluctuation-dominated
(FD) regime, the pairing correlations are qualitatively different than
in the bulk, superconducting regime. In this section we shall try to
make this statement more precise by analyzing the qualitative
differences between the two regimes, with regard to the $\lambda$- and
$d$-dependence of $E_{s}^\cond$, and the behavior of the occupation
probabilities $\bar v_j^2$.

\Fig{fig:exact-log-log}(a) shows, on a log-log plot, the
$d$-dependence of the even condensation energy $E_{0}^\cond (d)$.
Note that even on the log-log plot, the crossover of the exact
$E^\cond_{0}$ from the bulk to the FD regime is completely smooth.
According to Sierra and Dukelsky \cite{dukelsky99b}, the exact result
for $E_{0}^\cond (d)$ can be fitted very well to the form
\begin{eqnarray}
  \label{eq:E-cond-d-dependence}
  E_0^\cond (d) = -\tilde\Delta^2/(2d) - 
\eta_0  (\ln 2) \omegaD \lambda^2  + \gamma_0
 (\tilde\Delta d/2 \omegaD)   \log (2\omegaD/d) \; ,
\end{eqnarray}
where $\eta_0$ and $\gamma_0$ are  constants of order
unity \cite{dukelsky99b}.  The first term is \emph{extensive}
($\propto \Vol$) and dominates in the bulk limit; its standard
heuristic interpretation \cite{tinkham-book} is that roughly $\tilde
\Delta / d$ levels (those within $\tilde \Delta$ of $\eF$) are
strongly affected by pairing, with an average energy gain per level of
$-\tilde\Delta /2$.  The second term, which is \emph{intensive} and
dominates in the FD limit, is equal (up to the numerical factor
$\eta_0$) to the result from second-order perturbation theory
\cite{dukelsky99b}, namely $(\lambda d)^2 \sum_{i = 1}^n \sum_{j =
  n+1}^{2n} (2\varepsilon_i- 2\varepsilon_j)^{-1}$.  This subleading
term's $d$-independence (which was anticipated in
\cite{anderson59,muehlschlegel62}) may be interpreted by arguing that
in the FD regime, the number of levels that contribute significantly
to $E_0^\cond$ is no longer of order  $\tilde \Delta/d$: instead,
fluctuations affect \emph{all} $n \simeq 2\omegaD/d$ unblocked levels
within $\omegaD$ of $\eF$ (this is made more precise below), and
each of these levels 
contributes an amount of order $ -(\lambda d)^2/d$
(corresponding, in a way, to its selfenergy). Finally, the third term
%, whose logarithmic factor may
%be interpreted as arising by renormalizing the second term via
%$\lambda \to \tilde \lambda$ of \Eq{eq:renormalized-lambda}, 
 contains the small parameter $\tilde \Delta/ \omegaD$
and thus represents a very small correction.

The $\lambda$- and volume-dependencies of $E_0^\cond$ in
\Eq{eq:E-cond-d-dependence} strikingly illustrate the qualitative
differences between the bulk and FD regimes: in the bulk regime,
dominated by the first term, $E_0^\cond$ is nonperturbative in
$\lambda$ (since $\tilde \Delta \simeq 2 \omegaD e^{- 1/\lambda}$) and
extensive, as expected for a strongly-correlated state; in constrast,
in the FD regime, dominated by the second term, $E_0^\cond$ is
perturbative in $\lambda$ and practically intensive (up to the weak
$\log d$ dependence of the third term).

\label{sec:dmp-wavefunctions}

\begin{figure}[t]
  \begin{center}
  \epsfig{figure=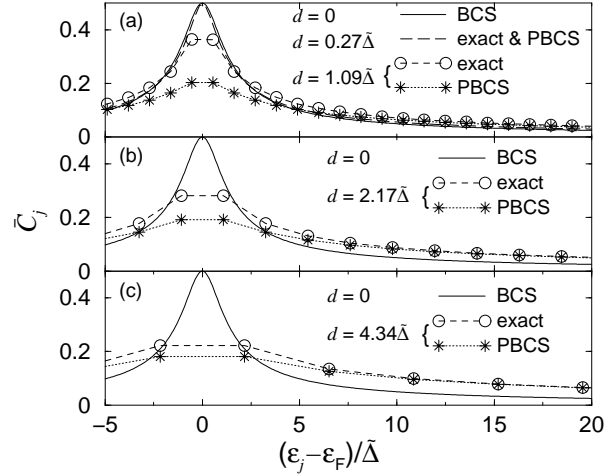,width=0.6\linewidth}
    \caption[Pairing amplitudes $\bar C_j$]{The occupation probabilities
      $\bar C_j$ of Eq.~(\ref{eq:C2_j}) for $d/ \tilde \Delta = 0$,
      0.27, 1.09, 2.17 and 4.34 \cite{braun-thesis}.  In all three
      figures, the thick solid lines give the $d=0$ bulk BCS result,
      whereas circles and stars represent $\bar C_j$-values evaluated
      for discrete $j$'s using the exact solution and PBCS method,
      respectively.  For $d=0.27\tilde\Delta$, the PBCS and exact
      results are indistinguishable, and are shown in (a) as a single
      long-dashed line, which is also virtually identical to the bulk
      curve. For small $d$, pairing correlations are evidently
      localized within a few $\tilde\Delta$ of $\eF$.  With increasing
      $d$ more and more weight is shifted away from $\eF$ into the
      tails; compared to the exact results, the PBCS method somewhat
      overemphasizes this delocalization, which is one of the reasons
      why it produces a somewhat too abrupt crossover.}
    \label{fig:exact-wf}
  \end{center}
\end{figure}

Perhaps the most vivid way of illustrating the
qualitative difference between the bulk and FD regimes
is to study properties of the ground state wavefunction. 
We shall consider here the correlators
  \cite{braun98}
\begin{equation}
  \label{eq:C2_j}
\bar   C^2_j (d) =
\langle b^\dagger_j b^\ds_j \rangle
\langle b^\ds_j b^\dag_j \rangle \; , 
\end{equation}
which measure the probability that a level can be ``both occupied and
empty'', and vanish identically for states without pairing
correlations.  For the \dbcsm\ $\bar C_j^2$ identically equals
$\langle b^\dagger_j b^\ds_j \rangle - \langle b^\dagger_j b^\ds_j
\rangle^2 = \bar v_j^2 - \bar v_j^4$ [by \Eqs{hard-core-boson-2} and
(\ref{eq:defineu2v2})], which measures the fluctuations
in the pair occupation number of level  $j$, 
and it vanishes for any blocked single-particle
level.  Note that $\bar C^2_j (d)$ also equals $\langle b^\dagger_j
b^\ds_j \rangle -\langle c^\dagger_{j+} c^\ds_{j+} \rangle\langle
c^\dagger_{j-} c^\ds_{j-}\rangle$; this form, which was used in
\cite{braun98} and corresponds to the diagonal terms under the sum in
\Eq{eq:canonical-order-parameter} for $\Delta_\can$, can be
interpreted as the probability enhancement for finding a \emph{pair}
of electrons instead of two uncorrelated electrons in a
single-particle level $|j,\pm\rangle$.

When evaluated using the grand-canonical BCS wavefunction, $(\bar
C_j^2)_\BCS$ is equal to $ u_j^2 v_j^2 = {1 \over 4} \tilde \Delta^2/
(\varepsilon_j^2+\tilde \Delta^2)$ [thick solid lines in
\Fig{fig:exact-wf}, the same function as that plotted in
\Fig{fig:v2u2-prb97}]. The $(\bar C_j)_\BCS$'s thus have a
characteristic peak of width $ \propto \tilde \Delta$ around $\eF$,
implying that pairing correlations are ``localized around $\eF$ in
energy space'', which may be taken to be the defining property of
``BCS-like correlations''.  Moreover, in the bulk regime $d \ll
\tilde\Delta$, the $(\bar C_j)_\BCS$ are virtually identical to $(\bar
C_j)_\ex$ [long-dashed line of \Fig{fig:exact-wf}(a)], \emph{vividly
  illustrating why the grand-canonical BCS approximation is so
  successful: not performing a canonical projection hardly affects the
  parameters $\bar u_j$ and $\bar v_j$ if $d \ll \tilde\Delta$, but
  tremendously simplifies their calculation}.

As one enters the FD regime $d\gtrsim \tilde \Delta$, the character of
the correlator $(\bar C_j)_\ex$ changes [\Fig{fig:exact-wf}(b),
circles]: weight is shifted into the tails far from $\eF$ at the
expense of the vicinity of the Fermi energy.  Thus \emph{pairing
  correlations become delocalized in energy space} (as also found in
\cite{mastellone98,dukelsky99a,dukelsky99b}), so that referring to
them as mere ``fluctuations'' is quite appropriate.  In the extreme
case $d \gg \tilde \Delta$, the $(\bar C_j)_\ex$ for all interacting
levels are roughly equal.

Richardson's solution can also be used to calculate,
for a given set $B$ of blocked levels, the $d$-dependence
of the canonical order parameter $\Delta^B_\can (d)$ of
\Eq{eq:canonical-order-parameter}.  Schechter has found
\cite{moshe-priv} that it can be fit to the form $\Delta^B_\can (d) =
\tilde \Delta (1 + \tilde \gamma_B d/\tilde \Delta)$, 
where $\tilde \gamma_B$ is a
positive numerical constant, and the linear term essentially reflects
the factor of $d$ in the definition of $\Delta^B_\can$. The fact that
$\Delta^B_\can $ is a strictly increasing function of $d$ is in very
striking contrast to the behavior of the grand-canonical pairing
parameters $\Delta_s (d)$ shown in \Fig{fig:pairing-parameter}(a).

%To conclude this section, let us point out that the contrast between
%extensive and practically intensive behavior of $E^\cond_{s}$ in the
%superconducting ($d \ll \tilde \Delta$) versus the FD ($d \gg \tilde
%\Delta$) regimes is a nice illustration of Anderson's famous dictum
%for strongly correlated systems, namely that ``more is different'': in
%the former regime (which is strongly correlated), but not the latter
%(which is not), adding \emph{more} particles gives a \emph{different}
%condensation energy.

\section{Effect of level statistics}
\label{sec:sc-level-statistics}

\begin{figure}[t]
\centerline{\epsfig{figure=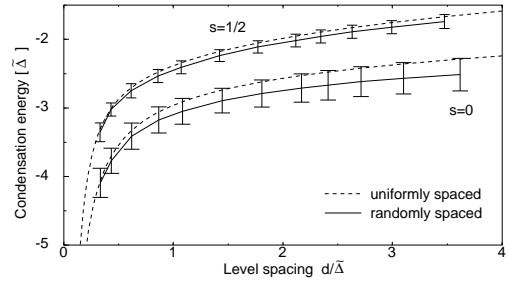,width=0.5\linewidth}} 
\caption{Exact even and odd condensation energies
  $E^\cond_s$ [in units of $\tilde \Delta$] for equally spaced levels
  (dashed line), and the ensemble-average $\langle E^\cond_s \rangle$
  for randomly-spaced levels (solid line) \cite{sierra99}. The height
  of the fluctuation bars gives the variances $\delta E^\cond_s$.}
\label{fig:randomly-spaced-levels}
\end{figure} 

Smith and Ambegaokar investigated the effect of level statistics on
the crossover between the bulk and FD regimes \cite{smith96}.  In
contrast to the uniform level spacing used in previous works, they
employed a random spacing of levels, distributed according to the
gaussian orthogonal ensemble.  Using a \gc\ mean-field BCS approach,
they found, interestingly, that \emph{randomness enhances pairing
  correlations:} compared to uniform spacings (u.s.)
\cite{vondelft96}, it (i) on average {\em lowers}\/ the condensation
energy $E^\cond_{s}$ to more negative values, $\langle E_s^\cond
\rangle < E_s^\cond \mbox{(u.s.)}$, but (ii) these still are parity
dependent, $\langle E_0^\cond \rangle < \langle E_{1/2}^\cond
\rangle$. These results can readily be understood intuitively: pairing
correlations become stronger the higher the density of levels around
$\eF$, where pair-mixing costs the least energy.  When determining the
amount of pairing correlations for a set of randomly-spaced levels,
fluctuations that increase the level density near $\eF$ are thus
weighted more than those that decrease it, so that randomness enhances
pairing correlations.

Although the \gc\ mean-field treatment of Smith and Ambegaokar breaks
down for mean level spacings much larger than $\tilde \Delta$, just as
was the case in \cite{vondelft96,braun97,braun99}, their main
conclusions (i) and (ii) are robust. Indeed, these were recently
confirmed by Sierra \etalia\ \cite{sierra99}, who used Richardson's
exact solution to calculate $E^\cond_{s}$ for ensembles of random
levels [\Fig{fig:randomly-spaced-levels}].  Moreover, they found that
the blocking effect responsible for (ii) manifests itself in the
fluctuations too, which likewise are parity dependent: for example,
\Fig{fig:randomly-spaced-levels} shows that both the variances $\delta
E^\cond_s \equiv [{\langle(E^\cond_s)^2\rangle - \langle
E^\cond_s\rangle ^2 } ]^{1/2}$ and the randomness-induced changes in
condensation energies $|\langle E^\cond_s \rangle - E^\cond_s {\rm
(u.c)}|$ were larger for even than for odd grains.

\section{Finite temperature parity effects} 
\label{sec:finite-T}

Although finite-temperature studies of the \dbcsm\ are not of direct
relevance for spectroscopic measurements of the BRT-type (a finite $T$
would simply smear out the discrete spectra, thereby blurring their
most interesting features), they are important in their own right for
extending our understanding of superconductivity in ultrasmall grains.
We hence review several recent finite-$T$ developments below.

To begin, let us note that parity effects are of course not restricted
to the $T = 0$ limit discussed so far. To be observable
\vphantom{\cite{Tuominen-92,Tuominen-93,Tinkham-95,Saclay,eiles93}}
\cite{Tuominen-92}-\cite{eiles93}, they only
require the temperature to be smaller than the free energy difference
$\delta {\cal F} \simeq \tilde \Delta - \kB T \ln[ N_{\rm eff} (T)]$
between an odd and even grain. Here $N_{\rm eff} (T)$ is the
effective number of states available for quasiparticle excitations at
temperature $T$, and for $d \ll \tilde \Delta$ is given by 
$N_{\rm eff} (T) = \sqrt{8
  \pi T \tilde \Delta /d^2}$ \cite{Tuominen-92}.  Below the
corresponding crossover temperature where $\delta {\cal F} = 0$,
determined by $\kB T^\ast_{\rm
  cr} = \tilde \Delta \ln[ N_{\rm eff} (T^\ast_{\rm cr})]$ and roughly
equal to $\tilde \Delta / \ln \sqrt{8 \pi \tilde \Delta^2 /d^2} $, the
single unpaired electron begins to matter: it causes a crossover from
$e$-periodicty to $2e$-periodicity in the $I$-$V$ characteristics of
mesoscopic superconducing SET's 
\cite{Tuominen-92,Tuominen-93,Tinkham-95,Saclay,eiles93}, due to
the ground state energy difference ${\cal E}_{1/2} - {\cal E}_{0}
\simeq \tilde \Delta$.  Since $T^\ast_{\rm cr}$ becomes of order
$\tilde \Delta$ in nanoscopic grains with $d \simeq \tilde \Delta$,
parity effects should survive to temperatures as high as the (bulk)
superconducting transition temperature $\Tc$ itself.

Regrettably, the canonical methods discussed in the preceding sections
become impractical at finite temperatures, since the number of states
that need to be considered increases rapidly for $T \gtrsim d, \tilde
\Delta$. On the other hand, \gc\ finite-$T$ methods, some of which we
review below, are, in principle, inherently unreliable for $d \gtrsim
\tilde \Delta$.  This applies in particular to the simplest of these,
parity-projected mean-field theory \cite{Janko-94,Golubev-94}
(\Sec{sec:sc-parity-projection}) and certain variational
generalizations thereof \cite{balian-short,balian-long}
(\Sec{sec:balian}): they yield the same sharp phase transition as
function of temperature for finite systems as for bulk systems,
whereas on general grounds no sharp transitions are possible in finite
systems. The reason for this problem is that they neglect fluctuations
in the order parameter, which become very important in the transition
region.  The sharp transition is smoothed out once fluctuations are
included. A rather efficient way of doing this is the so-called static
path approximation (\Sec{sec:SPA}). Its use is illustrated in
\Sec{sec:sc-susceptibility} for a calculation of the spin
susceptibility, which shows an interesting parity effect that should
be measurable in ensembles of ultrasmall grains.

\subsection{Parity-projected mean-field theory}
\label{sec:sc-parity-projection}

The simplest finite-$T$ approach that is able to keep track of parity
effects is parity-projected mean-field theory, first used in nuclear
physics by Tanabe, Tanabe and Mang \cite{Tanabe-81}, and,
indepedently, introduced to the condensed-matter community by Jank\'o,
Smith and Ambegaokar \cite{Janko-94} and Golubev and Zaikin
\cite{Golubev-94}.
% (who apparently were unaware of its prior use in 
%nuclear physics  \cite{Tanabe-81}.) 
One
projects the \gc\ partition function exactly onto a subspace of Fock
space containing only even or odd $(p=0,1)$ numbers of particles,
using the parity-projector $\hat P_p$:
\begin{eqnarray}
  \label{eq:parity-projection}
      Z_p^{\MF}  \equiv  \mbox{Tr}^\MF \hat P_p
      e^{- \beta (\hat H  - \mu_p \hat N)} \; , \qquad
      \hat P_{0,1} \equiv {\textstyle {1 \over 2}} [1 \pm (-1)^{\hat N}]
      \; .
\end{eqnarray}
One then makes the mean-field replacement $ b_j \to \{ b_j - \langle b_j
\rangle_p \} + \langle b_j \rangle_p \; , $ neglects terms quadratic in the
fluctuations represented by $\{ \quad \}$, and diagonalizes $\hat H$ in terms
of the Bogoljubov quasiparticle operators $\gamma_{j \sigma}$ of
\Eq{eq:Bogoljubov}. The self-consistency condition $\Delta_{p/2} \equiv
\lambda d \sum_j \langle b^\ds_j \rangle_p$, evaluated in a parity-projected
\gc\ ensemble according to \Eq{eq:parity-projection}, leads to a gap equation
of the standard form,
\begin{equation}
\label{gap}
   {1 \over \lambda} = d  \sum_{|\varepsilon_j| < \omegaD }
   {1 \over 2 E_{ j}} \left( 1 - \sum_\sigma f_{p j \sigma} \right) \; ,
 \quad E_{j} \equiv \sqrt{(\varepsilon_j - \mu)^2 
+ |\Delta_{p/2}|^2} \; , 
\end{equation}
which is parity-dependent, via the occupation function $ f_{p j\sigma}
= \langle \gamma_{j \sigma}^\dagger \gamma_{j \sigma} \rangle_p $ for
quasiparticles. Since their number parity is restricted to be $p$,
$f_{pj\sigma}$ differs from the usual Fermi function $f^0_{j\sigma}$.
The condition $2n + p = \langle \hat N\rangle_p$ fixes the chemical
potential $\mu$ to lie exactly half-way between the last filled and
first empty levels if $p=0$, and exactly on the singly-occupied level
if $p=1$, implying $\mu = 0$ in both cases [by \Eq{eq:ejs}].

 \begin{figure}[t]
\centerline{\epsfig{figure=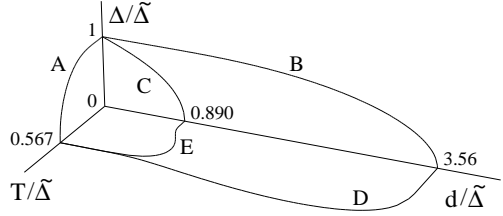,width=0.5\linewidth}}
\caption{\label{fig:T=0}
  $d$-and $T$-dependence of the pairing parameter $\Delta_{p/2}
  (d,T)$, calculated using parity-projected mean-field theory
  \cite{vondelft96}.  Curve A gives the bulk gap $\Delta (0,T)$, with
  $\Delta(0,0) \equiv \tilde \Delta$; curves B-E give $\Delta_{p/2}
  (d, T) / \tilde \Delta$ as a function of $d/\tilde \Delta$ and
  $T/\tilde \Delta$ for $p=0$ (B, D) and $p=1$ (C, E).  The critical
  spacings $d^\BCS_0 = 3.56 \tilde \Delta$ and $d^\BCS_{1/2} = 0.890
  \tilde \Delta$ given here differ somewhat from those in
  \Fig{fig:pairing-parameter}(a), because the present mean-field
  approach differs in minor details (via terms that vanish when $d\to
  0$) from the variational approach of \Sec{sec:generalnumerics}.
  \label{fig:3D}}
\end{figure}

von Delft \etalia\ \cite{vondelft96} applied this approach to the
\dbcsm\ with uniformly-spaced levels, and solved \Eq{gap} for the
parity parameter $\Delta_{p/2} (d,T)$ as function of both level
spacing and temperature.  \Fig{fig:3D} summarizes their results. At
zero temperature, $\Delta_{p/2} (d,0)$ corresponds to the spin-dependent
parity parameters $\Delta_{s= p/2}$ discussed in
\Sec{sec:generalnumerics} [cf. \Fig{fig:pairing-parameter}(a)], and
drops to zero at a critical level spacing $d^\BCS_{s}$.  The
$\Delta_{p/2} \to 0$ limit of Eq.~(\ref{gap}) defines the
parity-dependent ``critical temperature'' $T_{c,p} (d)$, which can be
viewed as another measure of how rapidly pair-mixing correlations
break down as function of level spacing (although ultrasmall grains of
course cannot undergo a sharp thermodynamic phase transition, which
can only occur if $n \to \infty$).  In both the even and odd cases,
the behavior of $T_{c,P} (d)$ shows direct traces of the parity
projection:

In the even case, $T_{c,0} (d)$ 
[\Fig{fig:3D}, curve D] is non-monotonic as function of
increasing $d$, initially increasing
slightly before dropping to zero very rapidly as $d \to d^\BCS_{0}$.
The intuitive reason for the initial increase is that the difference
between the actual and usual quasiparticle occupation functions is
$f_{pj\sigma} - f_{j \sigma}^0 < 0$ for an even grain (becoming
significant when $d \simeq \tilde \Delta$), reflecting the fact that
exciting quasiparticles two at a time is more difficult than one at a
time. Therefore the quasiparticle-induced weakening
of pairing correlations with increasing $T$ will set in at slightly higher
$T$ if $d \simeq \tilde \Delta$.

In the odd case, the critical level spacing $d_{1/2}^\BCS (T)$ 
[\Fig{fig:3D}, curve E] is
non-monotonic as a function of increasing $T$, first increasing to a
maximum before beginning to decrease toward $d^\BCS_{1/2} (T_c) = 0$.
The intuitive reason for this is that for $0 < \Delta_{1/2} \ll T,d$,
the odd $j=0$ function $f_{p0\sigma} (T)$ becomes somewhat smaller
than its $T=0$ value of $1/2$, because with increasing $T$
some of the probability for finding a quasiparticle in state $j$
``leaks'' from $j=0$ to higher states with $j \neq 0$, for which
$E^{-1}_{j} < E^{-1}_{0}$ in Eq.~(\ref{gap}). Thus, the
blocking-of-pair-scattering effect of the odd quasiparticle becomes
slightly less dramatic as $T$ is increased, so that $d^\BCS_{1/2}$
increases slightly.

It should be noted, however, that although the non-monotonicities of
$T_{c,0} (d)$ and $d_{1/2}^\BCS (T)$ are intuitively plausible within the
\gc\ framework in which they were derived, their physical significance
is doubtful, since they fall in the regime where $d / \Delta_s \gtrsim
1$ and the \gc\ approach is unreliable, due to its neglect of
fluctuations.
% This also applies
% to the generalization of the above results to finite magnetic fields
% by B$\!\not$\,onsager and MacDonald \cite{bonsager98}, using
% parity-projected mean-field theory.

\subsection{Variational extensions of BCS theory}
\label{sec:balian}

The above-mentioned results of von Delft \etalia\ \cite{vondelft96}
were reproduced and extended to finite magnetic fields by Balian,
Flocard and V\'en\'eroni, using a more general \gc\ variational BCS
approach \cite{balian-short,balian-long}.  It is designed to optimize
the characteristic function $\varphi (\xi) \equiv \ln \mbox{Tr} \hat
P_p \eer^{-\beta (\hat H - \mu \hat N)} \hat A (\xi)$, where $\hat
P_p$ is the parity projector of \Eq{eq:parity-projection} and $\hat
A(\xi) \equiv \exp (- \sum_\gamma \xi_\gamma \hat Q_\gamma)$, $\hat
Q_\gamma$ being observables of interest (\eg\ the total spin) and
$\xi_\gamma$ the associated sources (\eg\ the magnetic field).  This
approach goes beyond the usual minimization of the free energy
\cite{BCS-57}, since it optimizes not only thermodynamic quantities
but also equilibrium correlation functions, which can be obtained by
differentiating $\eer^{\varphi (\xi)}$ with respect to $\xi_\gamma$.
However, its \gc\ version also suffers from the drawback of yielding
abrupt, spurious phase transitions even though the systems are finite.
Presumably this problem would be cured if an exact projection to fixed
particle number were incorporated into this approach, but this is
technically difficult and has not yet been worked out.

\subsection{Static path approximation}
\label{sec:SPA}

\begin{figure}
\centerline{\epsfig{figure=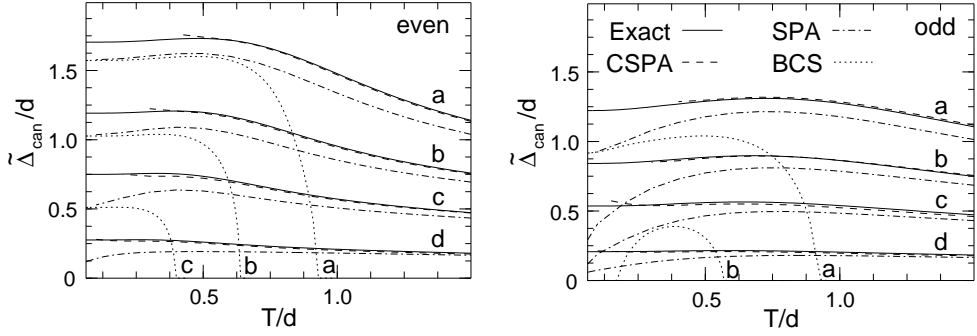,%
width=0.98\linewidth}}  
\caption[Pairing correlation energy at finite temperatures.]%
{Temperature dependence of the pairing correlation energy $\tilde
  \Delta_\can$ of \Eq{eq:Delta-rossignoli} [in units of level spacing
  $d$], as calculated in \cite{Rossignoli-99a} using parity-projected
  mean-field BCS theory (dotted lines), SPA (dash-dotted), CSPA (dashed)
  and exact diagonalization (solid lines).  A system of 10
  equally-spaced, doubly-degenerate, levels was studied, containing 10
  (left panel) or 11 (right panel) electrons.  $\tilde \Delta_\can
  (T)$ is shown at four fixed values of $d/\tilde \Delta$ (thus this
  figure elaborates \Fig{fig:3D}), namely 0.60, 0.91, 1.7, and 15,
  labeled by a,b,c and d, respectively (calculated using $\lambda =$
  0.55, 0.45, 0.35 and 0.2). CSPA data are shown only above the CPSA
  breakdown, which occurs at $T_{\rm CSPA} < \tilde \Delta/4$ for the
    cases considered.  The absence of dotted lines for the cases d
    (even) and c,d (odd) means that for these no nontrivial mean-field
    BCS solution exists. Of course, $(\tilde \Delta_\can)_{\rm exact}$ is
    nonzero nevertheless.  The abrupt BCS transition is completely
    smeared out for the SPA, CSPA and exact results, for which the
    asymptotic decay at $T \gg \tilde \Delta$ can be shown
    \cite{Rossignoli-99a} to be $\tilde \Delta_\can \sim \frac14
    (\lambda^3 d/T)^{1/2}$.}
\label{fig:rossignoli}
\end{figure}

For finite systems, in contrast to infinite ones, \emph{fluctuations}
of the order parameter about its mean-field value are very important
in the critical regime, causing the phase transition to be smeared
out; conversely, the spurious sharp transition found in the \gc\ 
approaches above is a direct consequence of the neglect of such
fluctuations.
%To remedy the occurence of sharp phase transitions in finite systems,
%fluctuations of the order parameter, which become important in the
%critical region, must be included. 
A rather successful way of including fluctuations is the so-called
\emph{static path approximation} (SPA), pioneered by M{\"u}hlschlegel,
Scalapino and Denton \cite{Muehlschlegel-72} and developed 
by various nuclear theorists 
\vphantom{\cite{Alhassid-84,Alhassid-92,Arve-88,Lauritzen-88,%
Rossignoli-92,Rossignoli-93,Rossignoli-94,Puddu-90,%
Lauritzen-90,Puddu-91,Puddu-92,Attias97,Rossignoli-97a,Rossignoli-97b,%
Rossignoli-95,Rossignoli-96a,Rossignoli-96b}}
\cite{Alhassid-84}-\cite{Rossignoli-96b},
while recently an exact parity projection has also been incorporated
\vphantom{\cite{Rossignoli-98,Rossignoli-99a,Rossignoli-99b,Rossignoli-00}}
\cite{Rossignoli-98}-\cite{Rossignoli-00}.  A
detailed and general discussion, including a complete list of relevant
references, was given very recently by Rossignoli, Canosa and Ring
\cite{Rossignoli-99a}. We therefore confine ourselves below to stating
the main strategies of the SPA and illustrating its capabilities by
showing its results [\Fig{fig:rossignoli}] for the quantity
\begin{eqnarray}
  \label{eq:Delta-rossignoli}
  \tilde \Delta_\can^2 = (\lambda d)^2
\sum_{ij} \left[ C_{ij} - (C_{ij})_{\lambda=0}\right] .
\end{eqnarray}
$\tilde \Delta_\can$ is reminiscent\footnote{The definitions of
  $\tilde \Delta_\can$ and $\Delta_\can$ differ by terms of order
  $(d/\omega_D)^{1/2}$; for example, when evaluating both
  using $|\BCS\rangle$ of \Eq{eq:BCSground} and comparing to
  $\Delta_\MF$ of \Eq{eq:BCS-gap}, one finds $( \Delta_\can)_\BCS =
  \Delta_\MF = (\tilde \Delta_\can)_\BCS + {\cal
    O}[(d/\omega_D)^{1/2}]$.}
%one finds  $(\tilde
%  \Delta_\can)_\BCS$ differs from $\Delta_\MF$ of \Eq{eq:BCS-gap by
%  terms of order $(d/\omega_D)^{1/2}$, whereas $( \Delta_\can)_\BCS =
%  \Delta_\MF$} of 
of $\Delta_\can$ of \Eq{eq:canonical-order-parameter}, and 
measures the increase in pairing correlation energy due to a nonzero
coupling strength $\lambda$.

One starts by decoupling the quartic interaction $H_\red$ of
\Eq{eq:hamiltonian} into a quadratic form using a Hubbard-Stratonovich
transformation with a complex auxiliary field $\Delta (\tau) =
\Delta^+ (\tau) + \iir \Delta^- (\tau)$, with  Matsubara-expansion 
$\Delta^\pm (\tau) = \sum_{n} \Delta^\pm_{n} \eer^{\iir 2 \pi n \tau /
  \beta}$ in the interval $\tau \in (0,\beta)$. The parity-projected
partition function of \Eq{eq:parity-projection}  then 
has the following path integral representation
(our notation is deliberately schematic; see \cite{Rossignoli-99a}
for a precise version): 
\begin{eqnarray}
  \label{eq:hubbard-stratonovich}
  Z_p & \propto &  \int \! \! \prod_n { \ddr} \Delta^+_n
{\ddr} \Delta^-_n \; {\cal Z}_p [\Delta] ,  \quad
{\cal Z}_p [\Delta] = 
%\int {\cal D} \left[ \Delta^+, \Delta^- \right] 
%    \eer^{- \beta |\Delta^2|/\lambda d} 
\mbox{Tr} \left\{ \hat P_p \hat {\cal T}
    \eer^{- \int_0^\beta {\ddr} \tau \, h[ \Delta(\tau)] } \right\}  , 
\\
  h[\Delta]  & = & 
  \sum_{j \sigma} (\varepsilon_j - \mu - \lambda d/2)
  c^\dagger_{j\sigma} c^\ds_{j \sigma}
  - \sum_j (b^\dagger_j \Delta + \Delta^\ast b^\ds_j) + 
  {|\Delta|^2 \over \lambda d} \; . 
\end{eqnarray}
The path integral can be treated at several levels of
sophistication:

(i) In the simplest, one uses a ``fixed-phase saddle-point
approximation'' for the ``static'' $n=0$ modes and neglects all
$n \neq 0$ modes, 
\ie\ one fixes the phase of $\Delta_0^+ + \iir \Delta_0^- =
|\Delta_0| \eer^{\iir \phi_0} $ by, say, setting $\phi_0=0$, so that 
$\int \!  {\ddr} \Delta^+_{0} {\ddr} \Delta^-_{0}$ is replaced by
$\int {\ddr} |\Delta_0|$,
%$|\Delta_0| = \Delta_0^+$, $\Delta_0^- = 0$), 
and approximates this integral by its saddle-point value.  The
saddle-point condition for maximizing ${\cal Z}_p[|\Delta_0|]$ then
yields the gap equation (\ref{gap}), thus this approach simply
reproduces the \emph{parity-projected mean-field} approach of
\Sec{sec:sc-parity-projection}, including its sharp phase transition
(\Fig{fig:rossignoli}, dotted lines).

(ii) The next-best approximation is obtained if one writes $\int {\ddr}
\Delta^+_{0} {\ddr} \Delta^-_{0}  = \int_0^{2 \pi} {\ddr} \phi_{0}$
$\int_0^\infty |\Delta_{0}|  {\ddr} |\Delta_{0}| $ and performs the phase
integral fully. Remarkably, ``liberating'' the phase degree of freedom
in this way already suffices to smooth out the phase transition
\cite{Rossignoli-95,Rossignoli-99a}, even if the $\int \ddr |\Delta_0 |$
integral is again replaced by its saddle-point value, provided that
the latter is found by now maximizing $|\Delta_0| {\cal Z}_p
[|\Delta_0|]$ (\ie\ including the factor $|\Delta_0|$ from the
integration measure).  This yields a modified gap equation with a
nontrivial solution for arbitrarily large $T$, \ie\ no abrupt
transition.

(iii) For finite systems, fluctuations about the saddle become large
in critical regions. To obtain an improved description of the latter
(\Fig{fig:rossignoli}, dash-dotted lines), the \emph{static path
  approximation} (SPA) 
\cite{Muehlschlegel-72},
\cite{Alhassid-84}-%,Alhassid-92,Arve-88,Lauritzen-88,%
%Rossignoli-92,Rossignoli-93,
\cite{Rossignoli-94} 
incorporates all
\emph{static} fluctuations exactly, via a (numerical) evaluation of
the full integral $\int_0^\infty |\Delta_{0}| \ddr |\Delta_{0}|$ $
\int_0^{2 \pi} \ddr \phi_{0}$ over all ``static paths''.

(iv) In the so-called \emph{correlated static path approximation}
(CSPA) (also called SPA\-+\-RPA), small-amplitude quantum fluctuations
around each static path are included too, by performing the remaining
$\int \!  \ddr \Delta^\pm_{n \neq 0}$ integrals in
the gaussian approximation 
\vphantom{\cite{Puddu-90,Lauritzen-90,%
Puddu-91,Puddu-92,Attias97,Rossignoli-97a,%
  Rossignoli-97b,Rossignoli-98,Rossignoli-99a,Rossignoli-99b,%
Rossignoli-00}}
\cite{Puddu-90}-\cite{Rossignoli-97b}, 
\cite{Rossignoli-98}-\cite{Rossignoli-00}. The CSPA yields
qualitatively similar but quantitatively more reliable results 
(\Fig{fig:rossignoli}, dashed lines) than
the SPA, but breaks down below a temperature $T_{\rm CSPA}$, below
which the fluctuations of the $\Delta^\pm_{n \neq 0}$ modes become
large at unstable values of $|\Delta_{0}|$, causing the gaussian
approximation to fail.

(v) Finally, in the so-called \emph{canonical CSPA} one projects the
partition function not only to fixed number parity (as done throughout
above) but also to fixed particle number, by performing an integration
over the chemical potential (\emph{before} performing any of the
$\Delta_n^\pm$ integrals)
\cite{Essebag-93,Rossignoli-92,Rossignoli-93,Rossignoli-94}.  However,
this too is usually done only in the gaussian approximation (and would
produce negligible corrections to the CPSA results for the quantities
shown in \Fig{fig:rossignoli}).

Comparisons with exact diagonalization results \cite{Rossignoli-96a}
(\Fig{fig:rossignoli}, solid lines) show that in its regime of
validity $(T > T_{\rm CSPA}$), the CSPA produces results that are
qualitatively completely similar and also quantitatively very close to
the exact ones, whereas the quantitative agreement is significantly
worse if only the SPA is used.  Since the CSPA is conceptually simple,
well-documented \cite{Rossignoli-99a} and straightforward to
implement, it seems to be the method of choice for not too low
temperatures.  A possible alternative is a quantum Monte Carlo
evaluation of the path integral (\ref{eq:hubbard-stratonovich})
\cite{Lang93,koonin}, but the numerics is much more demanding than for
the CPSA, while the convergence at low $T$ is in general rather poor,
due to the familiar sign problem of Monte Carlo methods.

The development of canonical finite-$T$ methods that remain
quantitatively reliable for $d \gtrsim \tilde \Delta$ and arbitrarily
small $T$ is one of the open challenges in this field. It would be
very interesting if progress in this direction could be made by
exploiting the integrability \cite{cambiaggio,sierra99b} of the model,
using Bethe Ansatz techniques. For the FD regime, another possibility
would be to develop a finite-$T$ generalization of the self-consistent
RPA approach of Dukelsky and Schuck \cite{dukelsky99c}.

\subsection{Re-entrant spin susceptibility}
\label{sec:sc-susceptibility}

For grains so small that $d \gg \tilde \Delta$, the spectroscopic
\emph{transport} measurements of BRT are not able, in principle, to
reliably detect the effect of pairing correlations, since in this
regime these cause only small changes to the eigenspectrum of a normal
metallic grain, whose spectrum is, however, irregular to begin with.
In contrast, \emph{thermodynamic} quantities do have the potential to
measurably reveal the existence of pairing correlations for $d \gg
\tilde \Delta$. Since very recently parity effects for the \emph{spin
  susceptibility} have been observed experimentally for an ensemble of
small, normal metallic grains \cite{Volotikin-96}, it is an
interesting and experimentally relevant question to investigate how
pairing correlations affect its behavior in superconducting grains.

\begin{figure}
\centerline{\epsfig{figure=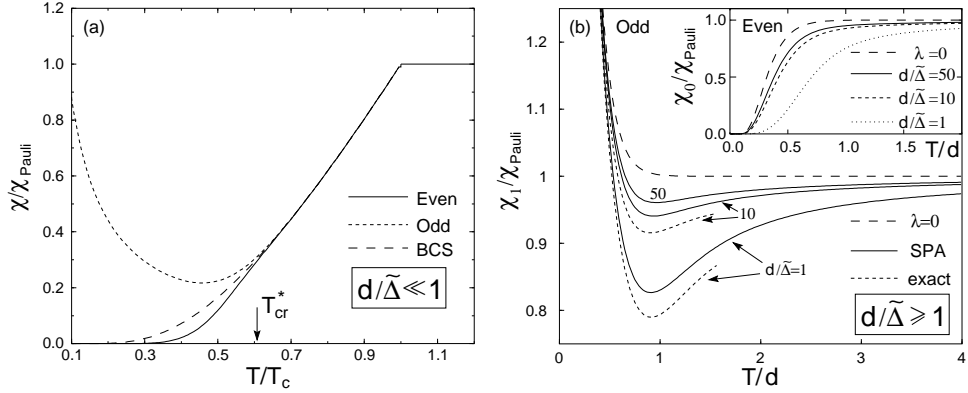,%
width=0.98\linewidth}}
\caption[Spin susceptibility of an even and odd superconducting grain]%
{Spin susceptibility $\chi_0$ ($\chi_1$) of an even (odd)
  superconducting grain as function of $T$, plotted in units of its
  bulk high-$T$ value $\chi_{\rm Pauli} = 2 \muB^2/d$.  (a) Bulk limit
  ($d/\tilde \Delta \ll 1$): the even (solid) and odd (short-dashed)
  curves were calculated using parity-projected mean-field theory, the
  long-dashed curve using standard  (unprojected) BCS theory.  (b)
  Fluctuation-dominated regime ($d/\tilde \Delta \gtrsim 1$) for
  $\chi_1$ (main figure) and $\chi_0$ (inset): All curves were
  calculated using the static path approximation, except the
  short-dashed curves in the main figure, calculated using
  Richardson's exact solution, and the long-dashed curves for the
  non-interacting case ($\lambda = 0$). 
\label{fig:susceptibilites}}
\end{figure}

This question was worked out in detail by Di Lorenzo \etalia\ 
\cite{dilorenzo99}, whose results are summarized in
\Fig{fig:susceptibilites}.  The spin susceptibility for an isolated
grain is defined as
\begin{equation}
  \label{eq:susceptibility}
  \chi_p (T) = - \left. {\partial^2 {\cal F}_p (T,H) \over \partial H^2}
  \right|_{H=0} \; , 
\end{equation}
where ${\cal F}_p = - \kB T \ln Z_p^\can$ is the free energy of a
grain with parity $p$ and $Z_p^\can$ is the canonical partition
function.

In the bulk limit [\Fig{fig:susceptibilites}(a)], it is well-known
\cite{Yosida-58} that the spin susceptibility decreases below its
Pauli value $\chi_\Pauli = 2 \muB^2/d$ once $T$ drops below the
superconducting transition temperature $T_{\rm c}$, since the
electrons tend to bind into Cooper pairs, which are spin singlets and
do not contribute to the spin susceptibility.  Interestingly, however,
the spin susceptibility becomes parity-dependent as $T$ is lowered
below the crossover temperature $T^\ast_{\rm cr}$ mentioned in the
opening paragraphs of \Sec{sec:finite-T}: In the even case, $\chi_0$
exponentially drops to zero for sufficiently small temperatures, $T
\ll \max(\tilde \Delta,d)$, for reasons that are intuitively obvious
in the two limits $\tilde \Delta \gg d$ (all electrons bound into
Cooper pairs) and $\tilde \Delta \ll d$ (no Cooper pairs, but all
levels doubly occupied). In contrast, in the odd case $\chi_1$ shows a
\emph{re-entrant} behavior, in that it increases as $\muB^2/T$ for low
temperatures, due to a Curie-like contribution from the unpaired odd
electron. As a result, $\chi_1 (T)$ has a \emph{minimum} somewhat
below $T^\ast_{\rm cr}$, which can be viewed as a ``smoking gun'' for
pairing correlations, since it is absent for odd normal grains.  For
these, $\chi_1 (T)$ [long-dashed $\lambda=0$ curve in
\Fig{fig:susceptibilites}(b)] also has the Curie-like increase at
very low $T$, but lacks the initial pairing-induced decrease as $T$ is
reduced below $T_{\rm c}$.

Remarkably, Di Lorenzo \etalia\ found that \emph{this reentrance of
  $\chi_1$ survives also for $d \gtrsim \tilde \Delta$}
[\Fig{fig:susceptibilites}(b)]: although pairing correlations survive
here only as fluctuations, these are evidently sufficiently strong to
still significantly reduce $\chi_1 (T)$ relative to $\chi_\Pauli$ [by
several percent even for $d /\tilde \Delta \simeq 50$ (!)], before the
Curie-like increase sets in at low $T$.  Di Lorenzo \etalia\ 
established this result by considering the limits $T \gg d$ and $T \ll
2 d$ analytically, using a static path approximation to capture the
crossover numerically, and checking the results for $T \lesssim d$
using Richardson's exact solution (they considered all eigenstates
with excitation energy up to a cutoff $\Lambda \sim 40 d$, for grains
with $N \le 100$ electrons).  This check shows that the static path
approximation somewhat underestimates the amount of pairing
correlations (its minima for $\chi_1 (T)$ are too shallow), but in
general is in good qualitative agreement with the exact results,
confirming that it is a useful and qualitatively reliable tool for
describing the crossover regime.

\section{Summary and outlook}
\label{sec:summary}

The technique of single-electron-tunneling spectroscopy on ultrasmall
metallic grains, applied to Al grains, has proved to be a very
fruitful way of probing electron pairing correlations, and the way in
which these are modified by level discreteness.  It has, in
particular, inspired theoretical attempts to quantitatively understand
how pairing correlations change during the crossover from the bulk the
limit of a few electrons.  Let us briefly summarize the main
conclusions reached in the preceding sections:

Part I: For largish Al grains, the observation of a distinct spectral
gap in even grains and its absence in odd grains is clear evidence for
the presence of \emph{superconducting pairing correlations}.  These
can be satisfactorily described using the simple \dbcsm\ introduced in
\Sec{sec:model}.  The blocking of some levels by unpaired electrons
leads to various measurable parity effects; among these, a
pairbreaking-energy parity effect should be observable in experiments
of the present kind, provided the grain size can be better controlled.
In ultrasmall grains, the effect of a magnetic field on orbital motion
is negligible.  The dominant mechanism by which a magnetic field
destroys pairing correlations in ultrasmall grains is Pauli
paramagnetism.  Decreasing the grain size softens the first-order
transition observed for thin films in a parallel field, by reducing
the number of spins flipped from being ma\-cros\-copically large for
$d \ll \tilde \Delta$ to being of order one for $d \simeq \tilde
\Delta$.  The grand-canonical variational BCS approach fails for $d
\gtrsim \tilde \Delta$; nevertheless, it yields a useful framework for
a qualitative analysis of the experiments, which had $d \lesssim
\tilde \Delta$.

Part II: The \emph{crossover} of the behavior of superconducting
pairing correlations from the bulk limit $(d \ll \tilde \Delta)$ to
the fluctuation-dominated regime $(d \gg \tilde \Delta)$ is parity
dependent and completely smooth. This remains true for systems with
non-uniform rather than uniform level spacings. -- Very remarkably,
the \dbcsm\ has an exact solution, due to Richardson, with which $T=0$
properties can be calculated rather easily.  Finite-temperature
properties for finite-sized systems can be calculated quite reliably
with the correlated static-path approximation (provided $T> T_{\rm
  CSPA}$).  However, the development of canonical finite-$T$ methods
that remain quantitatively reliable for $d \gtrsim \tilde \Delta$ is
still an open problem.  -- The spin susceptibility $\chi (T)$ of an
odd grain shows an interesting reentrant behavior even for $d \gg
\tilde \Delta$, which might be a way to detect remnants of pairing
correlations in the fluctuation-dominated regime.

Finally, we would like to mention two further aspects of pairing correlations
in ultrasmall metallic grains that have been discussed in the literature: the
effect on the measured excitation spectrum of nonequilibrium excitations
\cite{rbt97,agam97b}, and of spin-orbit interactions \cite{salinas99}.  These
topics have not been included here for lack of space, but have been reviewed
in detail in sections 6.2 and 7.6 of \Ref{PR-VDR}.

\emph{Prospects for future work:} 

\emph{Experiment:}
In the current generation of experiments, the grain's actual size and
shape cannot be determined very accurately.  It would be a great
advance if fabrication techniques could be developed to the point that
grains can be used which have been custom-made, by chemical
techniques, to have well-defined sizes and shapes (\eg\ spherical).
This would significantly reduce the uncertainties which one presently
encounters when estimating characteristic parameters of the grain,
such as the single-particle mean level spacing $d$
or the Thouless energy. Moreover, it would allow
systematic studies of the dependence of various quantities on grain
size or mean level spacing [for example, it would be interesting to
try to do this for the pairbreaking energies $\Omega_e, \Omega_o$ of
\Fig{fig:spectral-gap}(b)].  Encouragingly, the feasibility of using
chemically-prepared grains in SETs has already been demonstrated
several times \cite{Klein-97,sunmurray,Schmid97}, though the
resulting devices have not yet been used for single-electron-tunneling
spectroscopy.

\emph{Theory:} The behavior of superconducting pairing correlations in
an individual ultrasmall grain can now be regarded as a subject that
is well understood.  It would be interesting to try to use the
insights that have been gained for a single grain in order to now
study systems of several coupled grains: what, for example, is the
fate of the Josephson effect between two coupled grains as their sizes
are reduced to the point that $d \sim \tilde \Delta$?

\vspace*{0.25cm} \baselineskip=10pt{\small \noindent
It is a pleasure to thank all my colleagues
who actively collaborted in my research  on various aspects of 
the physics of ultrasmall grains:
C. Dobler,
S. Kleff,
J. Kroha,
M. Pirmann, 
G. Sch\"on,
W. Thimm, 
W. Tichy, and 
A. Zaikin
and in particular
F. Braun, all in Karlsruhe,
 and  
J. Dukelsky, 
G.  Dussel, 
D. Golubev, 
D. Ralph,
M. Schechter, 
G. Sierra, and
M. Tinkham 
from other parts of the world. 
I owe a great deal to D. Ralph for generous amounts of enthusiam, expertise
and unpublished data on ultrasmall grains, and for critically reading various
drafts of the entire review.
I also thank 
F. Braun,
P. Brouwer,
D. Davidovi\'c,
M. Greiter, 
R. Rossignoli,
M. Schechter,
G. Sch\"on, 
X. Waintal, 
for reading and commenting on various parts of the draft. 
Furthermore I acknowledge  helpful and stimulating discussions with
O. Agam,
I. Aleiner,
B. Altshuler,
V. Ambegaokar,
S. Bahcall,
C. Black,
Y. Blanter,
S. B\"ocker,
P. Brouwer,
C. Bruder,
T. Costi,
M. Deshmuck,
F. Evers,
G. Falci, 
R. Fazio,
P. Fulde,
A. Garg,
L. Glazman,
D. Golubev,
S. Gu\'eron,
B. Halperin,
B. Janko, 
J. K\"onig,
K. Likharev,
A. Mastellone,
K. Matveev,
A. Mirlin,
E. Myers,
Y. Oreg,
T. Pohjola, 
A. Rosch,
J. Siewert,
R. Smith,
M. Tinkham,
F. Wilhelm,
N. Wingreen,
P. W\"olfle,
G. Zar\'and,
F. Zawadowski,
and W. Zwerger. 
This work was supported in part by the Deutsche Forschungsgesellschaft through
Sonderforschungsbereich 195. I also acknowledge support from the
DFG-Pro\-gram, ``Semiconductor and Metallic Clusters'', and from the DAAD-NSF.
I thank the editors of Physics Reports and Annalen der Physik for the
permission to publish the present review, which is an exerpt of \Ref{PR-VDR},
separately, thereby enabling 
me to fulfil the habilitation requirements at the
University of Karlsruhe.
%
%The writing of this review would have taken twice the time it did, had it not
%been for the unwavering and loving support, in unimaginably numerous ways, of
%Nina and Lea. I thank them deeply.  
}

\end{document}